\documentclass{book}
\pdfoutput=1
\usepackage{geometry}
\usepackage{tensor}

\usepackage{amsthm}
\usepackage{amsmath}
\usepackage{amssymb}
\usepackage[english, shorthands=off]{babel}
\usepackage{natbib}
\usepackage{array}
\usepackage{physics}

\usepackage[bookmarks=True]{hyperref}
\usepackage{orcidlink}

\usepackage{xcolor}
\usepackage[savemem]{listings}
\definecolor{notebookbackground}{RGB}{220,220,220}

\lstdefinestyle{in}{
    language=Mathematica,
    basicstyle=\ttfamily\bfseries,
    frame=single,
    backgroundcolor=\color{notebookbackground},
    breaklines=true,
    mathescape
}

\lstdefinestyle{out}{
    language=Mathematica,
    basicstyle=\ttfamily,
    frame=single,
    breaklines=true,
    mathescape
}

\theoremstyle{definition}

\newtheorem{definition}{Definition}
\newtheorem{theorem}{Theorem}[chapter]
\newtheorem{example}{Example}[chapter]
\newtheorem{exercise}{Exercise}[chapter]

\theoremstyle{remark}
\newtheorem{remark}{Remark}

\newcommand{\e}{\mathbf{e}}
\newcommand{\Z}{\mathbb{Z}}
\newcommand{\Scal}{\mathcal{S}}
\newcommand{\half}{\frac{1}{2}}
\newcommand{\ihalf}{\frac{i}{2}}
\newcommand{\rthalf}{\frac{1}{\sqrt{2}}}
\newcommand{\irthalf}{\frac{i}{\sqrt{2}}}
\newcommand{\sutwo}{\emph{SU(2)}}
\newcommand{\suthr}{\emph{SU(3)}}
\newcommand{\code}[1]{\texttt{\textbf{#1}}}
\newcommand{\qtr}{\frac{1}{4}}
\newcommand{\thrqtr}{\frac{3}{4}}
\newcommand{\rtthrqtr}{\frac{\sqrt{3}}{4}}
\newcommand{\rtthrhalf}{\frac{\sqrt{3}}{2}}

\title{\Huge \textbf{Group Theory in Physics: An Introduction with Mathematica}}

\author{\Large{Balasubramanian {Ananthanarayan}\orcidlink{0000-0001-5955-2123}, Souradeep {Das}\orcidlink{0009-0006-3276-551X},  Amitabha {Lahiri}\orcidlink{0000-0001-8113-6345}}, \\ \vspace{0.1cm} \\ \Large{Suhas {Sheikh}\orcidlink{0000-0002-8290-9968}, Sarthak {Talukdar}\orcidlink{0000-0002-7127-2922}}\thanks{Corresponding Author}}

\begin{document}

\maketitle
\chapter*{Preface}

Over the years, Group Theory and its applications have become of immense value and use in the physics community. The history is now nearly a century old, with major contributions coming from theoretical physicists and fruitful interaction with mathematicians.  In the recent past, a large number of books have been written that introduce the subject.  Despite that, the subject remains quite forbidding to the beginning student.  In the recent past, Mathematica has become a widespread tool for research and education. While several packages and built-in tools have been developed in Mathematica, especially for Group Theory, these efforts remain too scattered to be used for educational purposes. Our efforts have been directed towards developing a concrete source that will help the beginning students understand the concepts of Group Theory and its applications from a physicist's perspective, as well as build familiarity with the symbolic language.  We have tried to put together several example notebooks based on the theory that is well-known and reproduce it in brief with many example notebooks that demonstrate specific concepts and can be, with minor modifications, used by the curious student.  Our work is in accordance with the recent resolutions of the EPJ publishers and the many collaborating European physics societies to bring out educational materials in a comprehensive and useful form.

\tableofcontents
\chapter*{About the Authors}
\chaptermark{About the Authors}
\addcontentsline{toc}{chapter}{About the Authors}

\textbf{B Ananthanarayanan: }
Ananthanarayan is a professor at the Centre for High Energy Physics, Indian Institute of Science, and an experienced teacher and researcher. \\
\textbf{Souradeep Das: }
At the time of writing this book, Souradeep was a 4th year undergraduate at the Indian Institute of Science. He works mainly on Astrophysics and Cosmology.\\
\textbf{Amitabha Lahiri: }
Amitabha is a senior professor at the S.N. Bose National Centre for Basic Sciences. He does research on theoretical physics and has nearly 30 years of graduate teaching experience.\\
\textbf{Suhas Sheikh: }
Suhas is a PhD student at the University of Illinois, Urbana-Champaign. He is broadly interested in Theoretical Condensed Matter Physics. \\
\textbf{Sarthak Talukdar: }
While this book was being written, Sarthak was a 4th-year Undergraduate student at the Indian Institute of Science. He broadly works on Quantum Gravity, Conformal Field Theory, and String Theory. Currently, he is a PhD student at King's College London.

\chapter*{How to Read This Book}
\chaptermark{How to Read This Book}
\addcontentsline{toc}{chapter}{How to Read This Book}

This document can be used as an introduction to the very vast subject of group theory. It is intended for Undergraduates and Graduates in Physics or Applied Physics who intend to do research in theoretical Condensed Matter or High Energy Physics. It is more of a course material and well-organized reference list rather than a complete book. The main objective is to make students realize the importance of Mathematica in modern theoretical physics so that they will be able to use it in their own research. \par
We do not claim the theory in this book to be an original work. We have used Mathematica to demonstrate the already existing works in standard books. Below, we have listed the main references we used to write chapters in this book. It is highly recommended for students to look at the references to have a complete understanding of the subject: 
\begin{itemize}
    \item For the chapters on finite groups, we have followed the book by \cite{Ramond} as a reference for the theory. We have further borrowed notation and examples from the lecture notes by \cite{Ramadevi}.
    \item For the Compact Lie group part, the main reference we have followed is the book by \cite{Cahn}. Some of the exercises are done here using Mathematica. It is advised to students to do the rest by themselves. For a quicker introduction to this subject, students may also refer to the book by Ramond. 
    \item For the Non-compact group part, we have mainly followed the book by \cite{Tung}. However, the content in this document is very little introduction to non-compact groups. There is a very large scope of further studies that is left for self-study. Students can learn about Conformal groups, Super-algebra, and Supersymmetry from the book of Ramond. Moreover, there are solved exercises at the end of the parts on Compact groups and Non-compact groups, mostly from the book by \cite{pollatsek2009lie} whose solutions are provided in the notebooks.
\end{itemize}
It is not possible to \textit{`refer'} to physics examples using group theory because it is used literally all over physics. A small demo can be seen in the book by Ramadevi and Dubey.\par
The reader can start reading this document with the GitHub repository open and run the relevant notebooks as soon as she/he comes across a particular code in this document. They should play around with the notebooks to get accustomed to the idea. They should also try to do the exercises mentioned in this document. \par
Students can be involved in group activities to create more content on the applicability of Mathematica in group theory and make them available to the community. One example topic for such can be \textit{Generalized Euler angles for different groups}. Refer to \cite{Cacciatori2009}.

\chapter*{List of Notebooks}
\chaptermark{List of Notebooks}
\addcontentsline{toc}{chapter}{List of Notebooks}
We provide a set of notebooks to help the readers understand the concepts in group theory. These notebooks are listed below:
\begin{enumerate}
    \item Finite Groups:
    \begin{enumerate}
        \item \texttt{Finite Groups.nb} -- Relevant to chapters 1 and 2
        \item \texttt{C4v Character Table.nb} -- Relevant to chapter 2
        \item \texttt{D3 Character Table.nb} -- Relevant to chapter 2
        \item \texttt{Q4 Character Table.nb} -- Relevant to chapter 2
        \item \texttt{T Character Table.nb} -- Relevant to chapter 2
        \item \texttt{Symmetric Group.nb} -- Relevant to chapter 3
        \item \texttt{DecomposeProductS3.nb} -- Relevant to chapter 3
    \end{enumerate}
    \item Compact Groups:
    \begin{enumerate}
        \item \texttt{Introduction to SU2.nb} -- Relevant to chapter 4, 7
        \item \texttt{SU2\_irrep.nb} -- Relevant to chapter 4
        \item \texttt{Introduction to SU3.nb} -- Relevant to chapter 5, 6
        \item \texttt{SU3\_products.nb} -- Relevant to chapter 5, 6
        \item \texttt{SU4\_Baryons.nb} -- Example Notebook
        \item \texttt{SU4\_Generators.nb} -- Example Notebook
        \item \texttt{so10\_groupmath\_branch\_rules\_slansky} -- Example Notebook 
        \item \texttt{so10\_cartan\_dynkin\_subalgebra} -- Example Notebook
        \item \texttt{so10\_lieart\_tensor\_prod\_combi} -- Example Notebook
    \end{enumerate}
    \item Non-compact Groups:
    \begin{enumerate}
        \item \texttt{Lorentz\_4D.nb} -- Relevant to chapter 11
        \item \texttt{Lorentz\_Irreps.nb} -- Relevant to chapter 11
        \item \texttt{Poincare.nb} -- Relevant to chapter 11
        \item \texttt{SL3C.nb} -- Relevant to chapter 11
    \end{enumerate}
\end{enumerate}

Please note that the above notebooks were compiled originally in Mathematica version 13.3. It is recommended to use the same version for convenience. 

We would like for the reader to have the relevant notebook open in front of them while going through a chapter. Some of the exercises are solved in the corresponding notebooks. We encourage the reader to learn the concepts from the interactive notebooks, and try solving the unsolved exercises on their own, using Mathematica.

The notebooks are available on GitHub at \url{https://github.com/iisc-ug-20/Group_Theory_for_Mathematica.git}

\chapter*{Introduction to Mathematica for Beginners}
\chaptermark{Introduction to Mathematica for Beginners}
\addcontentsline{toc}{chapter}{Introduction to Mathematica for Beginners}
This will be a naive introduction to the syntax of the language Mathematica. It is a symbolic manipulation language, which is similar to programming languages like Python or C++ in that it can be used to manipulate numbers or arrays of numbers. However, it has greater strengths in terms of the way it allows one to use human-readable analytical expressions instead of a fully numerical setup like say, Python. This is a great reason to use this language in a book like this, which focuses on theoretical development at par with the advances in modern technology.

In the chapters following this, the reader will be assumed to have ample knowledge of the basic syntax used in Mathematica, since most of the demonstrations will be done with the help of this language. Feel free to skip this chapter if you are already familiar with the language.

Let us introduce you to some of the terms and definitions associated with the Mathematica language, in a completely informal manner. Throughout this introductory chapter, we assume the reader has basic familiarity with any programming language. 

\begin{enumerate}
    \item \textbf{Notebook:} A file in Mathematica is called a notebook, and users usually make a single notebook to solve a particular problem or to demonstrate a bunch of closely related ideas. For example, you may choose to create a notebook to 
    \item \textbf{Cell:} It is the fundamental unit of operation in a notebook. It is a block of code. A cell usually contains a few lines of code, which are meant to be run together as a unit.
    \item \textbf{Types of cell:} A cell can be a text cell, which is used to display information besides the code. There are input cells where the code is typed in, and output cells which display the output generated by an input cell.
    \item \textbf{Suppressing an output:} Sometimes the output from an input is long and lacks presentability. It is useful to put a semicolon (;) at the end of that line in the input cell, which suppresses or hides the output of that line when the code is run.
    \item \textbf{Function:} A function in Mathematica is very similar to that in other programming languages. It is denoted by a specific symbol. When \textit{called} with a set of arguments (or parameters), it processes the arguments to produce a result. In Mathematica, the syntax is to put the arguments in square brackets following the name of the function. For example, \texttt{\textbf {Sqrt}} is a function that returns the square root of its argument. So, for the following input:
    \begin{lstlisting}[style=in]
Sqrt[36]
    \end{lstlisting}
    You will get this output:
    \begin{lstlisting}[style=out]
6
    \end{lstlisting}

    \item \textbf{User-defined function:} Mathematica allows users to write their own code for a custom function. The way a function is defined is pretty similar to how it is done in ordinary mathematics.
    \begin{lstlisting}[style=in]
MyFunction[x_]:=x^2-1
    \end{lstlisting}
    The function we defined above is named \texttt{\textbf{MyFunction}}. It takes an argument \texttt{\textit{x}} and returns the expression on the right-hand side: \texttt{\textit{x$^2$-1}}.

    Notice the syntax used in the above definition. On the left-hand side, the argument is stated with an underscore (\_) following it. The definition uses a colon before the equal sign (:=). Also note that this cell does not produce any output, because we haven't called this function yet. When called with a specific value for the input, it returns an output.
    \begin{lstlisting}[style=in]
MyFunction[11]
    \end{lstlisting}
    \begin{lstlisting}[style=out]
120
    \end{lstlisting}

    In fact, the argument does not have to be a numerical value. You can even put any valid symbol as the argument and get a result:
    \begin{lstlisting}[style=in]
MyFunction[p]
    \end{lstlisting}
    \begin{lstlisting}[style=out]
$p^2-1$
    \end{lstlisting}

    \item \textbf{Lists:} A list is an array of numbers. A list is bound by curly brackets, such as
    \[\{1, 4, 2, 7, 9\}\]
    One can create a list whose $i$-th element is known in terms of $i$, using the built-in function \texttt{\textbf{Table}}, as follows:
    \begin{lstlisting}[style=in]
list1=Table[i^3-i^2, {i, 1, 4}]
    \end{lstlisting}
    \begin{lstlisting}[style=out]
{0, 6, 18, 48}
    \end{lstlisting}
    The first argument is the expression for the $i$-th element in terms of $i$. The second element is a tuple, which denotes that $i$ goes from 1 to $n$. So this returns a list of $i^3-i^2$ for $i=1,2,3,4$. We could have used alternative notations for this tuple, such as \texttt{\textbf{ \{i, 4\} }} or \texttt{\textbf{ \{i, 1, 4, 1\} }}. In the former notation, the second argument 1 has been omitted, and Mathematica takes 1 as default for the seoncd argument. The latter notation has an extra argument (the 1 at the 4th position) which denotes the steps by which $i$ increments or decrements, which is 1 by default. That is, if it were replaced by say 0.5, the list would go as $1, 1.5, \ldots$. It could also be replaced by the tuple \texttt{\textbf{ \{i, list0\} }}, where \texttt{\textbf{list0}} is an arbitrary list of numbers (or symbols). This would return the value of $i^3-i^2$ for every element $i$ in \texttt{\textbf{list0}}.

    \item \textbf{Matrices:} A matrix is a nested list, or a list of lists. Hence it has two indices, the first denoting the row and the second, the column. The function \code{MatrixForm} can be used to display a matrix in the rectangular format. For example:
    \begin{lstlisting}[style=in]
mat={{1,0}, {2, 1}};
MatrixForm[mat]
    \end{lstlisting}
    \begin{lstlisting}[style=out]
$\pmqty{1&0\\2&1}$
    \end{lstlisting}
    Often the function \code{MatrixForm} is used in the suffix format, as \code{mat//MatrixForm} which is equivalent to \code{MatrixForm[mat]}. A matrix whose $(i,j)$-th element is known in terms of $i$ and $j$ can be created using the \code{Table} function. For example:
    \begin{lstlisting}[style=in]
Table[2i+j, {i, 1, 2}, {j, 1, 3}]//MatrixForm
    \end{lstlisting}
    \begin{lstlisting}[style=out]
$\pmqty{3&4&5\\5&6&7}$
    \end{lstlisting}
Note that the syntax now contains two tuples instead of 1, to denote the two indices. We could say that the nested list is of rank 2, as compared to an ordinary list, which has a rank of 1. Similarly one can build nested lists of higher rank (such as those used often to represent mathematical objects called tensors of higher rank) using the \code{Table} function.
\item \textbf{Module:} In Mathematica, a module is just like a function, in the sense that it takes arguments and returns a result, but has additional features compared to an ordinary function. It allows one to define local variables and execute multiple lines of code when the function is called, as opposed to a single line executed in a usual function. It is helpful in running a lengthy calculation at a function call. For example,
\begin{lstlisting}[style=in, language=Mathematica]
f[x_] := Module[{a = Log[x], b, c}, b = a x; c = b^2; x - c]
\end{lstlisting}
    Note the structure of the module. It contains two arguments separated by a comma. The first is a list of expressions or symbols where the variables local to the module are defined. These variables \code{a}, \code{b}, and \code{c} cannot be referred to outside the module. The second argument is a list of expressions separated by semicolons, where different steps of the calculation are performed. The last expression usually does not end with a semicolon, and only the result of the final line is returned as the output of the function.

    \item \textbf{Indices:} Indices are used to refer to a particular position in a list. For example, \code{list1[[2]]} refers to the 2nd item of the list \code{list1}. Note that indices in a list begin from 1 instead of the usual 0 in other programming languages. For the $(i,j)$-th element of a matrix, the syntax to use is \code{mat[[i, j]]}.

    \item \textbf{Non-alphanumeric symbols:} The basic arithmetic operators are very similar to that in other languages, with \code{+}, \code{-}, \code{*}, \code{/}, and \code{\^} denoting their usual actions. Please note that \code{//} is \textit{not} the modulo operator, rather the function \code{Mod[a,b]} is used for the modulo operation. \code{\%} returns the output of the last executed cell, and \code{\%3} for example returns the output of the cell in the notebook that is numbered as \texttt{Out[3]}. \code{\#} and \code{\&} are often used for defining functions in a shorthand notation. For example, 
    \begin{lstlisting}[style=in]
g = #^2 - # + 1 &;
g[9]
    \end{lstlisting}
    \begin{lstlisting}[style=out]
73
    \end{lstlisting}
    The first line defines the function \code{g} without using any arguments on the left-hand side, and using a simple \code{=} instead of \code{:=}. The right-hand side uses \code{\#} as a dummy variable (when there is more than one argument, they are denoted using \code{\#1}, \code{\#2} etc.) and the \code{\&} is used to mark the end of the definition. It is the same as defining $g(x)=x^2-x+1$ but without using the variable \code{x}.
\end{enumerate}

The above definitions are supposed to be a brief introduction for a beginner to the language. Please note that this chapter is not intended to be a complete tutorial on Mathematica. That purpose has to be served by specially designed professional tutorials that focus more on the structure, variations, errors, efficiency and other detailed aspects. What this chapter aims at is to help make the reader familiar with the code snippets that form an integral part of this text. For more details, readers are advised to learn Mathematica on their own from the official website\footnote{\url{https://www.wolfram.com/language/fast-introduction-for-math-students/en/}} and other tutorials available online. There are AI-based tools such as ChatGPT that too can help in understanding the syntax in case one gets stuck. We encourage the use of technology to get a better grasp of the concepts taught in the text.

\part{Finite Groups}

\chapter{Introduction to Finite Groups}\label{ch:group_basics}

\section{A brief review of Group Theory}

We begin our review by defining a group, the fundamental player in group theory.
\begin{definition}
A \textbf{Group} (say, $G$) is defined as a set equipped with a binary operation, called product, which has the following properties: 
\begin{enumerate}
    \item Closure: For all $a, b$ in $G$, $ab$ is in $G$
    \item Associativity: $(ab) c = a(bc)$ for all $a, b$, and $c$
    \item Existence of identity: There exists a unique element $e$ in $G$, such that $ae = ea = a$ for all $a$ in $G$
    \item Existence of inverse: For all $a$ in $G$, there exists a unique $b$ in $G$ such that $ab = e = ba$. This element $b$ is called the inverse of $a$.
\end{enumerate}
\end{definition}

A group is the association of the set forming the group and its associated `product'. Note that although the term `product' is used to refer to the group operation, it does not necessarily have to be an actual algebraic product of two numbers, as we shall see in our examples.

\begin{example}
    One of the easiest examples to understand is the group of two elements: $\{1, -1\}$. The group operation is the arithmetic product operation. It is easy to see that for every pair of elements chosen from this set, the product lies in the set as well. Similarly, the associativity property is satisfied. Here 1 is the identity element. The inverse of 1 is 1 itself - This property is true for the identity of any group. The inverse of $-1$ is also $-1$ itself. This group is often called $C_2$ or $\Z_2$.

    Now this group is physically equivalent to the flipping operation on a coin. Here the two group elements are: \textsc{do not flip}, and \textsc{flip}. The former is the identity element as it keeps the coin perfectly unchanged, while the latter is the element $-1$ which when applied twice, gives back the identity.
\end{example}

\begin{example}
    Let us now see an example of a set which is equipped with a binary operation, but is not a group. Consider the set of all integers, with the group operation being the multiplication of two integers. Then, although the set is closed under the operation, it is not a group since it violates the existence of an inverse for every element.
\end{example}

\begin{example}\label{ex:Z}
    Consider the set of all integers $\Z$, with the group operation being the addition operation on two integers. Again, the defining properties follow from the basic axioms of algebra on integers. The identity element of this group is 0, and the inverse of any element is its additive inverse, or negative. This group is called $\Z$.
\end{example}

\begin{example}
    Consider the set $\{0, 1, 2, 3, 4\}$, with the group operation being: Add the two numbers and get the remainder when divided by 5. It is easily seen that the set is closed under this operation, since the set itself contains all the possible remainders when an integer is divided by 5. Associativity again follows from the definition. 0 is the identity element and the inverses of other elements are easy to find - 1 and 4 are inverse of each other, and so are 2 and 3. This group is called $\Z_5$.

    This example can be seen in another light: Some groups can be viewed as a set of geometric operations. This particular example represents the rotations of a regular pentagon about its axis. Suppose you label the vertices of the pentagon as A, B, C, D, E. Now an element of the group corresponds to a rotation of the pentagon in the anticlockwise direction such that the vertices land exactly on to the previous position of some other (or the same) vertex. And as we rotate about the axis, we count the number of times the vertices coincided as we went on rotating. And this number corresponds to the group element in $\Z_5$. So $A\rightarrow B, B\rightarrow C,\ldots$ corresponds to 1, $A\rightarrow C, B\rightarrow D,\ldots$ corresponds to 2 and so on. You can verify how two such rotations can be combined to give another rotation. This has a one-to-one correspondence with the group discussed.
\end{example}

\begin{example}
    This example may not be that enlightening, but the singleton set $\{\e\}$ is also a group, whose operation is such that $\e\centerdot \e=\e$. This is the simplest group possible, since any group must contain at least the identity element.
\end{example}

If you take another look at the criteria in the definition of a group, you may find that it is very similar to the usual multiplication in arithmetic. Probably you will find a particular property missing (unless you are used to missing it, by virtue of studying topics like quantum mechanics), which \textit{does} hold true for ordinary multiplication of two numbers - commutativity, or the property that the order in which the two numbers appear on the two sides of the product operator does not change the result. But in group theory this property is not a necessity, and in fact the absence of commutation is what makes group theory much more interesting than ordinary arithmetic. In the above examples, the groups do obey commutativity, but that is just because they are a special kind of groups.

\begin{definition}[Abelian Group]
    A group $G$ such that for any two elements $a,b$ in $G$:
    \[a.b=b.a\]
    holds true is called an \textbf{Abelian group}.
\end{definition}
 All the examples we saw before were Abelian groups. But there are plenty of non-Abelian groups out there as well. They are more complex and richer than Abelian groups.
 \begin{example}
     Consider the group of all permutations of 3 distinguishable objects, say A, B, C. The group has $3!=6$ elements. Now consider the element $x$ which takes the first object to the second, second to the third, and third to the first. And consider the element $y$ that exchanges the first and second elements. So,
     \[x\centerdot\begin{pmatrix}A&B&C\end{pmatrix}=\begin{pmatrix}C&A&B\end{pmatrix}\]
     \[y\centerdot\begin{pmatrix}A&B&C\end{pmatrix}=\begin{pmatrix}B&A&C\end{pmatrix}\]
     Now check what happens if they are combined. Note that, whenever we write $x.y$ followed by the ordered tuple of objects to operate on, it means that we first operate on the object by $y$ and then by $x$ (order of proximity). So,
     \[xy\centerdot\begin{pmatrix}A&B&C\end{pmatrix}=\begin{pmatrix}C&B&A\end{pmatrix}\]
     \[yx\centerdot\begin{pmatrix}A&B&C\end{pmatrix}=\begin{pmatrix}A&C&B\end{pmatrix}\]
     Clearly, $xy\neq yx$. So the group of all permutations of three (or even more) objects is not an Abelian group. This group is called $\mathcal{S}_3$, the \textbf{Symmetric group} with index 3.
 \end{example}

 \begin{definition}[Finite groups]
     A group which has a finite number of elements is called a Finite Group. All the examples we saw above were finite groups, except for \autoref{ex:Z}.
 \end{definition}

In physics, we encounter finite as well as infinite groups on various occasions. For example, the rotations of a circular disk about its axis form an infinite group, while that of a square form a finite group. Again, infinite groups can be countable, like the group of integers under addition, or uncountable, such as the group of all real numbers under addition.
 
 \begin{definition}[Order of a group]
     The order of a finite group is defined as the number of elements in the group.
 \end{definition}
For a finite group, if we apply a group element repetitively, we get back the identity element at some point. This is easy to prove: Suppose $a$ is an element of the group, and we consider the sequence $\{a, a^2, a^3\ldots\}$. Here we have defined $a^2=a.a, a^3=a.a^2=a^2.a$ and so on. Now, if the group is finite, the sequence has to have repetitions at some point. Say the $m$-th and $n$-th terms in the sequence are the same ($m<n$). Then $a^{n-m}=\e$. This allows us to define the order of an element.
 \begin{definition}[Order of an element]
     The order of an element (in a finite group) is the lowest positive integer power to which the element must be raised to get the identity element.
 \end{definition}

 \begin{example}
     Again consider the group $\Z_5$. Here the order of each element (except the identity) is 5. For the group $\Z_2$, the order of $-1$ is 2. The order of $x$ in $\Scal_3$ is 3 while that of $y$ is 2. The order of the identity element is always 1.
 \end{example}

 Here we introduce the notion of multiplication tables. This represents the table showing the product of each possible pair of elements in a group. A noticeable property of a group multiplication table is that every element of the group appears exactly once in every row and exactly once in every column. Abelian groups look the same when their rows and columns are interchanged.

 \begin{example}\label{ex:S3}
     Here we will introduce a group defined by abstract symbols, without any explicit physical interpretation. The first element of the group is, of course, $\e$. Call the next non-trivial element $a$, and another element $b$. These two satisfy the following properties:
     \[a^3=\e,\qquad b^2=\e, \qquad ab=ba^2\]
     See that the order of the element $a$ is 3 while that of $b$ is 2. The group is non-Abelian. Now, if we multiply all possible combinations of $a$ and $b$ and their powers, keeping in mind the defining properties of this group, we get a total of 6 elements:
     \[G=\{\e, a, a^2, b, ab, a^2b\}\]
     It is a good exercise to obtain the multiplication table of this group yourself. In this, we will put the element on the row header at the left and the column header at the right. This gives the multiplication table as in \autoref{tab:multi-S3}.

     \begin{table}[h]
         \centering
         $\begin{array}{c|cccccc}
  & \e & a & a{}^2 & b  & ab  & a{}^2b  \\ \hline
 \e & \e & a & a^2 & b  & a b  & a^2 b  \\
 a & a & a^2 & \e & a b  & a^2 b  & b  \\
 a{}^2 & a^2 & \e & a & a^2 b  & b  & a b  \\
 b  & b  & a^2 b  & a b  & \e & a^2 & a \\
 ab  & a b  & b  & a^2 b  & a & \e & a^2 \\
 a{}^2b  & a^2 b  & a b  & b  & a^2 & a & \e \\
\end{array}$
         \caption{Multiplication table of the group described in \autoref{ex:S3}}
         \label{tab:multi-S3}
     \end{table}
 \end{example}

 \begin{exercise}
     Figure out which of the following are/aren't groups, and give a reason why:
         \begin{enumerate}
             \item The set of real numbers with addition as the group operation.
             \item The set of real numbers with multiplication as the group operation.
             \item The set of positive real numbers with multiplication as the group operation.
             \item The set of all permuting operations on a list of 5 objects (say A, B, C, D, E) such that C is always the first object.
         \end{enumerate}
 \end{exercise}

 \begin{exercise}\label{exc:multi-table}
     Write down the multiplication table of the following groups:
     \begin{enumerate}
         \item $\Z_3$
         \item $\Z_5$
         \item $\Scal_3$
     \end{enumerate}
     Notice how the multiplication table of $\Scal_3$ looks very similar to \autoref{tab:multi-S3}
 \end{exercise}

\section{Homomorphism and Representation}\label{sec:homomorph}
You noticed how the multiplication table of the group $G$ in \autoref{exc:multi-table} resembles that of the symmetric group $\Scal_3$. In fact, there exists a one-to-one correspondence between the elements of $\Scal_3$ and that of the group $G=\{\e, a, a^2, b, ab, a^2b\}$ under the given relations between $a$ and $b$. Such kind of correspondence is called a homomorphism between the two groups.

\begin{definition}[Homomorphism]
If two groups $G$ and $G'$ are related by a map $\phi:G\rightarrow G'$ such that $g_1 g_2=g_3$ in $G$ implies $\phi(g_1)\phi(g_2)=\phi(g_3)$ in $G'$, then $\phi$ is said to be a homomorphism from $G$ to $G'$.
 \end{definition}

 Note that the above definition is also applicable when the group operations of $G$ and $G'$ are quite different.

 \begin{definition}[Isomorphism]
     If a homomorphism $\phi:G\rightarrow G'$ is invertible, then it is called an isomorphism. The groups $G$ and $G'$ are said to be isomorphic to each other.
 \end{definition}

 \begin{exercise}
     It is claimed that the groups $(\mathbb{R}, +)$, \textit{i.e.,} the group of all real numbers under addition, and $(\mathbb{R}^+, \times)$, the group of all positive real numbers under multiplication, are isomorphic to each other. Find out one such isomorphism.
 \end{exercise}

 Now, we notice that a group element is often regarded as an operation on an entity, say a pentagon or a coin, for which there is a sense of direction, which the group element may change. This may remind you of matrices which act on columns (vectors) and cause a change. In fact, square matrices form a group under matrix multiplication, which can be proved easily. The proof is left as an exercise for you.

 We will use the space $Aut(V)$ in our definitions. Note that it denotes the space of automorphisms on $V$, which are maps from $V$ to itself. As long as we are dealing with finite groups, we will only consider linear automorphisms.

 \begin{definition}[Representation]
     A representation of a group $G$ is a homomorphism $\Gamma:G\to Aut(V)$, where $V$ is a linear space of dimension $n$, (hence matrices of $Aut(V)$ are of order $n\times n$) is called a representation of $G$. A representation is said to be \textit{faithful} if it is one-one, that is, no two distinct elements map to the same matrix. The dimension of the space $V$ corresponding to the representation is called the \textit{degree} of the representation.
 \end{definition}

\begin{example}
    Consider the group $\Z_2$ with elements $\{\e, a\}$ with $a^2=\e$. A representation in 2D of this group is:
    \[\e\rightarrow\begin{pmatrix} 1&0\\0&1\end{pmatrix}\equiv \mathbf{I_2}, \qquad a\rightarrow\begin{pmatrix} 0&1\\1&0\end{pmatrix}\equiv \sigma_x \]
    In any representation, the identity element always maps to the identity matrix of the correct dimension.
    
    Notice that mapping both the elements of the group to the identity matrix is also a homomorphism. This is a very general statement about groups - every group has a \textit{trivial} representation where every element is mapped to the identity matrix.  Of course, this representation is not faithful unless it is the singleton group.
\end{example}

\section{Subgroups}
\begin{definition}[Subgroup]
    Let $H$ be a subset of a group $G$. If $H$ is itself a group under the same multiplication operator as $G$, then $H$ is said to be a subgroup of $G$.
\end{definition}

\begin{exercise}\label{exc:subgr}
    Determine which of the following are subgroups of the group $G$ as defined in example \ref{ex:S3}:
    \begin{enumerate}
        \item $\{\e, a, a^2\}$
        \item $\{\e, b\}$
        \item $\{\e, a, b\}$
        \item $\{b, ab, ab^2\}$
    \end{enumerate}
\end{exercise}

\begin{example}\label{ex:mtmc1}
    We will now use Mathematica to check whether a subset of a group is a subgroup. For this, we will need to define a group in the notebook. Our definition is in the form of a set of matrices which form a representation of the group. Once we define the group, we set a map called \texttt{MatRep} that maps the element name to the corresponding matrix. Thus, putting $a$ into the function \texttt{MatRep} gives the matrix corresponding to $a$ in the output. We are not going to elaborate much on the definitions, these are available in the Mathematica notebooks as part of the online resources. The matrix representations that we used are as follows:
    \begin{lstlisting}[style=out]
        $\{e\to \begin{pmatrix}
 1 & 0 \\
 0 & 1 \\
\end{pmatrix},
C_3\to \begin{pmatrix}
 -\frac{1}{2} & -\frac{\sqrt{3}}{2} \\
 \frac{\sqrt{3}}{2} & -\frac{1}{2} \\
\end{pmatrix},
C_3^2\to \begin{pmatrix}
 -\frac{1}{2} & \frac{\sqrt{3}}{2} \\
 -\frac{\sqrt{3}}{2} & -\frac{1}{2} \\
\end{pmatrix}$,
$\sigma \to \begin{pmatrix}
 1 & 0 \\
 0 & -1 \\
\end{pmatrix},
C_3 \sigma \to \begin{pmatrix}
 -\frac{1}{2} & \frac{\sqrt{3}}{2} \\
 \frac{\sqrt{3}}{2} & \frac{1}{2} \\
\end{pmatrix},
C_3^2 \sigma \to \begin{pmatrix}
 -\frac{1}{2} & -\frac{\sqrt{3}}{2} \\
 -\frac{\sqrt{3}}{2} & \frac{1}{2} \\
\end{pmatrix}\}$
    \end{lstlisting}

Once we define the map, we can plug in any subset of the group into the following function and determine whether it is a subgroup:
    \begin{lstlisting}[style=in]
    (* Function to determine whether a subset of the group is a subgroup or not *) 
IsSubGroup[subset_] := 
 Module[{y = Flatten[Table[
      Table[MatRep[i] . MatRep[j] // MatrixForm, {i, subset}], {j, 
       subset}], 1]},
   SubsetQ[Table[ MatRep[i] // MatrixForm, {i, subset}], y] && Length[subset] != 0]
    \end{lstlisting}

    A few words on how this function works: First, the function iterates over all elements of the subset and generates a multiplication table of the subset. Now, if this subset is a group in itself, then every resulting element in the table must belong to the subset itself. So the module flattens\footnote{To flatten a table means converting it into a one-dimensional array or list. This function joins the rows of the matrix end-to-end to give a single list.} the table to look at every element and returns \texttt{True} if the flattened table is a subset of the subset that we put in as an argument.

    So, for the example of the group $G$ which is also called $\Scal_3$ or $C_{3v}$ depending on the context, we have a few examples. Here we have renamed $a$ to $C_3$ and $b$ to $\sigma$, the reason being geometrical - $a$ is very similar to a 120\textdegree ~rotation while $b$ is similar to a reflection. We will come to this analogy (in fact, this is another isomorphism) later.
    
    \begin{lstlisting}[style=in]
        IsSubGroup[{e, $\sigma$}]
    \end{lstlisting}
    \begin{lstlisting}[style=out]
        True
    \end{lstlisting}
    While the following is not a subgroup:

    \begin{lstlisting}[style=in]
        IsSubGroup[{e, $C_3$, $\sigma C_3$}]
    \end{lstlisting}
    \begin{lstlisting}[style=out]
        False
    \end{lstlisting}
\end{example}

Now, we can use this to find all the subgroups of the given group. First, we find a list of all the subsets of the group using Mathematica's inbuilt function \texttt{Subsets}, and then select subsets out of the list based on whether they are a subgroup of the group:

\begin{lstlisting}[style=in]
    SubSetList = Subsets[GroupNames];
    SubGroupList = Select[SubSetList, IsSubGroup[MatRep[#]] &]
\end{lstlisting}

\begin{lstlisting}[style=out]
    {{e}, {e, $\sigma$}, {e, $\sigma C_3$}, {e, $\sigma C_3^2$}, {e, $C_3$, $C_3^2$}, 
    {e, $C_3$, $C_3^2$, $\sigma$, $\sigma C_3$, $\sigma C_3^2$}}
\end{lstlisting}

\section{Normal Subgroups and Conjugacy Classes}

\begin{definition}[Normal Subgroup]
    Let $K$ be a subgroup of $G$. If, for every element $k\in K$ and every $g\in G$, $g^{-1}kg\in K$ holds true, then $K$ is said to be a normal subgroup of $G$.
\end{definition}
\begin{exercise}
    Find out which are normal subgroups out of the subgroups of the group $G$ for each of the sets in \autoref{exc:subgr}.
\end{exercise}

\begin{example}\label{ex:mtmc2}
    We continue on the example of the group $C_{3v}$ in Mathematica. We will define a function \texttt{IsNormalSubGroup} that detects whether a subset is a normal subgroup of the group.
    \begin{lstlisting}[style=in]
        (*Function to determine whether a subset is a normal subgroup*)
IsNormalSubGroup[subset_] := 
 Module[{y = Flatten[Table[
      Table[i . j . Inverse[i] // MatrixForm, {i, Group}], {j, subset}], 1]},
   IsSubGroup[subset] && SubsetQ[Table[ i // MatrixForm, {i, subset}], y]]
    \end{lstlisting}
    This function computes the set $i.j.i^{-1}$ for every element $i$ in the parent group, and every element $j$ in the subset. The table is then flattened, and then it checks whether the resulting set is still a subset of the original subset we started with. Added to that, we do the additional check of whether the subset is actually a subgroup.

    This time, we directly find out the list of all normal subgroups of the group $C_{3v}$ using the \texttt{SubGroupList} we defined earlier in \autoref{ex:mtmc1}:
    \begin{lstlisting}[style=in]
        NormalSubGroupList = Select[SubGroupList, IsNormalSubGroup[MatRep[#]] &]
    \end{lstlisting}
    \begin{lstlisting}[style=out]
        {{e}, {e, $C_3$, $C_3^2$}, {e, $C_3$, $C_3^2$, $\sigma$, $\sigma C_3$, $\sigma C_3^2$}}
    \end{lstlisting}
    
\end{example}

\begin{definition}[Conjugacy Class]
    The conjugacy class $C(x)$ of any element $x$ in a group $G$ is defined as the set of all $g^{-1}xg$ for all $g\in G$.
    \[y\in C(x) \iff y=g^{-1}xg ~\mathrm{for~some}~g\in G\]

    When $y$ belongs to the conjugacy class of $x$, we also say that $y$ is conjugate to $x$.
\end{definition}

\begin{exercise}
    Prove that conjugacy is an equivalent relation.
\end{exercise}
\begin{exercise}
    Find out all the normal subgroups of the group $G$ in \autoref{ex:S3}. Also find out the conjugacy classes of each of its elements.
\end{exercise}

\begin{example}
    We will continue the demonstration of \autoref{ex:mtmc2}. We define the following function to return the conjugacy class corresponding to a particular element of the group:
     \begin{lstlisting}[style=in]
 (* Function to return the conjugacy class containing the given element *)
     ConjClassNames[elementName_] :=  Table[(g . MatRep[elementName] . Inverse[g] // MatrixForm) /. Thread[Group -> GroupNames], {g, Group}] // DeleteDuplicates
 \end{lstlisting}
 This function generates the list of all elements of the form $g.x.g^{-1}$ where $x$ is the given element and $g$ is an element of the group, followed by a mapping from the matrix representation to the element names using the \texttt{Thread} function, and removal of repeated elements from the list.

 Here are a few examples of the implementation:

\begin{lstlisting}[style=in]
    ConjClassNames[e]
\end{lstlisting}
 \begin{lstlisting}[style=out]
      {e}
 \end{lstlisting}
 
\begin{lstlisting}[style=in]
    ConjClassNames[$C_3$]
\end{lstlisting}
 \begin{lstlisting}[style=out]
      {$C_3$, $C_3^2$}
 \end{lstlisting}

\begin{lstlisting}[style=in]
    ConjClassNames[$\sigma C_3$]
\end{lstlisting}
 \begin{lstlisting}[style=out]
      {$\sigma$, $\sigma C_3$, $\sigma C_3^2$}
 \end{lstlisting}

Suppose we want to find out all the conjugacy classes of a particular group. We just obtain the classes corresponding to each element, followed by clearing up, that is, sorting and removing repetitions.

\begin{lstlisting}[style=in]
    ConjugacyClassesNames = DeleteDuplicatesBy[Table[ConjClassNames[i], {i, GroupNames}], Sort]
\end{lstlisting}
\begin{lstlisting}[style=out]
    {{e}, {$C_3$, $C_3^2$}, {$\sigma$, $\sigma C_3$, $\sigma C_3^2$}}
\end{lstlisting}
 
\end{example}

We can already see that conjugacy classes divide the group into a bunch of disjoint sets. The identity element always forms a singleton conjugacy class. The proof is easy and left as an exercise. Also, notice how every non-null normal subgroup is a union of one or more conjugacy classes. That is, if one element from a conjugacy class belongs in a normal subgroup, it immediately implies that all other elements from the class must belong to the normal subgroup as well.

\begin{example}\label{ex:order}
    We can use a simple iterative program in Mathematica to obtain the order of an element in a group. The following function begins with the identity matrix and repetitively multiplies it by the matrix in hand until it reaches the identity again.

    \begin{lstlisting}[style=in]
        OrderGr[x_] := Module[{k = matE, n = 0, y = MatRep[x]},
  While [(k != matE || n == 0), k = k . y; n++];  n]
    \end{lstlisting}

    For example,
    \begin{lstlisting}[style=in]
    OrderGr[$\sigma C_3$]
\end{lstlisting}
\begin{lstlisting}[style=out]
    2
\end{lstlisting}
\begin{lstlisting}[style=in]
    OrderGr[$C_3^2$]
\end{lstlisting}
\begin{lstlisting}[style=out]
    3
\end{lstlisting}
\end{example}

\section{Direct Product}
\begin{definition}[Direct Product]\label{def:dirpr}
    A direct product of two groups $\mathcal{G}$ and $\mathcal{H}$, denoted by $\mathcal{G\times H}$, is a group whose elements are ordered pairs $(g, h)$ such that $g\in\mathcal{G}$ and $h\in\mathcal{H}$. In addition, their multiplication rule is as follows:
\begin{equation}
    (g_1, h_1).(g_1, h_2)=(g_1.g_2, h_1.h_2)
\end{equation}
\end{definition}
 An important point to note above is that the products $g_1.g_2$ and $h_1.h_2$ are performed using the multiplication rules of the groups $\mathcal{G}$ and $\mathcal{H}$ respectively, and these rules may be different from each other. Note that the two groups could be the same as well. For example, consider the product $\Z_2\times \Z_2$. Its elements are:
 \[\e=(\e,\e), \qquad p=(\e, a),\qquad q=(a, \e),\qquad pq=(a, a)=qp\]
 It is easy to see that $p^2=q^2=(pq)^2=\e$.

\chapter{Representations of Finite Groups}

In this chapter, we will deal with the properties of representations of finite groups, and shine some light on their importance in the physical context. We begin with a classification of representations into reducible and irreducible. Before that, we introduce a few tools we need to understand representations.

\section{Matrices and Direct Sums}
First we will define a similarity transformation that plays an important role in classifying representations. 
\begin{definition}
    A similarity transformation on a square matrix is defined as $\Psi: Aut(V)\to Aut(V)$, such that \[\Psi(M)=A^{-1}MA\]
    where $A\in Aut(V)$ is some invertible matrix.
\end{definition}
Note that $Aut(V)$ is the space of linear automorphisms on $V$, as defined earlier in \autoref{sec:homomorph}.

Similarity transformations have the property that they preserve both the trace and determinant of the matrix.

Now, we define a particular class of matrices that take a vector space to its subspace, or in other words, project a higher-dimensional vector onto a lower-dimensional space.
\begin{definition}
    A projection matrix (or operator) is an idempotent matrix (operator) whose non-zero eigenvectors span a subspace of the total vector space.

    Idempotency means that the matrix applied twice has the same effect as it applied once.
    \[P^2=P\]
\end{definition}

A very simple example of projection operators is an identity matrix with some 1s removed from the diagonal. It is easy to show that the only possible eigenvalues of an idempotent matrix are 0 and 1. A projection operator is associated with a subspace of the given vector space. When applied on any vector $\mathbf{a}$ on the vector space, it returns a vector $\mathbf{a}'$ which is obtained by subtracting the component of $\mathbf{a}$ in the orthogonal subspace of the projection operator. The subspace of a projection operator is the span of its eigenvectors, and the orthogonal subspace is obtained using methods such as Gram-Schmidt orthogonalization.

\begin{definition}
    A Block-diagonal matrix is of the form:
    \[\begin{pmatrix}
        A_1&0&0 &\ldots &0\\
        0&A_2&0&\ldots&0\\
        0&0&A_3&\ldots&0\\
        \vdots&\vdots&\vdots&\ddots&\vdots\\
        0&0&0&\ldots&A_n
    \end{pmatrix}\]
    where $A_1, A_2, \ldots A_n$ are square matrices and the zero entries are null matrices of the appropriate dimensions. The matrix is also expressed as $A_1\oplus A_2\oplus\ldots\oplus A_n$. The blocks can be viewed as subspaces of the linear space, and taking projections of the total matrix into the subspaces gives the individual blocks (with other rows and columns identically zero). To take the projection of a matrix $A$ into a subspace with projection operator $P$, we obtain $PAP$.
\end{definition}

This was all about matrices. Now we can relate them to groups in general. For groups, we can define a conjugation in an identical way as we define similarity transformation for matrices. Now, we define reducible representations.

\section{Reducible and Irreducible Representations}
\begin{definition}
    A representation is said to be reducible if each of the representative matrices can be expressed in non-trivial block-triangular form (with the same shape, that is, same number of block dimensions for each element).
\end{definition}

An important theorem called Maschke's theorem proves that any reducible representation can be put into block-diagonal form.

Now, whenever two block diagonal matrices of the same shape are multiplied, the different blocks multiply individually and the resulting matrix is a block-diagonal of the products of corresponding blocks, quite similar to how two diagonal matrices multiply element by element on the diagonal. This means that if a particular reducible representation $\Gamma:G\to Aut(V)$  breaks down as $\Gamma(g)=\Gamma_1(g)\oplus\Gamma_2(g)\oplus\ldots\oplus\Gamma_n(g)$, with $\Gamma_i:G\to Aut(V_i)$, where $V_i$ is the subspace of $V$ corresponding to the block $\Gamma_i(g)$, then each of the $\Gamma_i$ is also a representation of the same group $G$. Note here that $V_i$ has dimensions $N_i$ (which is the same for all $g$), such that $V=\bigoplus_{i=1}^n V_i$ and $\dim(V)=\sum_{i=1}^n N_i$. This is the reason this kind of representation is called reducible. It is important to remember that all matrices have the same block-diagonal forms, that is, the size of corresponding blocks are the same across the group.

Those representations which cannot be broken down into such a sum of more than one matrix is called an \textbf{irreducible representation} or \textit{irrep}. A more concrete way to define irreps is that they are representations that do not have a non-trivial \textit{invariant subspace}. To clarify the above, an invariant subspace is a subspace of $V$ that maps to itself under the transformation $\Gamma$. However, since the spaces ${0}$ or $V$ itself are trivially invariant under any linear transformation, we frame the definition excluding them from the criterion. 

The task of finding out all the irreps of a group is an important task with regards to both mathematical satisfaction and physical applications.

An important property about representations is the trace of the matrix. This is because this is preserved in similarity transformations, which are an isomorphism.

\begin{definition}
    The trace of the representing matrix of a group element under a particular representation is called the \textit{character} of the element under that representation.
\end{definition}

A few properties to note about representations:
\begin{enumerate}
    \item Similarity transformations preserve the character of an element. Since all elements in a conjugacy class are related by a similarity transformation, this means that the character of every element in a conjugacy class is the same for a given representation.
    \item The character of any element under a reducible representation is the sum of its characters under the smaller representations that add up to it.
    \item Any group must have at least one irreducible representation - the trivial representation in one dimension. The trivial representation can be expanded to more dimensions by simply using higher dimensional identity matrices, even though those would be reducible representations of the form $\bigoplus_{i=1}^{l} A$, where $A$ is the trivial representation.
    \item There are an infinite number of reducible representations of any group. Just take an arbitrary number of representations and get their direct sum to obtain one of them.
\end{enumerate}

\section{Character Tables and How to Find Them}\label{sec:char_table}

\begin{definition}
    The multiplicity of an irrep in a reducible representation is defined as the number of times the irrep occurs in the block-diagonal form of the reducible representation.
\end{definition}

The above definition requires more introspection. When we say `number of times' an irrep occurs, it means we have a way to distinguish an irrep from another. To make this stricter, we do not distinguish between two irreps if they are related by a similarity transformation. Thus, it is not exactly the matrices but the traces and dimensions of the matrices that distinguish between irreps. Again, traces are not a property of individual elements but of conjugacy classes. At this point, it is safe to assume that any irrep is uniquely defined by the set of characters corresponding to each conjugacy class, and the dimension of the irrep.

This brings us to the concept of a character table. A character table lists all the irreps of a finite group, along with their dimensions, and the characters corresponding to each conjugacy class. For example, the character table for the group $C_{3v}$ that we discussed in the previous chapter is given in\autoref{tab:char-C3v}. 

\begin{table}[ht]
    \centering
    \begin{tabular}{c|ccc}
\hline
                    & 1 $\e$ & 2 $C_3$ & 3 $\sigma$ \\ \hline
$A_1$ & 1   & 1       & 1          \\ \hline
$A_2$ & 1   & 1       & -1         \\ \hline
$E$   & 2   & -1      & 0          \\ \hline
\end{tabular}
    \caption{Character Table for the group $C_{3v}$}
    \label{tab:char-C3v}
\end{table}

In this table, the first row represents the trivial representation $A_1$,  where each element is represented by an identity matrix of degree 1. Thus, its trace is one for all elements. The second representation is also a 1-dimensional representation, but it differs from the trivial one in the sense that the three reflection elements $\sigma$, $\sigma C_3$, $\sigma C_3^2$ are represented by the number -1 (since it is a 1D matrix). The third representation has degree 2, and is named $E$. The identity element is represented by the $2\times2$ identity matrix, while $C_3$ and $C_3^2$ are matrices with trace $-1$, and the reflection elements have trace 0. In fact, it is not difficult to figure out these $2\times2$ matrices. It suffices to just find the matrices corresponding to the generators $\sigma$ and $C_3$, which satisfy the algebra of the group. There is actually a class of such representations all related by similarity transformations, but we focus on this one:
\[\e\to \begin{pmatrix}1&0\\0&1\end{pmatrix},\qquad C_3\to \begin{pmatrix}\frac{-1}{2}&\frac{-\sqrt{3}}{2}\\\frac{\sqrt{3}}{2}&\frac{-1}{2}\end{pmatrix},\qquad \sigma \to \begin{pmatrix}1&0\\0&-1\end{pmatrix}\]

The above example is more enlightening if we look at the vector spaces on which the representing matrices act. For the representation $A_2$, we look at the 1D vector space, and mark two points $A\equiv 1$ and $B\equiv -1$. Now, applying either $\e$ or $C_3$ keeps the point as it is, while $\sigma$ interchanges $A$ and $B$.

To get more insight into the 2D representation, consider an equilateral triangle with vertices $A\equiv (1,0)$, $B\equiv\qty(\frac{-1}{2}, \frac{\sqrt{3}}{2})$ and $C\equiv\qty(\frac{-1}{2}, \frac{-\sqrt{3}}{2})$. The matrix $C_3$ is related to the transformation $A\to B, B\to C, C\to A$, while $\sigma$ does $A\to A, B\to C, C\to B$. This demonstrates why $C_3$ is similar to a rotation and $\sigma$ to a reflection. It also gives brilliant insight into why the group $C_{3v}$ is isomorphic to the symmetric group $\Scal_3$. Clearly, all we are doing using the $E$ representation is permuting between 3 points on a triangle.

\subsection{How to get a character table}
Building a character table is easy if all the irreps of the group are known. But we do not need to do the tedious work of building up all the irreps from first principles. Schur's lemma comes to the rescue.

The exact statement of Schur's lemma is not important in the context of this text. Our requirement is satisfied by a very useful corollary of this lemma which states that matrices of different irreducible representations must be orthogonal to each other in the following sense: Say the two irreps are $\Gamma$ and $\Theta$, with degrees $l_\Gamma$ and $l_\Theta$ respectively. Then,
\begin{equation}
    \sum_{g\in G} \Gamma(g)_{ik} \overline{\Theta(g)}_{lm}=\frac{|G|}{l_\Gamma}\delta_{\Gamma\Theta}\delta_{il}\delta_{km}
\end{equation}
Here $\Gamma(g)_{ik}$ stands for the $i,k$-th element of the matrix $\Gamma(g)$, and the overbar denotes complex conjugation. Now, if we want to obtain a relation between their traces, we add over the diagonal elements of the matrices, that is, sum over $i$ with $i=k$ and over $l$ with $l=m$. Then we get:
\[\sum_{g\in G} \sum_i\sum_l\Gamma(g)_{ii} \overline{\Theta(g)}_{ll}=\frac{|G|}{l_\Gamma}\delta_{\Gamma\Theta}\sum_i\sum_l\delta_{il}\delta_{il}\]
The character of $g$ is the representation $\Gamma$ is denoted as $\chi^{\Gamma}(g)$. The sum on the right-hand side reduces since $\delta_{il}\delta_{il}=\delta_{il}$, and $\sum_l \delta_{il}=1$. When we do the second summation over the index $i$, the term getting added each time is just 1, so the sum is the number of values that $i$ takes. This number is $l_{\Gamma}$ since $i$ numbers the rows of the matrix $\Gamma(g)$ whose order is $l_{\Gamma}\times l_{\Gamma}$. So we get:
\begin{equation}\label{eq:ortho-char-element}
    \sum_{g\in G}\chi^{\Gamma}(g) \overline{\chi^{\Theta}(g)}=|G|\delta_{\Gamma\Theta}
\end{equation}
 
Now, we already saw that the character is the same for every element in a given conjugacy class. This means the term in the summation is identical for all elements $g$ in a particular conjugacy class $C$. So we can rather write the above expression:
\begin{equation}\label{eq:ortho-char}
    \sum_{C\subset G}n_C \chi^{\Gamma}(C) \overline{\chi^{\Theta}(C)}=|G|\delta_{\Gamma\Theta}
\end{equation}
where $n_C$ is the number of elements in the conjugacy class $C$, sometimes also referred to as its multiplicity in the group.

Now, the above statement puts a limit on the number of distinct irreducible representations a group can have. Suppose a group has $s$ conjugacy classes $C_1, C_2, \ldots C_s$. And it has $r$ distinct irreps $\Gamma_1, \Gamma_2 \ldots \Gamma_r$. Then consider the set of $r$ distinct $s$-vectors $X_i$ whose $j$-th component is given by:
\[X_i^j=\sqrt{\frac{n_{C_j}}{|G|}}\chi^{\Gamma_i}(C_j) ~,j=1,2,\ldots s\]
for every $i$ from 1 to $r$. Then \autoref{eq:ortho-char} states that the vectors $X_i$ form an orthonormal set. But we know that a vector space with $s$-dimensions cannot have more than $s$ mutually orthogonal vectors. This means that $r$ cannot be greater than $s$.

Another relation holds between the degrees $l_{\Gamma_i}$ of the irreducible representations of a finite group. Their sum of squares add up to the group order:
\[\sum_{i=1}^{r} (l_{\Gamma_i})^2=|G|\]
Moreover, each of these degrees must divide the group order. The most well-known proofs of this statement use algebraic number theory, which is beyond the scope of this book.\footnote{The proof can be found in \textit{Representation Theory} by Fulton and Harris, \textit{Linear representations of finite groups} by Serre, and \textit{Representations of finite and compact groups} by Simon among others.}

Additionally, at least one of these degrees must be 1, since the trivial representation is a valid representation of every group. We take it to be the first irrep by convention. 

Using these criteria we can first obtain the list containing degrees of the irreps of the group. Note that, we do not know the number of distinct irreps beforehand, but only after solving for the list of irrep degrees. So, for example, if we are trying to find the character table of $C_{3v}$, the list of degrees that satisfies the sum-of-squares and other criteria is found to be $(1,1,2)$. Note that we may not always find a unique list of degrees as we found here. In that case, we proceed to the next step to see if some of those lists can be eliminated using some other criteria.

\begin{example}
    We will use Mathematica to continue the example from the previous chapter and try to obtain the character table using basic properties derived from group theory and representation theory. Our first step will be to solve for the list of degrees of all possible irreps of the group.

    Recall from the previous chapter that we obtained the list of conjugacy classes:
    \begin{lstlisting}[style=in]
    ConjugacyClassesNames = DeleteDuplicatesBy[Table[ConjClassNames[i], {i, GroupNames}], Sort]
\end{lstlisting}
\begin{lstlisting}[style=out]
    {{e}, {$C_3$, $C_3^2$}, {$\sigma$, $\sigma C_3$, $\sigma C_3^2$}}
\end{lstlisting}

Now let us name the conjugacy classes according to their first element (or the generator):
\begin{lstlisting}[style=in]
    ConjNames = Table[class[[1]], {class, ConjugacyClassesNames}]
\end{lstlisting}
\begin{lstlisting}[style=out]
    {e, $C_3$, $\sigma$}
\end{lstlisting}

Then we obtain the multiplicities $n_C$ of each class:
\begin{lstlisting}[style=in]
    ConjMultiplicities = 
 Table[Length[class], {class, ConjugacyClassesNames}]
\end{lstlisting}
\begin{lstlisting}[style=out]
    {1, 2, 3}
\end{lstlisting}

Each element in a conjugacy class has the same order. Using a previously defined function to find the order of an element (See \autoref{ex:order} of the previous chapter), we obtain:
\begin{lstlisting}[style=in]
    Orders = 
 Table[OrderGr[MatRep[i[[1]]] // First // Normal], {i, 
   ConjugacyClassesNames}]
\end{lstlisting}
\begin{lstlisting}[style=out]
    {1, 3, 2}
\end{lstlisting}

To find the list of dimensions/degrees, we start with the set of possible dimensions, that is, what numbers to fill in the list of degrees. Such numbers must obey two rules - first, they must be divisors of the group order $N$, and second, each of them must not be greater than $\sqrt{N}$, since the squares of the degrees add up to $N$.

\begin{lstlisting}[style=in]
    PossibleDims = Select[Divisors[n], #^2 <= n &]
\end{lstlisting}
\begin{lstlisting}[style=out]
    {1, 2}
\end{lstlisting}

As we figured out already, the number of distinct irreps must not be greater than the number of conjugacy classes, which is labelled by $s$:
\begin{lstlisting}[style=in]
    s = Length[ConjNames]
\end{lstlisting}
\begin{lstlisting}[style=out]
    3
\end{lstlisting}

Now that we have the numbers to fill in the list, we actually try to find the list:
\begin{lstlisting}[style=in]
    PossibleDimSets = DeleteDuplicatesBy[
  Select[
  Flatten[Table[Tuples[PossibleDims, j], {j, 1, n}], 1],
  (Sum[i^2, {i, #}] == n && MemberQ[#, 1] && Length[#] <= s) &],
      Sort]
\end{lstlisting}
\begin{lstlisting}[style=out]
    {{1, 1, 2}}
\end{lstlisting}

The above code needs a bit of elaboration. It is easy to see from the inside out of the expression. In the third line, we use the \texttt{Tuples} function, which generates a list of all tuples (repetitions allowed) of length $j$ from the list \texttt{PossibleDims}. Here $j$ runs over 1 to $n$, giving rise to a list of list of tuples, which we flatten by one degree to get a list of tuples. In the second and fourth line, we use the \texttt{Select} function to select from these list of tuples, only those which obey the criteria we noted before - their squares must add up to $N$, the list must contain 1 (for the trivial irrep), and its length $r$ must be less than $s$. In the first and fifth lines, we remove permutations of the same list, such as $(1,2,1)$ and $(2,1,1)$.

Note that the output is a nested list. This is because there are other cases where the list of degrees is not uniquely determined by the above criteria, and we get multiple lists of degrees, which we then try to eliminate based on the criteria obtained in the next part of the solution. Similarly in those cases we obtain a list of possible values of $r$, although in this case the list is singleton.
\begin{lstlisting}[style=in]
    rList = Table[Length[tuple], {tuple, PossibleDimSets}]
\end{lstlisting}
\begin{lstlisting}[style=out]
    {3}
\end{lstlisting}
 \end{example}

Once we have the list of $l_{\Gamma_i}$ for $i=1,2 \ldots r$, we proceed to solve for the characters. We have a matrix of unknowns of degree $rs$. To find these unknowns, we solve \autoref{eq:ortho-char}, which is a set of $\frac{r(r-1)}{2}$ linear equations. However, we also keep in mind that not all the characters are really unknown. The first row is reserved for the trivial representation, whose character is 1 for every element. The first column is reserved for the identity element, which is always represented by an identity matrix, so its character in any representation $\Gamma$ is always $l_{\Gamma}$. Moreover, we have to pay special attention to other 1D representations, since they are just complex numbers with absolute value unity, and hence their characters are highly constrained. Additionally, the character of any element raised to its order in the group must give 1, which follows from the definition of order. Utilizing these constraints in most cases gives a unique character table (sometimes this uniqueness is up to some permutations between irreps of same degree or conjugacy classes of same multiplicity).

\begin{example}
    We will continue with the previous example to obtain the full character table of the group. Since we may get a number of different lists of degrees, it is useful to solve the character table for them one-by-one. So we define the following function:
    \begin{lstlisting}[style=in]
    ProduceCharTable[index_] := 
 Module[{xArr = Array[x, {rList[[index]], s}]},
  unitDim = 
   Select[Range[rList[[index]]], PossibleDimSets[[index, #]] == 1 &];
   equations = 
   Table[Sum[x[i1, j]*x[i2, j]*ConjMultiplicities[[j]], {j, 1, s}] == 
     n*KroneckerDelta[i1, i2], {i1, 1, rList[[index]]}, {i2, 1, 
     rList[[index]]} ];
  knowns1 = Thread[xArr[[1]] == 1];(* 
  All matrices in trivial irrep have trace 1*)
  knowns2 = 
   Thread[Transpose[xArr][[1]] == PossibleDimSets[[index]]]; (* 
  In all irreps, identity matrix has trace = dimension *)
  knowns3 = 
   Thread[Table[
     xArr[[ unitDim, t]]^Orders[[t]] == Table[1, {m, unitDim}] , {t, 
      1, s}]];(* In every 1D irrep, trace^order must be equal to 1 *)
  char = 
   Solve[Join[Flatten[equations], knowns1, knowns2, knowns3], 
    Flatten[xArr]];
  If[char == {}, char, xArr /. char]]
\end{lstlisting}

The above may look overwhelming, but it is easy to see that it is just the paragraph preceeding this example, translated into symbols. Our input is an index, which determines which of the possible lists of degrees we want to test out. In this case, it is obviously going to be put as 1 when we run the function.

We first define a variable \texttt{xArr} which is an array of variables, with dimensions $r\times s$. In the next line, \texttt{UnitDim} is the list of those irreps (or more accurately, the index of those irreps in the list) whose degree is 1.

The variable \texttt{equations} stores the constraint equations on the characters obtained from the orthogonality relation in \autoref{eq:ortho-char}. The three \texttt{knowns} variables state those characters which are already known from elementary rules. In the end, we solve the set of equations for \texttt{xArr} keeping in mind the knowns. The last line instructs the program to return a blank list if no solution is found. We can then run the function for all possible indexes in the list of $r$'s (here it runs only for index 1):
\begin{lstlisting}[style=in]
    CharTables =  Flatten[Table[ProduceCharTable[i], {i, Length[rList]}], 1]
\end{lstlisting}
\begin{lstlisting}[style=out]
    {{{1, 1, 1}, {1, 1, -1}, {2, -1, 0}}}
\end{lstlisting}
Note that here the matrix (or the table) is embedded inside a list, since there can be (but aren't there for this case) multiple forms of the same character table, as mentioned before.

\end{example}

\section{Decomposing Reducible Representations}
One of the main applications of character tables arises in the context of expressing a reducible representation as a direct sum of irreducible representations of the group. We use the orthogonality relations to obtain the multiplicity of any irrep in a reducible representation. To obtain the multiplicities, we use the property that the trace of a block diagonal matrix is the sum of traces of the individual blocks. This statement is trivial to prove. Thus, if a representation $\Gamma$ can be reduced into irreps $\Gamma_1, \Gamma_2, \ldots$ with multiplicities $m_1, m_2, \ldots$ respectively, then the characters are given by:
\begin{equation}\label{eq:block-diag-trace}
    \chi^\Gamma(g)=\sum_{i} m_i\chi^{\Gamma_i}(g)
\end{equation}
Now we use \autoref{eq:ortho-char-element} along with \autoref{eq:block-diag-trace} to obtain:
\begin{equation}\label{eq:multiplicity}
    \sum_{g\in G} \chi^{\Gamma}(g)\overline{\chi^{\Gamma_i}(g)}=|G|m_i
\end{equation}
The above equation can be used to determine the multiplicity of each irrep of the group in any representation. The base cases are when $\Gamma$ itself is irreducible, in which case the multiplicity of one of the irreps is one, the others zero, because of the orthogonality condition.
\begin{example}\label{ex:red-rep_S3}
    We look at an example of a reducible representation of the group $\Scal_3$. 

    Consider the representation given by:

    \begin{lstlisting}[style=in]
        RmatE = IdentityMatrix[3];
RmatC3 = {{5/8, -(Sqrt[3]/4), -((3 Sqrt[3])/8)}, {Sqrt[3]/4, -(1/2), 
    3/4}, {-((3 Sqrt[3])/8), -(3/4), -(1/8)}};
RmatSigma = {{-(1/2), 0, Sqrt[3]/2}, {0, -1, 0}, {Sqrt[3]/2, 0, 1/
    2}};
RmatC3sq = RmatC3 . RmatC3;
RmatC3Sigma = RmatC3 . RmatSigma;
RmatC3sqSigma = RmatC3sq . RmatSigma;
    \end{lstlisting}
    We embed all these matrices into a list as:
    \begin{lstlisting}[style=in]
        ReducibleRep = {RmatE, RmatC3, RmatC3sq, RmatSigma, RmatC3Sigma, 
   RmatC3sqSigma};
Table[GroupNames[[i]] -> MatrixForm[ReducibleRep[[i]]], {i, 1, n}]
    \end{lstlisting}
    \begin{lstlisting}[style=out]
        $\{\e\to \pmqty{
 1 & 0 & 0 \\
 0 & 1 & 0 \\
 0 & 0 & 1
}, C_3\to \pmqty{
 \frac{5}{8} & -\frac{\sqrt{3}}{4} & -\frac{3\sqrt{3}}{8}  \\
 \frac{\sqrt{3}}{4} & -\frac{1}{2} & \frac{3}{4} \\
 -\frac{3\sqrt{3}}{8} & -\frac{3}{4} & -\frac{1}{8} \\}
 C_3^2\to \pmqty{
 \frac{5}{8} & \frac{\sqrt{3}}{4} & -\frac{3\sqrt{3}}{8}  \\
 -\frac{\sqrt{3}}{4} & -\frac{1}{2} & -\frac{3}{4} \\
 -\frac{3\sqrt{3}}{8}& \frac{3}{4} & -\frac{1}{8} \\
},$


$\sigma \to \pmqty{
 -\frac{1}{2} & 0 & \frac{\sqrt{3}}{2} \\
 0 & -1 & 0 \\
 \frac{\sqrt{3}}{2} & 0 & \frac{1}{2} \\
},
C_3 \sigma \to \pmqty{
 -\frac{7}{8} & \frac{\sqrt{3}}{4} & \frac{\sqrt{3}}{8} \\
 \frac{\sqrt{3}}{4} & \frac{1}{2} & \frac{3}{4} \\
 \frac{\sqrt{3}}{8} & \frac{3}{4} & -\frac{5}{8} \\
},C_3^2 \sigma \to \pmqty{
 -\frac{7}{8} & -\frac{\sqrt{3}}{4} & \frac{\sqrt{3}}{8} \\
 -\frac{\sqrt{3}}{4} & \frac{1}{2} & -\frac{3}{4} \\
 \frac{\sqrt{3}}{8} & -\frac{3}{4} & -\frac{5}{8} \\
}\}$    \end{lstlisting}
\end{example}

These matrices form a reducible representation of the group, as we will demonstrate here. In fact, since the matrices are 3 by 3, it is already clear that it cannot be an irreducible representations, by looking at the character table. So, we take the trace of each of the matrices:
\begin{lstlisting}[style=in]
    ReducibleTraces = 
 Table[GroupNames[[i]] -> Tr[ReducibleRep[[i]]], {i, n}]
\end{lstlisting}
\begin{lstlisting}[style=out]
    $\left\{\e\to 3,C_3\to 0,C_3^2\to 0,\sigma \to -1,C_3 \sigma \to -1,C_3^2 \sigma \to -1\right\}$
\end{lstlisting}
As usual, the elements in the same conjugacy class have the same characters. So, we just consider the characters associated with each conjugacy class in the representation:
\begin{lstlisting}[style=in]
    ConjugacyTraces = ConjNames /. ReducibleTraces;
ConjugacyTraces = Table[ConjugacyTraces[[i]] // First, {i, s}]
\end{lstlisting}
\begin{lstlisting}[style=out]
    {3, 0, -1}
\end{lstlisting}

Now, we are ready to use the character table of this group to find the multiplicity of each irrep ($A_1$, $A_2$ and $E$) in this representation using \autoref{eq:multiplicity}. We will consider the character table (in fact, we choose the first character table from the list of possible ones, in the general case):
\begin{lstlisting}[style=in]
    CharTable = CharTables[[1]];
IrrepMultiplicities = 
 Table[Rest[rowNames][[i]] -> 
   Dot[ConjMultiplicities*CharTable[[i]], ConjugacyTraces]/n, {i, 1, 
   rList[[1]]}]
\end{lstlisting}
\begin{lstlisting}[style=out]
    $\{A_1\to 0,A_2\to 1,E\to 1\}$
\end{lstlisting}
Here, in the second line of input, \code{Rest[rowNames][[i]]} marks the name of the row, that is, of the irrep, and maps it to its multiplicity in the representation. The multiplicity is given by \code{Dot[ConjMultiplicities*CharTable[[i]], ConjugacyTraces]/n}, which is simply the multiplicity $m_i$ from \autoref{eq:multiplicity}.
\section{Direct Products of Representations}

\begin{definition}[Direct Product Representation]
    A direct product of two representations is defined as one where the matrix corresponding to an element is the direct product (or Kronecker product) of matrices corresponding to that element in the two representations. That is, if $\Gamma$ and $\Theta$ are two representations of the group $G$, then their product, say $\Xi=\Gamma\otimes\Theta$ is defined as :
    \[\Xi(g)=\Gamma(g)\otimes\Theta(g)\]

    It is easy to show that the matrices $\Xi(g)$ also form a representation of the group $G$, that is:
    \[\Xi(g_1.g_2)=\Xi(g_1).\Xi(g_2)\]
    whenever $g_1, g_2\in G$. It follows from the theorem:
    \[(A_1\otimes B_1).(A_2\otimes B_2)=(A_1.A_2)\otimes(B_1.B_2)\]
    whenever $A_1$ and $A_2$ have the same shape, and so do $B_1$ and $B_2$.

\end{definition}

Now recall the different irreducible representations of the group $\Scal_3$ that we discussed in \autoref{sec:char_table}. We can similarly obtain direct products of pairs of them to obtain (perhaps) new representations.

A few points to note about direct product representations are:
\begin{enumerate}
    \item For two matrices $A_{m\times n}$ and $B_{p\times q}$, the direct product is defined as \[(A\otimes B)_{(i-1)p+k, (j-1)q+l}=A_{i,j} B_{k,l}\]
    \item It follows from the above that, if $A_1$ is the trivial representation, then $A_1\otimes \Gamma=\Gamma\otimes A_1=\Gamma$.
    \item $\Tr(A\otimes B)=\Tr(A).\Tr(B)$
\end{enumerate}

\subsection{\texorpdfstring{Decomposition Rules for $\Scal_3$}{Decomposition Rules for S3}}

First of all, recall that irreps $A_1$ and $A_2$ of $\Scal_3$ were 1-dimensional representations. For them, the direct products are nothing but the products of the numbers that represent the elements. So, 
\[A_1\otimes A_1=A_1\]
\[A_1\otimes A_2=A_2\otimes A_1=A_2\]
\[A_2\otimes A_2=A_1\]
The third result follows from the fact that `matrices' of $A_2$ are just either 1 (for $\e, C_3, C_3^2$) or $-1$ (for $\sigma, C_3\sigma, C_3^2\sigma$). So when a product is obtained between a matrix and itself (since we want to evaluate $A_2\otimes A_2$), we always get a 1, which is nothing but the trivial representation.

For products with the 2-dimensional representation $E$, things become a bit more complicated. We have the simple relations:
\[A_1\otimes E=E\otimes A_1=E\]

Now we look at the product $A_2\otimes E$ which demands slightly more attention. If we look at the representing matrices, we can see that they are not exactly the same as in any of the irreducible representations. However, the matrices are just those of $E$ with an extra factor of $-1$ for three of them. We know one fact about the representation - that is, it is either an irrep or expressible as a direct sum of irreps of the group. So, we go back to \autoref{eq:multiplicity} and use it to determine the multiplicities of the irreps $A_1$, $A_2$ and $E$ in it. We find out that the multiplicity of $E$ is 1 while those of $A_1$ and $A_2$ are zero each. This basically means that the product \textit{is} the representation $E$, upto a unitary transformation. 

Similarly, we look at the product $E\otimes E$. This is now a 4-dimensional representation, and using \autoref{eq:multiplicity} we see that here the multiplicities of $A_1$, $A_2$ and $E$ are 1 each. (See following example.) So we have the result:
\begin{equation}\label{eq:EE}
    E\otimes E=A_1\oplus A_2 \oplus E
\end{equation}

\begin{example}
    In this example, we are going to decompose the product $E\otimes E$ into a direct sum of irreps of the group $\Scal_3$. We start by defining the irreps:
    \begin{lstlisting}[style=in]
GroupNames = {e, Subscript[C, 3], Subscript[C, 3]^2, \[Sigma], 
   Subscript[C, 3] \[Sigma], Subscript[C, 3]^2 \[Sigma]};
n = Length[GroupNames];
IrrepA1 = Table[{{1}}, {i, n}];
IrrepA2 = Join[Table[{{1}}, {i, n/2}], Table[{{-1}}, {i, n/2}]];
IrrepE = {IdentityMatrix[2], RotationMatrix[2 Pi/3], 
    RotationMatrix[4 Pi/3], PauliMatrix[3], 
    RotationMatrix[2 Pi/3] . PauliMatrix[3], 
    RotationMatrix[4 Pi/3] . PauliMatrix[3]};
    \end{lstlisting}
    Then we define the functions \code{NameToIrrepA1}, \code{NameToIrrepA2}, \code{NameToIrrepE} which map the list \code{GroupNames} to the lists \code{IrrepA1}, \code{IrrepA2} and \code{IrrepE} respectively.
    Similarly, the functions \code{IrrepToNameA1} etc do the reverse mapping.

    Next, we define the lists of the characters, that is, the lists of the traces of each of the elements in the group, for every irrep. So we have:
    \begin{lstlisting}[style=in]
CharA1 = Table[Tr[mat], {mat, IrrepA1}]
CharA2 = Table[Tr[mat], {mat, IrrepA2}]
CharE = Table[Tr[mat], {mat, IrrepE}]
    \end{lstlisting}
    \begin{lstlisting}[style=out]
{1, 1, 1, 1, 1, 1}
{1, 1, 1, -1, -1, -1}
{2, -1, -1, 0, 0, 0}
    \end{lstlisting}

Using this structure, we can decompose any product of the group irreps. For example, let us start with the product $E\otimes E$. 

\begin{lstlisting}[style=in]
    RepEE = Table[KroneckerProduct[mat, mat], {mat, IrrepE}];
Table[GroupNames[[i]] -> (RepEE[[i]] // MatrixForm), {i, n}]
\end{lstlisting}
\begin{lstlisting}[style=out]
$\{\e\to \pmqty{1&0&0&0\\ 0&1&0&0\\ 0&0&1&0\\ 0&0&0&1}, C_3\to \pmqty{\qtr&\rtthrqtr&\rtthrqtr&\thrqtr\\ -\rtthrqtr&\qtr&-\thrqtr&\rtthrqtr\\  -\rtthrqtr&-\thrqtr&\qtr&\rtthrqtr\\ \thrqtr&-\rtthrqtr&-\rtthrqtr&\qtr}, C_3^2\to \pmqty{\qtr&-\rtthrqtr&-\rtthrqtr&\thrqtr\\ \rtthrqtr&\qtr&-\thrqtr&-\rtthrqtr\\  \rtthrqtr&-\thrqtr&\qtr&-\rtthrqtr\\ \thrqtr&\rtthrqtr&\rtthrqtr&\qtr},  $

   $\sigma\to \pmqty{1&0&0&0\\ 0&-1&0&0\\ 0&0&-1&0\\ 0&0&0&1}, \sigma C_3\to \pmqty{\qtr&-\rtthrqtr&-\rtthrqtr&\thrqtr\\ -\rtthrqtr&-\qtr&\thrqtr&\rtthrqtr\\  -\rtthrqtr&\thrqtr&-\qtr&\rtthrqtr\\ \thrqtr&\rtthrqtr&\rtthrqtr&\qtr}, \sigma C_3^2\to \pmqty{\qtr&\rtthrqtr&\rtthrqtr&\thrqtr\\ \rtthrqtr&-\qtr&\thrqtr&-\rtthrqtr\\  \rtthrqtr&\thrqtr&-\qtr&-\rtthrqtr\\ \thrqtr&-\rtthrqtr&-\rtthrqtr&\qtr}\}$
\end{lstlisting}
Then we obtain the traces of the matrices:
\begin{lstlisting}[style=in]
CharEE = Table[Tr[mat], {mat, RepEE}]
\end{lstlisting}
\begin{lstlisting}[style=out]
{4, 1, 1, 0, 0, 0}  
\end{lstlisting}

Now we use \autoref{eq:multiplicity} to obtain multiplicities of the three irreps:
\begin{lstlisting}[style=in]
MultA1 = (1/n)*Sum[Tr[IrrepA1[[i]]]*Tr[RepEE[[i]]], {i, n}]
\end{lstlisting}
\begin{lstlisting}[style=out]
1
\end{lstlisting}
\begin{lstlisting}[style=in]
MultA2 = (1/n)*Sum[Tr[IrrepA2[[i]]]*Tr[RepEE[[i]]], {i, n}]
\end{lstlisting}
\begin{lstlisting}[style=out]
1
\end{lstlisting}
\begin{lstlisting}[style=in]
MultE = (1/n)*Sum[Tr[IrrepE[[i]]]*Tr[RepEE[[i]]], {i, n}]
\end{lstlisting}
\begin{lstlisting}[style=out]
1
\end{lstlisting}

Thus, we see that the decomposition rule is:
\[E\otimes E=A_1\oplus A_2 \oplus E\]

\end{example}

Once we have the elementary products of every pair of irreps, we can find more complicated products by simply using the algebraic properties of the direct sum and direct product operators. These properties are - commutativity, distributivity and associativity, much like the arithmetic sum and product operations. So we have, for example:
\begin{align*}
    E\otimes E\otimes E&=(E\otimes E)\otimes E\\
    &=(A_1\oplus A_2 \oplus E)\otimes E\\
    &=(A_1\otimes E) \oplus (A_2\otimes E) \oplus (E\otimes E)\\
    &=E\oplus E \oplus (A_1\oplus A_2\oplus E)\\
    &=A_1\oplus A_2 \oplus 3 E
\end{align*}

\begin{exercise}
    Verify the above decomposition for the product $E\otimes E\otimes E$ using \autoref{eq:multiplicity}.
\end{exercise}

\subsection{Clebsch-Gordan Coefficients}
When we wrote down \autoref{eq:EE}, the equality sign did not mean that the matrices on both sides are equal. That is, the matrices of $E\otimes E$ are not the same as those of $A_1\oplus A_2\oplus E$, which are just block-diagonal matrices with one block of each of $A_1$, $A_2$ and $E$. This is easily visible in the previous example where we saw that the matrices of $E\otimes E$ are definitely not block-diagonal. \autoref{eq:EE} then states that the matrices on the right-hand side are related to the corresponding matrices on the left-hand side by a similarity transformation defined by a unitary matrix. That is, there exists a unitary matrix $U$ such that
\begin{equation}\label{eq:CGcoeff}
    U. \Gamma_{E\otimes E}(g). U^{-1}=\Gamma_{A_1\oplus A_2\oplus E}(g)\qquad \forall g\in \Scal_3
\end{equation}

We can also look at this in terms of vector spaces forming the representation. The irrep $E$ represents the group $S_3$ as a group of transformations on the vector space $\mathbb{R}^2$, and hence $E\otimes E$ does so on the space $\mathbb{R}^4$. Decomposing a representation into a direct sum is equivalent to splitting the vector space into a subspace and its complementary subspace in such a manner that the same transformation in the larger vector space can represent the group in the smaller subspaces. (An complementary subspace to a subspace $S$ of a vector space $V$ is defined as another subspace $S^\perp$ such that every vector from $S^\perp$ is orthogonal to every vector in $S$. For example, the $xy$-plane and the $z$ axis are two complementary subspaces in the vector space $\mathbb{R}^3$.) In this case, the vector space $\mathbb{R}^4$ is split into $\mathbb{R}$, $\mathbb{R}$ and $\mathbb{R}^2$. But this splitting does not take place along the basis vectors of the original space. Rather, the new basis is rotated with respect to the old basis, the rotation defined by the unitary matrix $U$. So, this matrix describes the new basis vectors in terms of the old basis.

The elements of this matrix are called \textit{Clebsch-Gordan coefficients}.

\begin{example}
    In this example, we delve deeper into the details of the decomposition of the product representation $E\otimes E$ into the direct sum $A_1\oplus A_2\oplus E$. We already obtained the matrices of the product representation in the previous section. Let us now obtain the sum representation:

\begin{lstlisting}[style=in]
RepEESum = 
  Table[BlockDiagonalMatrix[{IrrepA1[[i]], IrrepA2[[i]], 
      IrrepE[[i]]}] // Normal, {i, n}];
Table[GroupNames[[i]] -> (RepEESum[[i]] // MatrixForm), {i, n}]
\end{lstlisting}
\begin{lstlisting}[style=out]
$\{\e\to \pmqty{1&0&0&0\\ 0&1&0&0\\ 0&0&1&0\\ 0&0&0&1}, C_3\to \pmqty{1&0&0&0\\ 0&1&0&0\\ 0&0& -\half &-\rtthrhalf\\ 0&0& \rtthrhalf& -\half}, C_3^2\to \pmqty{1&0&0&0\\ 0&1&0&0\\ 0&0& -\half &\rtthrhalf\\ 0&0& -\rtthrhalf& -\half},  $

   $\sigma\to \pmqty{1&0&0&0\\ 0&-1&0&0\\ 0&0&1&0\\ 0&0&0&-1}, \sigma C_3\to \pmqty{1&0&0&0\\ 0&-1&0&0\\ 0&0& -\half &\rtthrhalf\\ 0&0& \rtthrhalf& \half}, \sigma C_3^2\to \pmqty{1&0&0&0\\ 0&-1&0&0\\ 0&0& -\half &-\rtthrhalf\\ 0&0& -\rtthrhalf& \half}\}$
\end{lstlisting}

As we can clearly see, the matrices are not the same as those in the product. However, staring carefully at them tells us that they have the same traces as the product matrices. Thus, we can hope to solve \autoref{eq:CGcoeff} for the elements of the matrix $U$. For this task, we first define a matrix of unknowns $U_{i,j}$:
\begin{lstlisting}[style=in]
    U = Table[Subscript[u, i, j], {i, 4}, {j, 4}];
\end{lstlisting}
We solve \autoref{eq:CGcoeff} for these unknowns:
\begin{lstlisting}[style=in]
USolution = 
 Solve[Table[(U . RepEE[[i]] . Inverse[U]) == RepEESum[[i]], {i, n}], 
  Flatten[U], Complexes]    
\end{lstlisting}
\begin{lstlisting}[style=out]
$\{\{u_{1,2}\to 0,u_{1,3}\to 0,u_{1,4}\to u_{1,1},u_{2,1}\to 0,u_{2,3}\to -u_{2,2},u_{2,4}\to 0,u_{3,2}\to 0,u_{3,3}\to 0, $

$u_{3,4}\to -u_{3,1},u_{4,1}\to 0,u_{4,2}\to -u_{3,1},u_{4,3}\to -u_{3,1},u_{4,4}\to 0\}\}$
\end{lstlisting}
These list of rules, when applied to the original matrix $U$, gives the matrix $U_{solved}$ with less number of undetermined variables.

\begin{lstlisting}[style=in]
USolved = U /. USolution[[1]];
USolved // MatrixForm
\end{lstlisting}
\begin{lstlisting}[style=out]
    $\pmqty{
 u_{1,1} & 0 & 0 & u_{1,1} \\
 0 & u_{2,2} & -u_{2,2} & 0 \\
 u_{3,1} & 0 & 0 & -u_{3,1} \\
 0 & -u_{3,1} & -u_{3,1} & 0}$
\end{lstlisting}

Now note that any matrix of this form satisfies \autoref{eq:CGcoeff}, but we want a unique solution, so we impose the conditions that the matrix must be unitary and special (have unit determinant). On a side note, this matrix belongs to the group $SO(4)$ that we are going to study later in the text.

\begin{lstlisting}[style=in]
    USolvedSO4Solution = 
 Solve[{Inverse[USolved] == ConjugateTranspose[USolved], 
   Det[USolved] == 1}, {u$_{1,1}$, u$_{2,2}$, 
   u$_{3,1}$, Reals],
\end{lstlisting}
\begin{lstlisting}[style=out]
    $\{ \{ u_{1,1}\to -\frac{1}{\sqrt{2}}, u_{2,2}\to \frac{1}{\sqrt{2}},u_{3,1}\to -\frac{1}{\sqrt{2}} \}, \{ u_{1,1}\to -\frac{1}{\sqrt{2}},u_{2,2}\to \frac{1}{\sqrt{2}},u_{3,1}\to \frac{1}{\sqrt{2}}\},$

    $\{ u_{1,1}\to \frac{1}{\sqrt{2}},u_{2,2}\to -\frac{1}{\sqrt{2}},u_{3,1}\to -\frac{1}{\sqrt{2}}\}, \{ u_{1,1}\to \frac{1}{\sqrt{2}},u_{2,2}\to -\frac{1}{\sqrt{2}}, u_{3,1}\to \frac{1}{\sqrt{2}} \}\}$
\end{lstlisting}

We get 4 different real solutions for the unitary matrix of transformations. We can choose to go with any one as our set of Clebsch-Gordan coefficients. We choose the first one, and hence the unitary matrix is obtained as:
\begin{lstlisting}[style=in]
USolvedSO4 = USolved /. USolvedSO4Solution[[1]];
USolvedSO4 // MatrixForm
\end{lstlisting}
\begin{lstlisting}[style=out]
$\pmqty{-\rthalf & 0 & 0 & -\rthalf \\ 0 & \rthalf & -\rthalf & 0 \\ -\rthalf & 0 & 0 & \rthalf \\ 0 & \rthalf & \rthalf & 0}$
\end{lstlisting}

One can go on to check that this matrix indeed satisfies \autoref{eq:CGcoeff}.

\end{example}

\begin{exercise}
    Find the Clebsch-Gordan coefficients for the product $E\otimes E\otimes E=A_1\oplus A_2\oplus 3 E$.
\end{exercise}

\chapter{Symmetric Groups}

In Chapter 1, we already saw an example of the class of groups called Symmetric Groups. In this chapter, we delve deeper into this class and study an essential application to representation theory that arises from these groups.

\begin{definition}[Symmetric Groups]
    The group of all permutations of an ordered list of $n$ distinguishable objects is called the symmetric group $\Scal_{n}$.
\end{definition}

\section{Notations of a Symmetric Group}

There are several useful ways to represent or denote the elements in a symmetric group. We will study the list notation, cycle notation, and the matrix representation of the group.

\subsection{List Notation}

We begin with the ordered list $L_0=(1, 2, \ldots n)$. We know that this ordered list has $n!$ permutations. They all have the exact set of elements from 1 to $n$ but in a different order. Suppose we denote one such permuted list by $L_g$. Then the group element $g$ corresponding to the list $L_g$ is the permutation transformation that takes the list from the usual ordered $L_0$ to the new $L_g$. And $L_g$ is said to be the permutation list corresponding to the element $g$. The symbolic representation is often easily visualized by writing the element as a table $g\equiv\begin{pmatrix} L_0\\L_g\end{pmatrix}$. Each element goes to the element just below itself upon applying the permutation. The identity element corresponds to the list $L_0$.

\begin{remark}[Note on Convention]
    We can have different conventions on how a permutation is defined from a list. Say we are considering the group $\Scal_4$ and the permutation list $L_g=(1, 3, 2, 4)$. We see that 2 and 3 are being interchanged. It could mean:
    \begin{itemize}
        \item Interchange the elements labeled 2 and 3, or
        \item Interchange the elements at positions 2 and 3.
    \end{itemize}
    While these two conventions are the same when we apply the element on the usually ordered list $L_0$, they can give different results when applied on other ordered lists. In this text, we follow the first convention that elements labeled 2 and 3 are interchanged.
\end{remark}

\begin{example}
    Consider the ordered lists $L_0=(1,2,3,4)$, $L_1=(1,2,4,3)$ and $L_2=(1,3,4,2)$. Their corresponding elements are denoted as:
    \[\e\equiv \pmqty{1&2&3&4\\1&2&3&4},\quad g_1\equiv\pmqty{1&2&3&4\\1&2&4&3},\quad g_2\equiv\pmqty{1&2&3&4\\1&3&4&2}\]
    We can now obtain their products $g_1g_2$ and $g_2g_1$. Note that we will follow the same convention we used earlier, that group elements are always applied from right to left. That is, $g_1g_2$ is defined as the element formed by applying $g_2$ first and then $g_1$ on the list $L_0$.

    We have $g_2L_0=L_2=(1,3,4,2)$. Now we apply $g_1$ on it, that is, we map $1\to 1, 2\to2, 3\to4, 4\to 3$ in the list $L_2$. This gives $L_{g_1g_2}=(1,4,3,2)$. Similarly, $g_1L_0=(1,2,4,3)$. We apply the map from $g_2$ that is $1\to1, 2\to3, 3\to4, 4\to 2$ to get the new list $L_{g_2g_1}=(1, 3, 2, 4)$. Thus:
    \[g_1g_2\equiv\pmqty{1&2&3&4\\1&4&3&2}, \quad g_2g_1\equiv\pmqty{1&2&3&4\\1&3&2&4}\]
\end{example}
\begin{exercise}
    Write down all the elements of the group $\Scal_3$ and find its multiplication table by combining each pair of elements. Find out the conjugacy classes.
\end{exercise}
\begin{exercise}\label{exc:S4}
    Write down all the elements of the group $\Scal_4$. Divide them into classes on the basis of the order of each element. Also, find out the conjugacy classes.
\end{exercise}
\begin{exercise}
    Show that the group $\Scal_m$ is a subgroup of $\Scal_n$ whenever $m<n$. Is it always a normal subgroup too?
\end{exercise}

\subsection{Cycle Notation}

If you solved \autoref{exc:S4}, you may have observed that elements of the form $\spmqty{1&2&3&4\\1&2&4&3}$, $\spmqty{1&2&3&4\\1&3&2&4}$, $\spmqty{1&2&3&4\\4&2&3&1}$, etc. form a conjugacy class, where every element is just formed by interchanging any two objects in the list and keeping everything else unmoved. These are often called \textit{two-cycles}. Similarly, there is a conjugacy class of \textit{three-cycles} such as $\spmqty{1&2&3&4\\2&3&1&4}$ where the effective mapping is $1\to2, 2\to3$ and back to $3\to 1$, while keeping other elements unchanged. In fact, any element of any symmetric group is always either a cycle or a combination of cycles of disjoint objects. For example, in the group $\Scal_7$, the element $\spmqty{1&2&3&4&5&6&7\\1&2&4&3&6&7&5}$ is a combination of a two-cycle between 3 and 4, and a three cycle $5\to6\to7\to5$. Besides, there are trivial one-cycles acting on objects 1 and 2 separately. This motivates the cycles notation, where we denote every element by a product of cycles of disjoint sub-lists of the list.

The algorithm to construct a cycle notation is simple: Begin with the list notation $\spmqty{L_0\\L_g}$. Start from the object 1 and find the number written just below 1 in $L_g$. Then, find that number in the list $L_0$ and note the number below it in $L_g$. Continue this process until you reach 1 again. The ordered list of all numbers you encountered in the process forms the first cycle. Now mark off all the columns of $\spmqty{L_0\\L_g}$ that you already visited. Repeat the above procedure with the first number in the remaining list $L'_0$ instead of 1 this time and form the second cycle. Note that this cycle is completely disjoint from the first one since we struck off all the numbers in the first cycle. Continue this algorithm until the entire list is exhausted. It is customary to ignore the 1-cycles while writing the element. The identity element is indicated by empty parentheses `()'.

\begin{example}
    We will construct the cycles notation for the elements $g_1=\spmqty{1&2&3&4\\1&2&4&3}$ and $g_2=\spmqty{1&2&3&4\\2&4&3&1}$ of $\Scal_4$.

    Start with the first object (1) of $L_0$ in $g_1$. After just one step, we return to itself. This adds the one-cycle (1). Similarly, with the first object (that is, 2) of the remaining list (the first column removed from the list notation of $g_1$) we get another 1-cycle (2). For the remaining table $\spmqty{3&4\\4&3}$, we start with 3. The element below it is 4. So we move to 4 on the first row which has 3 below it. So we get a two-cycle $(3~4)$. Hence our cycle notation for this element is $g_1\equiv(1)(2)(3~4)\equiv(3~4)$. 

    For $g_2$, follow the same steps. We begin with 1, which has 2 below it. Move to 2, which now has 4 below it. 4 has 1 below it, so it completes the first cycle $(1~2~4)$. The remaining table is a single column so it is a one-cycle $(3)$. The cycle notation is $g_2\equiv(1~2~4)(3)\equiv(1~2~4)$.
\end{example}

\begin{example}
    We will now look at the element $\spmqty{1&2&3&4&5&6&7\\3&7&2&6&4&5&1}$ of the group $\Scal_7$. Start with the first object (that is, 1) on the top row and follow where it maps to. It goes to 3, 3 goes to 2, 2 to 7, and 7 back to 1, giving the 4-cycle $(1~3~2~7)$. Removing these columns, we are left with $\spmqty{4&5&6\\6&4&5}$ which is a 3-cycle $(4~6~5)$. So the element is equivalent to the cycle notation $(1~3~2~7)(4~6~5)$.
\end{example}

\begin{example}
    In this example, we look at two built-in functions in Mathematica that can help interconvert between cycle notation and list notation. The first function, \code{PermutationList} converts a cycle notation into list notation:
    \begin{lstlisting}[style=in]
PermutationList[Cycles[{{1, 3, 2}}]]
    \end{lstlisting}
\begin{lstlisting}[style=out]
{3, 1, 2}
\end{lstlisting}

The other function is \code{PermutationCycles}, which takes a list notation and converts it into a set of disjoint cycles. For example:

\begin{lstlisting}[style=in]
PermutationCycles[{1, 2, 3, 5, 4}]
\end{lstlisting}
\begin{lstlisting}[style=out]
Cycles[{{4, 5}}]
\end{lstlisting}

\end{example}

\begin{exercise}
    Recall the conjugacy classes of the group $\Scal_4$ that you derived in \autoref{exc:S4}. Find out the cycle notation for all the elements of this group and group them according to their conjugacy classes again. Do you notice a pattern?
\end{exercise}

A very important property of the cycle notation is that all elements in a group with the same cycle structure belong to the same conjugacy class. That is, if two elements in a group have the same cycle structure, which means that they have the same number of 1-cycles, the same number of 2-cycles, etc. then they must be related by a similarity transformation. We do not give proof of this statement, since the proof is available in standard textbooks on finite group theory.

\begin{example}
    In this example, we will obtain the cycle notations of all elements of the symmetric groups $\Scal_3$ and $\Scal_4$  in Mathematica.

    Mathematica has a built-in function \code{SymmetricGroup} to define the symmetric group formed by the permutations of any number of elements. So, we can get the list of elements for $\Scal_3$ as:
\begin{lstlisting}[style=in]
S3Elements = GroupElements[SymmetricGroup[3]]
\end{lstlisting}
\begin{lstlisting}[style=out]
{Cycles[{}], Cycles[{{2, 3}}], Cycles[{{1, 2}}],
    Cycles[{{1, 2, 3}}], Cycles[{{1, 3, 2}}], Cycles[{{1, 3}}]}
\end{lstlisting}

The first element is the identity transformation, and the others are the 2-cycles and 3-cycles that belong to the group.

Similarly, we can find the elements of $\Scal_4$:
\begin{lstlisting}[style=in]
S4Elements = GroupElements[SymmetricGroup[4]]
\end{lstlisting}
\begin{lstlisting}[style=out]
{Cycles[{}], Cycles[{{3, 4}}], Cycles[{{2, 3}}],
Cycles[{{2, 3, 4}}], Cycles[{{2, 4, 3}}], Cycles[{{2, 4}}],
Cycles[{{1, 2}}], Cycles[{{1, 2}, {3, 4}}], Cycles[{{1, 2, 3}}], 
 Cycles[{{1, 2, 3, 4}}], Cycles[{{1, 2, 4, 3}}],
 Cycles[{{1, 2, 4}}], Cycles[{{1, 3, 2}}], Cycles[{{1, 3, 4, 2}}],
 Cycles[{{1, 3}}], Cycles[{{1, 3, 4}}], Cycles[{{1, 3}, {2, 4}}], 
 Cycles[{{1, 3, 2, 4}}], Cycles[{{1, 4, 3, 2}}],
 Cycles[{{1, 4, 2}}], Cycles[{{1, 4, 3}}], Cycles[{{1, 4}}],
 Cycles[{{1, 4, 2, 3}}], Cycles[{{1, 4}, {2, 3}}]}
\end{lstlisting}
A new feature visible in the group $\Scal_4$ is the presence of composition of cycles, for example the element \code{Cycles[\{\{1, 4\}, \{2, 3\}\}]} represents the element $(1~4)(2~3)$.
\end{example}

\subsection{Matrix Notation}

Any finite group has a representation, that is, a set of matrices with the same multiplication rules as the corresponding elements. For symmetric groups, it is much simpler to construct a matrix notation when we consider the idea of the group elements being represented as permutations of the basis vectors of a linear space. That is, the list of objects (that we permute) is identified as being the list of basis vectors of the space. So the permutation refers to an interchange of the basis vectors. For example, the group element $\spmqty{1&2&3\\1&3&2}$ is identified as the linear transformation that maps the first basis vector to itself, the second basis vector to the third, and the third vector to the second. Clearly, in this basis, this transformation has the matrix form:
\[\pmqty{1&0&0\\0&0&1\\0&1&0}\]

Producing a matrix similarly for every element of the symmetric group gives the matrix notation. It is noteworthy that this is also a representation of the group since the matrices follow the same multiplication rules as the group elements. This should be obvious from the way we defined the matrix notation as a permutation of the basis vectors. Some important points to note about this representation:
\begin{itemize}
    \item Each matrix row has all zeroes except for a 1 at exactly one position. The same property holds for each column.
    \item This is an $n$-dimensional representation for the group $\Scal_n$.
    \item This is not necessarily an irreducible representation (irrep). For example, we saw that for the group $\Scal_3\equiv C_{3v}$, there is no irrep of dimension 3, so the matrix notation of permutations is definitely a reducible representation (See \autoref{ex:red-rep_S3} of the previous chapter for instance).
    \item The trace of the matrix is either zero or a positive integer less than or equal to $n$. It is zero only for those permutations where no object in the list maps to itself, and $n$ for the identity element. In fact, the trace of this matrix is simply the number of distinct 1-cycles in the transformation. This is consistent with the fact that all elements with the same cycle structure form a conjugacy class and hence have the same character in any representation.
\end{itemize}

The algorithm for constructing a matrix from a permutation list is quite simple: Observe each element of the table $\spmqty{L_0\\L_g}$. For every column of this table, say $\spmqty{x_i\\y_i}$ for $i=1,2,\ldots n$ (where $x_i,y_i\in \{1,2,3,\ldots n\}$), the element of the matrix representation in the $x_i$-th column and $y_i$-th row is marked as 1, for all $i$. All other elements are marked 0.

Notice that there are exactly $n!$ matrices of $n$ dimensions that satisfy the first two criteria. So all matrices that fall into this category form the representation. However, it is also important to note that this is not the unique representation for this group. Any representation formed by a similarity transformation acting on this set of matrices does not satisfy these two properties in general.

\begin{example}
    In this example, we will look at a Mathematica function that converts a list notation into its corresponding matrix. This function is named \code{PermutationMatrix}. For example

    \begin{lstlisting}[style=in]
Perm1 = PermutationMatrix[{3, 1, 2, 4}] // Normal // Transpose;
Perm1 // MatrixForm
    \end{lstlisting}
    \begin{lstlisting}[style=out]
$\pmqty{
 0 & 1 & 0 & 0 \\
 0 & 0 & 1 & 0 \\
 1 & 0 & 0 & 0 \\
 0 & 0 & 0 & 1 }$
    \end{lstlisting}
    We took a transpose after applying the function since Mathematica uses a slightly different convention - the resultant matrix is the transpose of what we want to obtain.
\end{example}

\section{The Regular Representation}
So far we have seen how any symmetric group $\Scal_n$ has a representation derived from the permutation of axes in an $n$-dimensional space. There are $n!$ such matrices corresponding to all the elements of the group. Now we will discuss an important theorem that helps generalize this idea to all finite groups.

\begin{theorem}[Cayley's Theorem]
    Every finite group is isomorphic to a subgroup of some symmetric group.
\end{theorem}

The statement is easy to verify intuitively. Every group element $x$ in a group $G$ can be seen as a map from one group element $g$ to another element $x.g$. That this map is always bijective follows from group axioms, and we will not show the proof here. Any bijective map from a finite set to itself is, by definition, a permutation. This may not always include all possible permutations (if there are $n$ elements in the group, then there are $n!$ possible permutations, which is almost always larger than $n$.). Thus, this group is isomorphic to a subgroup of the symmetric group $\Scal_{|G|}$.

So the procedure to construct this representation is as follows: Begin by considering the group as an ordered list of elements. Build the multiplication table following this order and choose the row corresponding to any specific element, say $x$. The elements appearing in the row state the action of the element $x$ on the other elements, and give a permutation of the ordered list considered before. This permuted list can be thought of as the list notation discussed in the previous section. Then follow the steps described before, to construct the matrix notation.

The representation of a finite group constructed in this manner is called the \emph{Regular Representation} of the group. One can define left-regular and right-regular representations based on whether rows or columns of the multiplication table are considered as the permuted list.

\begin{example}
    In this example we will construct the regular representation of the familiar group $\Scal_3$. For this we first define a representation of the group elements in the same manner as inn \autoref{ex:mtmc1}. These are defined so that the list \code{Group} contains the matrices representing the elements, \code{GroupNames} contains the names of the elements $\e$, $\sigma$ etc. We also define a few mappings such as \code{GrpToName} that map a matrix to its corresponding element name.

    Our desired task is to find the permutation of the ordered set \code{GroupNames} that is obtained when an element of the group is applied on it. For example, applying $\sigma$ gives:

\begin{lstlisting}[style=in]
Table[matSigma . elem /. GrpToName, {elem, Group}]
\end{lstlisting}
\begin{lstlisting}[style=out]
$\{\sigma, C_3^2\sigma, C_3\sigma, \e, C_3^2,C_3\}$
\end{lstlisting}

We can similarly obtain a table of all such permutations. This is the group multiplication table itself, which gives:
\begin{lstlisting}[style=in]
Thread[GroupNames -> multTableC3v] // TableForm
\end{lstlisting}
\begin{lstlisting}[style=out]
$\{\e\to \{\e,C_3,C_3^2,\sigma ,C_3\sigma ,C_3^2\sigma \},
C_3 \to \{C_3,C_3^2,\e,C_3\sigma ,C_3^2\sigma ,\sigma \},
C_3^2 \to \{C_3^2,\e,C_3,C_3^2\sigma ,\sigma ,C_3\sigma \},$
$\sigma \to \{\sigma ,C_3^2\sigma ,C_3\sigma ,\e, C_3^2,C_3\},
C_3\sigma \to \{C_3\sigma ,\sigma ,C_3^2\sigma ,C_3,\e,C_3^2\},
C_3^2\sigma \to \{C_3^2\sigma ,C_3\sigma ,\sigma ,C_3^2,C_3,\e\}\}$
\end{lstlisting}
Thus we see that each of the elements of the group corresponds to a unique permutation of all the group elements. Now we want to see them in terms of permutations of the numbers 1 to 6. This is simple, if we map $\e$ to 1, $C_3$ to 2 etc. Then for example, the element $\e$ corresponds to the permutation $(1,2,3,4,5,6)$ and $C_3$ corresponds to $(2,3,1,6,4,5)$. These are nothing but list notations of the elements of the group $\Scal_6$. Thus, we can obtain the regular representation of the group $\Scal_3$ as being that of a subgroup of the group $\Scal_6$.

So we define the following table that is effectively the multiplication table, with the group elements replaced by an index from 1 to 6:
\begin{lstlisting}[style=in]
PermListsC3v = 
  Table[Position[Group, i][[1, 1]], {i, 
    Table[elem1 . elem2, {elem2, Group}]},
    {elem1, Group}];
\end{lstlisting}

Then we map each of the elements in the group to the corresponding row:
\begin{lstlisting}[style=in]
ElemToPermList = Thread[GroupNames -> PermListsC3v];
ElemToPermList // TableForm
\end{lstlisting}
\begin{lstlisting}[style=out]
$\{\e\to \{1,2,3,4,5,6\},C_3\to \{2,3,1,6,4,5\},C_3^2\to \{3,1,2,5,6,4\},$
$\sigma \to \{4,5,6,1,2,3\},C_3\sigma \to \{5,6,4,3,1,2\},C_3^2\sigma \to \{6,4,5,2,3,1\}\}  $
\end{lstlisting}

Once we have the list notation for each element, it is easy to obtain the regular representation using the function \code{PermutationMatrix} that we saw earlier:
\begin{lstlisting}[style=in]
    RegRep = 
  Table[PermutationMatrix[PermListsC3v[[i]]] // Transpose, {i, n}];
  RegRepMapping = 
 Table[GroupNames[[i]] -> (RegRep[[i]] // MatrixForm), {i, n}]
\end{lstlisting}
\begin{lstlisting}[style=out]
    $\{\e\to \pmqty{
 0 & 0 & 1 & 0 & 0 & 0 \\
 1 & 0 & 0 & 0 & 0 & 0 \\
 0 & 1 & 0 & 0 & 0 & 0 \\
 0 & 0 & 0 & 0 & 1 & 0 \\
 0 & 0 & 0 & 0 & 0 & 1 \\
 0 & 0 & 0 & 1 & 0 & 0}, 
 C_3\to \pmqty{
 1 & 0 & 0 & 0 & 0 & 0 \\
 0 & 1 & 0 & 0 & 0 & 0 \\
 0 & 0 & 1 & 0 & 0 & 0 \\
 0 & 0 & 0 & 1 & 0 & 0 \\
 0 & 0 & 0 & 0 & 1 & 0 \\
 0 & 0 & 0 & 0 & 0 & 1},
 C_3^2\to \pmqty{
 0 & 1 & 0 & 0 & 0 & 0 \\
 0 & 0 & 1 & 0 & 0 & 0 \\
 1 & 0 & 0 & 0 & 0 & 0 \\
 0 & 0 & 0 & 0 & 0 & 1 \\
 0 & 0 & 0 & 1 & 0 & 0 \\
 0 & 0 & 0 & 0 & 1 & 0},$
 
 $
 \sigma\to \pmqty{
 0 & 0 & 0 & 1 & 0 & 0 \\
 0 & 0 & 0 & 0 & 1 & 0 \\
 0 & 0 & 0 & 0 & 0 & 1 \\
 1 & 0 & 0 & 0 & 0 & 0 \\
 0 & 1 & 0 & 0 & 0 & 0 \\
 0 & 0 & 1 & 0 & 0 & 0},
 C_3\sigma\to \pmqty{
 0 & 0 & 0 & 0 & 1 & 0 \\
 0 & 0 & 0 & 0 & 0 & 1 \\
 0 & 0 & 0 & 1 & 0 & 0 \\
 0 & 0 & 1 & 0 & 0 & 0 \\
 1 & 0 & 0 & 0 & 0 & 0 \\
 0 & 1 & 0 & 0 & 0 & 0},
 C_3^2\sigma\to \pmqty{
 0 & 0 & 0 & 0 & 0 & 1 \\
 0 & 0 & 0 & 1 & 0 & 0 \\
 0 & 0 & 0 & 0 & 1 & 0 \\
 0 & 1 & 0 & 0 & 0 & 0 \\
 0 & 0 & 1 & 0 & 0 & 0 \\
 1 & 0 & 0 & 0 & 0 & 0}
 \}$
\end{lstlisting}
\end{example}

\begin{exercise}
    Prove that the regular representation derived above can be decomposed into $A_1\oplus A_2\oplus 2E$.
\end{exercise}

\part{Compact Groups}
\chapter{Lie Groups and Lie Algebras}

Finite groups were all about discrete symmetries, but it is not difficult to find a continuous symmetry, \textit{e.g.,} rotation of a circle (this group is called $SO(2)$). Lie groups are tools to analyze these kinds of symmetries. It is trivial to check that all points on the unit circle on the complex plane form a group, with inversion operation as $z \rightarrow \frac{1}{z}$ for and binary operation as $z_1 \cdot z_2 =z_1z_2$, which is the usual multiplication of complex numbers. This group is called $U(1)$ and is isomorphic to the group $SO(2)$ of rotations of the circle. Moreover, there is a homomorphism from the additive group of real numbers onto $U(1)$, i.e., $t \rightarrow e^{i \omega t}$, where $\omega$ is any non-zero real constant and $t$ varies over $\mathbf{R}$. If $z(t)=z(t_1)z(t_2)=z_1z_2$, then the function $t=f(t_1,t_2)$ must be $(t_1+t_2)$. Likewise, the function connecting the elements $z$ and $z^{-1}$ is $g(t)=-t$.\par
The above parametrization of $U(1)$ is a one-dimensional \textit{representation} (We call it a representation because complex numbers can be considered to be $1 \times 1$ matrices). In higher dimensional representations, the $\omega$ in the above example is replaced by \textbf{Generators} of the group (which are $n \times n$ matrices, $n \in \mathbb{Z}$). Say $\{X_i\}$ are the generators for a particular group, then they satisfy the commutation relations $[X_a, X_b]= C^c_{ab}X_c$, where $ C^c_{ab}$ are called \textbf{Structure constants}, and the above relation is called Lie algebra of the group.\par
A \textbf{Lie Algebra} $\mathbf{g}$ is a vector space on which is defined a binary operation [,], having the following properties: 
\begin{enumerate}
    \item $\forall x, y \in \mathbf{g}$, $[x, y] \in \mathbf{g}$. 
    \item $\forall x,y, z \in \mathbf{g}$, and scalars $\lambda, \mu$, $[\lambda x+ \mu y, z]=\lambda [x, z]+ \mu [y, z]$. 
    \item $[x, y]=-[y, x]$. 
    \item $[x, [y, z]]+[y, [z, x]]+[z, [x, y]]=0$
\end{enumerate}
The last one is known as the \textbf{Jacobi Identity}. \par
In the rest of this chapter, we will be working with the example of $SU(2)$ group to demonstrate these above-explained concepts. $SU(2)$ is the group of $2\times 2$ unitary matrices, the rotation group in three dimensions.  It is characterized by the three operators $T_{1,2,3}$ satisfying the commutation relations:
\begin{equation}
[T_i,T_j]=i \epsilon_{ijk} T_k
\end{equation}
Note that the group of $2\times 2$ unitary matrices is obtained by 
\begin{equation}
\exp{i \theta_i T_i}.
\end{equation}
This group can be realized by Pauli matrices in $2 \times 2$ dimensions as: 
$t_1 \rightarrow \frac{1}{2}\sigma_x$, $t_2 \rightarrow \frac{1}{2}\sigma_y$ and $t_3 \rightarrow \frac{1}{2}\sigma_z$.  

\begin{lstlisting}[style=in]
ElementName = {$t_x, t_y, t_z$};
Elements = {1/2 {{0, 1}, {1, 0}}, 1/2 {{0, -I}, {I, 0}}, 
   1/2 {{1, 0}, {0, -1}}};
NameToElem = Thread[ElementName -> Elements];
MatRep[x_] := x /. NameToElem;
ElemToName = Thread[Elements -> ElementName];
NameElem[x_] := x /. ElemToName;
Commutator[x_, y_] := (MatRep[x] . MatRep[y] - MatRep[y] . MatRep[x]);
\end{lstlisting}
The commutation relations can be checked as follows: 
\begin{lstlisting}[style=in]
NameElem[Commutator[$t_x, t_y$]/I]
NameElem[Commutator[$t_y, t_z$]/I]
NameElem[Commutator[$t_z, t_x$]/I]
\end{lstlisting}
\begin{lstlisting}[style=out]
$t_z$
$t_x$
$t_y$
\end{lstlisting}
The Jacobi identity can be checked as follows: 
\begin{lstlisting}[style=in]
    Commutator[Commutator[[$t_x, t_y$], $t_z$]+ Commutator[Commutator[$t_y, t_z$], $t_x$] + 
 Commutator[Commutator[$t_z, t_x$], $t_y$]
\end{lstlisting}
\begin{lstlisting}[style=out]
    {{0, 0}, {0, 0}}
\end{lstlisting}
It is customary to define the combinations $T_{\pm}= T_1\pm T_2$, which are the familiar raising and lowering operators.  It is possible to diagonalize only one of the operators $T_i$, and customarily, it is chosen to be $T_3$.  
\begin{lstlisting}[style=in]
    NewElementName1 = Insert[ElementName, $t_+$, 4];
NewElementName2 = Insert[NewElementName1, $t_-$, 5];
NewElements1 = 
  Insert[Elements, MatRep[$t_x$ + I $t_y$], 4];
NewElements2 = 
  Insert[NewElements1, MatRep[$t_x$ - I $t_y$], 5];
ElemToName = Thread[NewElements2 -> NewElementName2];
NameToElem = Thread[NewElementName2 -> NewElements2];
NameElem[x_] := x /. ElemToName;
MatRep[x_] := x /. NameToElem;
\end{lstlisting}
Checking the commutation relations for these newly introduced elements as follows: 
\begin{lstlisting}[style=in]
    NameElem[Commutator[$t_+$, $t_-$]/2]
    NameElem[Commutator[$t_z$, $t_+$]/2]
    NameElem[-Commutator[$t_z$, $t_-$]/2]
\end{lstlisting}
\begin{lstlisting}[style=out]
    $t_z$
    $t_+$
    $t_-$
\end{lstlisting}
i.e., $[t_+, t_-]=2t_z$, $[t_z, t_+]=t_+$ and $[t_z, t_-]=-t_-$. These $t_+$ and $t_-$ are analogous to \textbf{Creation} and \textbf{Annihilation} operators, and $t_z$ is analogous to the Hamiltonian operator for the Quantum Harmonic Oscillator system. \par  
It can be shown that only one $SU(2)$ generator can be diagonalized at a time using a similarity transformation. Conventionally, it is chosen to be $t_z$. In general, $t_z$ should have more than one eigenvalue, and its eigenvalues are in general half-integers or integers, separated by intervals of 1. The proof for this statement can be found in standard textbooks for group theory. Let us say that the highest eigenvalue is $j$. Then, a particular eigenstate of $t_z$ can be written in general as $\vert j; m \rangle$, where $m$ is the $t_z$ eigenvalue of that particular state. Hence,  $t_z \vert j; m \rangle = m \vert j; m \rangle$ and $t_z \vert j; j \rangle = j \vert j; j \rangle$. The ladder operators $t_+$ and $t_-$ can be used to skim over all the $m$'s for a particular $j$, using the following relation that can be derived from the commutation relations between them: $t_z t_{\pm} \vert j, m \rangle= (m \pm 1) t_{\pm} \vert j, m \rangle$. 
\begin{example}{(Two-level System)}
    Denote a general in the 2-d vector space as $(a, b)$. The highest weight state will give zero upon applying the creation operator on it:
\begin{lstlisting}[style=in]
    MatRep[$t_+$] . {a, b}
\end{lstlisting}
\begin{lstlisting}[style=out]
    {b, 0}
\end{lstlisting}
Hence, the highest weight state is (1, 0). 
$t_z$ eigenvalue for this state is $\frac{1}{2}$: 
\begin{lstlisting}[style=in]
    MatRep[$t_z$] . {1, 0}
\end{lstlisting}
\begin{lstlisting}[style=out]
    {$\frac{1}{2}$, 0}
\end{lstlisting}
Now, applying the annihilation operator to this state will give us another state in the Hilbert space. 

\begin{lstlisting}[style=in]
    MatRep[$t_-$] . {1, 0}
\end{lstlisting}
\begin{lstlisting}[style=out]
    {0, 1}
\end{lstlisting}
With $t_z$ eigenvalue as $-\frac{1}{2}$. 
\begin{lstlisting}[style=in]
    MatRep[$t_z$] . {0, 1}
\end{lstlisting}
\begin{lstlisting}[style=out]
    {0, -$\frac{1}{2}$}
\end{lstlisting}
And except for these two states, there are no other states in the Hilbert space that are independent to both of these because we are considering the 2-dimensional representation. As a consistency check, we can show that the annihilation operator on the lowest weight state yields 0. 
\begin{lstlisting}[style=in]
    MatRep[$t_-$] . {0, 1}
\end{lstlisting}
\begin{lstlisting}[style=out]
    {0, 0}
\end{lstlisting}
The \textbf{Casimir operator}, which will have the same eigenvalue of j(j+1) for all the states where j is the \textbf{weight of the highest weight state}, is analogous to the total angular momentum operator $L^2=L_x^2+L_y^2+L_z^2$ in Quantum Mechanics. In our case, $j=1/2$, so the eigenvalue for Casimir Operator should be 3/4, as we can see below:
\begin{lstlisting}[style=in]
    $t_x.t_x+t_y.t_y+t_z.t_z$ /. NameToElem
\end{lstlisting}
\begin{lstlisting}[style=out]
    {{$\frac{3}{4}$, 0},{0, $\frac{3}{4}$}}
\end{lstlisting}
\end{example} 
The following stanzas introduce a bunch of terminologies as well as provide a short summary of this chapter: 
\begin{enumerate}
    \item For non-abelian groups, generators do not usually commute with each other, but some particular non-linear combination of generators can commute with all the generators, which are called \textbf{Casimir operators}. A semi-simple lie algebra has $l$ independent Casimir operators, where $l$ is the rank of the algebra. For example, $SU(N)$ is of rank $(N-1)$. The lowest-order Casimir operator is the sum of squares of all the generators. For $SU(2)$, the Casimir operator is $t_1^2+t_2^2+t_3^2$. 

    \item As you will observe as we go on, some of the generators are diagonal and, hence, mutually commuting. This maximal subset of such generators is called \textbf{Cartan Subalgebra}, and the number of such generators is called the \textit{Rank} of the algebra. For example, $SU(2)$, $t_z$ alone forms the Cartan subalgebra. 
    
    \item Eigenvalues for members of the Cartan subalgebra are called \textbf{Weights}, which may be assembled together as a \textbf{Weight Vector}. For an algebra of rank $l$, the weight vectors are $l$-dimensional. Cartan subalgebra provides $l$ Eigenvalue equations: 
    \begin{equation}
        H_i\vert j, \{m_i\} \rangle = m_i \vert j, \{m_i\} \rangle
    \end{equation}
    $i$ taking integer values from 1 to $l$. Here $j$ is the label necessary to specify which representation is being referred to, and $\{m_i\}$ is the weight vector of $\vert j, \{m_i\} \rangle$. The weight vector will be written as $(m_1, m_2, \cdots, m_l)$. An $l$-dimensional plot of all weights for a representation is called \textbf{Weight Diagram} for that representation. Weights in the \textbf{Adjoint Representation} (a representation in the same dimension as the number of generators of the algebra) are called \textbf{Roots}. The \textbf{Multiplicity} or \textbf{Degeneracy} of weight is the number of eigenvectors with the same weight in that representation. For $SU(2)$, the weight diagram will be simply a straight line, with multiplicity 1 for each state, as there is only one generator in the Cartan subalgebra here. 
    \item All other generators that are not in the Cartan subalgebra are called the \textbf{Ladder Operators}. For $SU(2)$, they are $t_+$ and $t_-$. They transform between different weight states corresponding to an irrep. Since all of the ladder operators commute with the Casimir operator, the Casimir operator has the same Eigenvalue for all weight states in the irrep. Thus we can label an irrep simply by its Casimir eigenvalue. In our above example, the states will be labeled as $\vert \frac{1}{2}, m \rangle$, where $m= \pm \frac{1}{2}$ (As Casimir eigenvalue for $SU(2)$ is $j(j+1)$, labeling by $j$ is equivalent to labeling by $j(j+1)$, with the condition that $j$ is the weight of the highest weight state). 
\end{enumerate}
\begin{example}{(Building the Adjoint Representation of SU(2))}
    As mentioned before about the Structure Constants, we have the expression: 
    \begin{equation}
        [X_a, X_b]= C^c_{ab}X_c
    \end{equation}
These constants form a particular representation of a group of the same dimension as that of the number of generators of the group. This is called \textbf{Adjoint Representation}, in which the matrix elements are defined as: 
\begin{equation}
    (X_a)_{bc}=C^b_{ac}
\end{equation}
Implementing this idea in Mathematica as follows: 
\begin{lstlisting}[style=in]
    Coeff[i_, j_] := 
 Solve[Commutator[ElementName[[i]], ElementName[[j]]] == 
   Sum[  a[i, j, k] MatRep[ElementName[[k]]], {k, 1, 3}], {a[i, j, 1],
    a[i, j, 2], a[i, j, 3]}]
ElemToIdx[x_] := x /. Thread[ElementName -> Range[1, 3]]
StrCoeff[i_, j_, k_] := a[i, j, k] /. (Coeff[i, j] // Flatten)
AdjRep[x_] := 
 Table[StrCoeff[ElemToIdx[x], k, j], {j, 1, 3}, {k, 1, 3}]
 AdjRep[$t_x$] // MatrixForm
 AdjRep[$t_y$] // MatrixForm
 AdjRep[$t_z$] // MatrixForm
\end{lstlisting}

\begin{lstlisting}[style=out]
    $\begin{pmatrix}
    0 & 0 & 0\\
    0 & 0 & -i\\
    0 & i & 0
    \end{pmatrix}$

    $\begin{pmatrix}
    0 & 0 & i\\
    0 & 0 & 0\\
    -i & 0 & 0
    \end{pmatrix}$

    $\begin{pmatrix}
    0 & -i & 0\\
    i & 0 & 0\\
    0 & 0 & 0
    \end{pmatrix}$
\end{lstlisting}
We can check that it is a correct representation of $SU(2)$ by showing that the above solves the correct commutation relations: 
\begin{lstlisting}[style=in]
    Commutator[AdjRep[$t_x$], AdjRep[t_y]] - I AdjRep[$t_z$] 
\end{lstlisting}
\begin{lstlisting}[style=out]
    {{0,0,0},{0,0,0},{0,0,0}}
\end{lstlisting}
and so on. 
\end{example}

\section{\texorpdfstring{Clebsch-Gordon Decomposition of $SU(2)$ Algebra}{Clebsch-Gordon Decomposition of SU(2) Algebra}}

In this section we will see how the Clebsch-Gordan coefficients of the $SU(2)$ group can be obtained practically. If you have taken a course in quantum mechanics, you may already be familiar with how these coefficients are calculated for any specific case, by identifying certain states to one another and then using the $L_+$ and $L_-$ operators. We aim to show new aspects of the problem, especially how these calculations can be easily automated and related to the idea of decomposing direct product representations that you learned earlier in this text, with regard to finite groups.

In our Mathematica-based approach, the first priority in working with a group has been to build a representation. For $SU(2)$ this task is quite easy. You may be familiar with the idea that an $SU(2)$ irrep exists for any positive integer dimension. The 3 generators of SU(2): $L_x$, $L_y$ and $L_z$ act on the basis vectors and return a linear combination of them. It is simpler to work with the generators in the form $L_\pm=L_x\pm iL_y$ and $L_z$.

In the physical context of angular momentum, the basis vectors are simultaneous eigenvectors of the total angular momentum operator $\vec{L}^2$ and the $z$-component of the angular momentum $L_z$. The eigenvalue corresponding to $\vec{L}^2$ is $l(l+1)$ while that corresponding to $L_z$ is $l_z$. The `magnetic quantum number' $l_z$ can take values starting from $-l$ to $l$ separated by intervals of 1. So the representation has $2l+1$ basis states. While the quantization of orbital angular momentum requires $l$ to be a non-negative integer, $SU(2)$ admits a more general set of representations, allowing $l$ to take half-integer values as well. So $l=0, \frac{1}{2}, 1, \ldots$ are the representations of the group. 

We will use the bra-ket notation introduced by Dirac. A basis state is represented as $\ket{l, l_z}$. This state has following properties when acted upon by the angular momentum operators (or the generators of the $SU(2)$ group):
\begin{align}
    L_+\ket{l, l_z}&=\sqrt{l(l+1)-l_z(l_z+1)}\ket{l, l_z+1}\label{eqs:su2-gen-p}\\
    L_-\ket{l, l_z}&=\sqrt{l(l+1)-l_z(l_z-1)}\ket{l, l_z-1}\label{eqs:su2-gen-m}\\
    L_z\ket{l, l_z}&=l_z\ket{l, l_z}\label{eqs:su2-gen-z}
\end{align}

This property is sufficient to obtain the representation for these three matrices in the basis spanned by $\ket{l, l_z}$ for $l_z=-l, -l+1, \ldots l$.

\begin{example}[Representations with $l=0, \half, 1$]
We can use Equations \ref{eqs:su2-gen-p} to \ref{eqs:su2-gen-z}, to produce any representation of the group $SU(2)$. We do this for the first 3 representations.

Define a function \texttt{Irrep} in Mathematica to take a value of the parameter $l$ and produce a list of three matrices $L_x, L_y, L_z$ that produce the representation corresponding to $l$, which is also interpretable as a spin-$l$ irrep of the algebra \sutwo.

\begin{lstlisting}[style=in]
Irrep[l_] := Module[{n = 2 l + 1},
 L_p = 
 Table[If[j == i + 1, Sqrt[l (l + 1) - (l - i) (l + 1 - i)], 0],
  {i, 1, n}, {j, 1, n}];
 L_m = 
 Table[If[j == i - 1, Sqrt[l (l + 1) - (l + 2 - i) (l + 1 - i)], 0],
  {i, 1, n}, {j, 1, n}];
  L_1 = (L_p + L_m)/2;
  L_2 = (L_p - L_m)/(2 I);
  L_3 = DiagonalMatrix[Table[l + 1 - i, {i, n}]];
  Table[L_i, {i, 3}]]
    \end{lstlisting}

Let us see what this function does, in action. It begins by defining the generators $L_p$ and $L_m$ which stand for $L_+$ and $L_-$ respectively. The matrix elements are obtained from the equations \ref{eqs:su2-gen-p} and \ref{eqs:su2-gen-m}, taken in the following form:
\begin{align}
    \bra{l, m_1}L_+\ket{l, m_2}=\sqrt{l(l+1)-m_1m_2}\delta_{m_1, m_2+1}\\
    \bra{l, m_1}L_-\ket{l, m_2}=\sqrt{l(l+1)-m_1m_2}\delta_{m_1, m_2-1}
\end{align}
Since a matrix can have only positive integer indices, we rename the indices to $(i,j)$ instead of $(m_1, m_2)$ so that $i=1$ corresponds to $m_1=l$, $i=2$ corresponds to $m_1=l-1$ and so on up to $i=2l+1$ pointing to $m_1=-l$. Similar correspondence holds for $j$ and $m_2$. Applying these rules, we obtain the matrices $L_+$ and $L_-$, after which we take a linear combination of them and name those as $L_1$ and $L_2$. For the diagonal matrix $L_3$, we use the relation
\begin{equation}
    \bra{l, m_1}L_3\ket{l, m_2}=m_1\delta_{m_1, m_2}
\end{equation}

Now that we have the function to generate any irrep of the group \sutwo, we will use it for this purpose:
\begin{enumerate}
    \item The spinor representation \textbf{2}
       \begin{lstlisting}[style=in]
irrephalf = Irrep[1/2];
Table[x // MatrixForm, {x, irrephalf}]
    \end{lstlisting}
    \begin{lstlisting}[style=out]
{$\pmqty{0&\half\\\half&0}, \pmqty{0&-\ihalf\\ \ihalf&0}, \pmqty{\half&0\\0&-\half}$}
    \end{lstlisting}

As expected, the generators for this representation are half times the Pauli matrices.

\item The vector representation \textbf{3}. This is also an Adjoint Representation, other than that one built in the last section. 
\begin{lstlisting}[style=in]
irrepone = Irrep[1];
Table[x // MatrixForm, {x, irrepone}]
    \end{lstlisting}
    \begin{lstlisting}[style=out]
{$\pmqty{0&\rthalf&0\\ \rthalf&0&\rthalf\\ 0&\rthalf&0}, \pmqty{0&-\irthalf&0\\ \irthalf&0&-\irthalf\\ 0&\irthalf&0}, \pmqty{1&0&0\\0&0&0\\0&0&-1}$}
    \end{lstlisting}
\end{enumerate}
Similarly we can use this function to build any irrep of the group algebra.
\end{example}

\begin{exercise}
    Find the generators of the representation \textbf{4} of \sutwo, and demonstrate that they obey the correct commutation relations.
\end{exercise}

Once we have the matrices for irreducible representations, we can build more representations (perhaps, reducible) by taking direct products of these, as we did for finite groups. It may be tempting to take a direct product of the matrices representing the generators, but that will be wrong. Direct products of matrices representing group elements, and \textit{not generators}, give rise to new representations.

To simplify our problems, we can use the infinitesimal limit of the elements. An element close to the identity in a representation $\Gamma$ is given in the infinitesimal limit by:
\[g^{\Gamma}=I^{\Gamma}+i\sum_\alpha \epsilon^\alpha J^{\Gamma}_\alpha\]
Similarly in another representation $\Theta$ it is given by:
\[g^{\Theta}=I^{\Theta}+i\sum_\beta\epsilon^\beta J^{\Theta}_\beta\]
In their direct product representation $\Gamma\otimes\Theta$ it is given by:
\[g^{\Gamma\otimes\Theta}=\qty(I^{\Gamma}+i\sum_\alpha \epsilon^\alpha J^{\Gamma}) \otimes \qty(I^{\Theta}+i\sum_\beta\epsilon^\beta J^{\Theta}_\beta)\]
The right hand side, upto first order in $\epsilon$, is:
\[I^{\Gamma}\otimes I^{\Theta}+i\sum_\alpha \epsilon^\alpha \qty(J_\alpha^{\Gamma}\otimes I^{\Theta}+I^{\Gamma}\otimes J^{\Theta}_\alpha)\]

Thus the representing matrix for any generator $J$ in a product representation
is given by:
\begin{equation}\label{eq:ten-pro-gen}
    J^{\Gamma\otimes\Theta}=J^{\Gamma}\otimes I^{\Theta}+I^{\Gamma}\otimes J^{\Theta}
\end{equation}

We will begin with the representing matrices of the generators in two different representations of the group, using which we find the matrices in the product representation, and then break it down into block diagonal form.

\begin{example}
    In this example, we consider two irreps of the group $SU(2)$ and take their direct product. Then we will decompose this product into the direct sum of two or more irreducible representations of the group.

    We will begin with the product $\mathbf{2}\otimes\mathbf{2}$. Here $\mathbf{2}$ refers to the 2-dimensional representation of the group (or of the Lie algebra). This has $l=1/2$ is often called the spinor representation because this is the algebra associated with spin-half particles (like electrons), and the two basis states represent the two possible orientations of the spin with respect to the $z$-axis. You may be familiar with the result of combining two spin-half particles. It results in two possibilities -- a spin-1 particle with three possible spin states (for $l=1$ the number of basis states is 3, also called a triplet) and a spin-0 particle with only one possible spin state (called a singlet). We will show how exactly this comes about.

    We begin with the function \texttt{Irrep} that we defined in the previous example to build a 2-dimensional representation of the algebra. The results are as follows:
    \begin{lstlisting}[style=in]
SpinorIrrep = Irrep[1/2];
SpxSpRep = 
  Table[KroneckerProduct[SpinorIrrep[[i]], IdentityMatrix[2]] + 
    KroneckerProduct[ IdentityMatrix[2], SpinorIrrep[[i]]], {i, 3}];
Table[mat // MatrixForm, {mat, SpxSpRep}]
    KroneckerProduct[ IdentityMatrix[2], VectorIrrep[[i]]], {i, 3}];
    \end{lstlisting}
    \begin{lstlisting}[style=out]
{$\pmqty{0&\half& \half& 0\\ \half&0&0&\half\\ \half&0&0&\half\\ 0&\half& \half& 0} $, $\pmqty{0&-\ihalf& -\ihalf& 0\\ \ihalf&0&0&-\ihalf\\ \ihalf&0&0&-\ihalf\\ 0&\ihalf& \ihalf& 0} $, $\pmqty{1&0&0&0\\ 0&0&0&0\\ 0&0&0&0\\ 0&0&0&-1} $}
    \end{lstlisting}

    The matrices that are produced above are the ones corresponding to the representation $\mathbf{2\otimes 2}$. The operation that produces this is quite simple to see: The \texttt{KroneckerProduct} function takes the direct product of one of the generators in the first irrep with the identity matrix in the other and adds the two combinations, according to \autoref{eq:ten-pro-gen}.

    We now want to represent these matrices in the block diagonal form. To do so, we need to be able to know what the blocks are, as we did in the case of products of finite groups. Then we can obtain a unitary matrix that takes these matrices to the expected block-diagonal form. You may have learned in a physics course how we know what the blocks will be -- they will be the generators in the representations $\mathbf{3}$ for the triplet and $\mathbf{1}$ for the singlet. We can then just proceed in the same manner as we did while block-diagonalizing products of finite group representations.

    However, we have a simpler and clearer way in this case. We make use of the fact that the resultant states must be eigenstates of the Casimir operator -- the squared angular momentum operator $\vec{L}^2$. So we first evaluate this operator:
    \begin{lstlisting}[style=in]
Lsq = Sum[i . i, {i, SpxSpRep}];
Lsq // MatrixForm
    \end{lstlisting}
    \begin{lstlisting}[style=out]
$\pmqty{2&0&0&0\\ 0&1&1&0\\ 0&1&1&0\\ 0&0&0&2} $
    \end{lstlisting}

    Since the resultant states in the direct sum are also eigenstates of this operator, we can state that: \textit{The unitary transformation that diagonalizes the Casimir operator also block-diagonalizes the generators.}

    Hence we find the eigenvectors and eigenvalues of the matrix $\vec{L}^2$:
    \begin{lstlisting}[style=in]
vecs = Eigenvectors[Lsq // Normal]
    \end{lstlisting}
    \begin{lstlisting}[style=out]
{{0, 0, 0, 1}, {0, 1, 1, 0}, {1, 0, 0, 0}, {0, -1, 1, 0}}
    \end{lstlisting}

    \begin{lstlisting}[style=in]
Eigenvalues[Lsq // Normal]
    \end{lstlisting}
    \begin{lstlisting}[style=out]
{2, 2, 2, 0}
    \end{lstlisting}

    We know, when a state is an eigenstate of the $\vec{L}^2$ operator, its eigenvalue is $l(l+1)$. Thus, the vectors in the representation $\mathbf{2\otimes 2}$ can be rotated in such a manner that 3 of the resulting vectors have $l=1$ (so that $l(l+1)=2$) and the remaining vector has $l=0$ (so that $l(l+1)=0$). Now, we demand that these vectors rotate through a unitary matrix. This matrix is obtained from the matrix \texttt{vecs} which has the eigenvectors of $\vec{L}^2$ as its rows. Unitarity demands that the row vectors be normalized. This gives us the matrix $\mathbf{U}$:
    \begin{lstlisting}[style=in]
U = Table[vec // Normalize, {vec, vecs}];
U // MatrixForm
    \end{lstlisting}
    \begin{lstlisting}[style=out]
$\pmqty{0&0&0&1\\ 0&\frac{1}{\sqrt{2}}&\frac{1}{\sqrt{2}}&0\\ 1&0&0&0\\ 0&-\frac{1}{\sqrt{2}}&\frac{1}{\sqrt{2}}&0} $
    \end{lstlisting}

    This matrix has a meaningful interpretation -- The rows represent the eigenvectors of the total angular momentum squared. The first three rows correspond to the 3-dimensional representation of $SU(2)$, while the fourth is a vector in the zero-dimensional representation. In fact, it can be shown that the first three rows are exactly the three eigenvectors of the matrix $L_z$ obtained in the product representation. The columns represent the direct product states of the two representations that were multiplied. Since the two factor representations were identical, we labeled them with 1,2. Then the direct product states, in the correct order, are:
    \[\ket{\half, \half}\otimes\ket{\half, \half},\qquad\ket{\half, \half}\otimes\ket{\half, -\half},\qquad\ket{\half, -\half}\otimes\ket{\half, \half},\qquad\ket{\half, -\half}\otimes\ket{\half, -\half}\]

    We can check that the unitary matrix \textbf{U} indeed block-diagonalizes the matrices representing the generators in the representation $\mathbf{2\otimes 2}$. We obtain the rotated matrices $\mathbf{U.L_i.U^{-1}}$:
    
    \begin{lstlisting}[style=in]
Table[U . mat . Inverse[U] // MatrixForm, {mat, SpxSpRep}]
    \end{lstlisting}
    \begin{lstlisting}[style=out]
{$\pmqty{0&\rthalf&0& 0\\ \rthalf&0&\rthalf&0\\ 0&\rthalf&0&0\\  0&0&0&0}$, $\pmqty{0&-\irthalf&0& 0\\ \irthalf&0&-\irthalf&0\\ 0&\irthalf&0&0\\  0&0&0&0}$, $\pmqty{1&0&0&0\\ 0&0&0&0\\ 0&0&-1&0\\ 0&0&0&0} $}
    \end{lstlisting}
    Clearly the matrix is in a block diagonal form, with the first block with three rows and columns representing the matrices in the 3-dimensional representation, and the last row and column representing the matrices in the 1-dimensional representation.

    This gives us the decomposition:
    \begin{equation}
        \mathbf{2\otimes2=3\oplus1}
    \end{equation}

    This replicates the physical example of combining two spin-half angular momenta (such as two fermions) to get a triplet and a singlet.

    Thus the 4 rows of the unitary matrix correspond to the eigenstates of the total angular momentum, in the correct order:
    \[\bra{1, 1}, \qquad \bra{1, 0}, \qquad \bra{1, -1}, \qquad \bra{0,0}\]

    The elements of the matrix are $U_{i,j}=\braket{i}{j}$ where $\bra{i}$ is the $i$-th bra from the above list, and $\ket{j}$ is the $j$-th ket from among the 4 direct product kets listed before. The matrix elements are therefore the overlap amplitudes between the direct product kets and the eigenbras of the decomposed representation. These amplitudes are called \textbf{Clebsch-Gordan coefficients}.

    The decomposition of direct products into direct sums plays a vital role not only in the realm of angular momentum but also in other fields where continuous symmetries are present, for example, the formation of baryons and mesons by combination of quarks (and antiquarks) can be similarly analyzed, and their Clebsch-Gordan coefficients can be obtained from a similar analysis of diagonalizing Casimir operators. This involves more complicated algebra as the working group in that case is \suthr. To give an idea about the complexity of this group algebra, it has 8 generators, and 2 Casimir operators, and analyzing even the simplest combination of them (the mesons) involves block-diagonalizing matrices of order $9\times9$.

    To conclude the above discussion, we put forward the relation between the eigenstates of $\mathbf{3\oplus1}$ and those of $\mathbf{2\otimes2}$:

    \begin{align}
        \ket{1,1}&=\ket{\half,\half}\label{eq:ex-sum-prod-begin}\\ 
        \ket{1,0}&=\rthalf\ket{\half, -\half}+\rthalf\ket{-\half, \half}\\
        \ket{1,-1}&=\ket{-\half,-\half}\\
        \ket{0,0}&=-\rthalf\ket{\half, -\half}+\rthalf\ket{-\half, \half}\label{eq:ex-sum-prod-end}
    \end{align}

    \end{example}

\begin{exercise}
    Carry out a similar analysis as in the above example to show that the following reduction relations hold true for irrep products in \sutwo. Also find out the Clebsch-Gordan coefficients and write down the states in the RHS in terms of products of states in the LHS as in equations \ref{eq:ex-sum-prod-begin} to \ref{eq:ex-sum-prod-end}.
    \begin{enumerate}
        \item $\mathbf{2\otimes4=5\oplus3}$
        \item $\mathbf{3\otimes3=5\oplus3\oplus1}$
        \item $\mathbf{2\otimes2\otimes2=4\oplus2\oplus2}$ (Note: This one is tricky, as there are two subspaces with the same $l$ in the decomposed space.)
    \end{enumerate}
\end{exercise}

\section{The Isospin Algebra: Subgroups in Lie Group}

Isospin is the realization of neutron and proton being two different spin states in the same abstract space. This intuition came from their mass being almost equal $(m_n = 939.56 MeV/c^2; m_p = 938.27 MeV/c^2)$. Introduce the creation and annihilation operators for proton and neutron as $P_i^{\dagger}, P_i$ and $N_i^{\dagger}, N_i$. These particles being fermions, follow the anti-commutator algebra: $\{P_i, P_j^{\dagger}\}= \{N_i, N_j^{\dagger}\}= \delta_{ij}$, $\{P_i, P_j\}= \{N_i, N_j\}= 0$. \par
Bilinear combinations that conserve the total number of nucleons are $N_i^{\dagger}N_i$, $P_i^{\dagger}P_i$, and the mixed ones. Also, it can be seen that the total charge is $Q = P_i^{\dagger}P_i$. With these nucleon-conserving operators, construct the following linear combinations of them: 
$B = N_i^{\dagger}N_i+P_i^{\dagger}P_i$, $T_+=P_i^{\dagger}N_i$, $T_- = N_i^{\dagger}P_i$ and $T_3 = \frac{1}{2}(P_i^{\dagger}P_i-N_i^{\dagger}N_i)= Q - \frac{1}{2}B$. \par
It is straightforward to check the following commutation relations: 
\begin{equation}
    [T_+, T_-] = 2T_3 \qquad [T_3, T_{\pm}]= \pm T_{\pm} \qquad [B, T_{\pm}] = [B, T_3]=0
\end{equation}

It can be clearly seen that $(T_3, T_{\pm})$ and B are separately closed under commutation, and the former is the set of generators for SU(2), whereas the latter is a one-parameter U(1). A feasible representation for this algebra  would be: 
\begin{equation*}
    \tau_0 = \begin{pmatrix}
        1 & 0 \\
        0 & 1
    \end{pmatrix}
    \qquad
    \tau_1 = \begin{pmatrix}
        0 & 1 \\
        1 & 0
    \end{pmatrix}
    \qquad
    \tau_2 = \begin{pmatrix}
        0 & -i \\
        i & 0
    \end{pmatrix}
    \qquad
    \tau_3 = \begin{pmatrix}
        1 & 0 \\
        0 & -1
    \end{pmatrix}
\end{equation*}
All group elements $U = e^{\frac{i}{2}\tau_i \theta_i}$ are unitary because all the matrices are hermitian. However, the unitary matrix associated with $\tau_0$ is not \textbf{Special} Unitary because its determinant is not 1. This overall group is called U(2), of which U(1) and $SU(2)$ are two subgroups.

\section{Exercises}
\textit{Hints for * marked questions are given in the notebook \texttt{Exercise 1.nb}}
\begin{enumerate}
    \item (*) Verify that the following functions $\gamma : \mathbb{R} \rightarrow GL(2, \mathbb{R})$ are actually group homomorphisms. 
    \begin{itemize}
        \item $\gamma(t) = \begin{pmatrix}
            \cos{t} & -\sin{t}\\
            \sin{t} & \cos{t}
        \end{pmatrix}$

        \item $\gamma(t) = \begin{pmatrix}
            \cosh{t} & \sinh{t} \\
            \sinh{t} & \cosh{t}
        \end{pmatrix}$

        \item $\gamma(t) = \begin{pmatrix}
            e^t & 0 \\ 
            0 & e^{-t}
        \end{pmatrix}$

        \item $\gamma(t) = \begin{pmatrix}
            1 & t\\
            0 & 1
        \end{pmatrix}$
    \end{itemize}

    \item (*) Show that $\gamma$ is a one-parameter subgroup of $SL(2,\mathbb{R})$, where 
    \begin{equation*}
        \gamma(t) = \begin{pmatrix}
            e^t & 0 & 0\\
            0 & e^t & 0\\
            0 & 0 & e^{-2t}
        \end{pmatrix}
    \end{equation*}

    \item Show that $\gamma$ is a one-parameter subgroup of $O(3,\mathbb{R})$, where 
    \begin{equation*}
        \gamma(t) = \begin{pmatrix}
            \cos{t} & -sin{t} & 0\\
            \sin{t} & \cos{t} & 0\\
            0 & 0 & 1
        \end{pmatrix}
    \end{equation*}

    \item (*) Let $A = \begin{pmatrix}
        0 & 1\\
        1 & 0
    \end{pmatrix}$. 
    \begin{itemize}
        \item Find $tA$, $(tA)^2, \cdots$ 
        \item Find $e^{tA}$. 
        \item Check that $\gamma(t)=e^{tA}$ implies $\gamma'(0)=A$ in this case. 
        \item Check that $det(e^{tA})=e^{tr(tA)}$ in this case. 
    \end{itemize}

    \item Let $A = \begin{pmatrix}
        0 & -1\\
        1 & 0
    \end{pmatrix}$. 
    \begin{itemize}
        \item Find $tA$, $(tA)^2, \cdots$ 
        \item Find $e^{tA}$. 
        \item Check that $\gamma(t)=e^{tA}$ implies $\gamma'(0)=A$ in this case. 
        \item Check that $det(e^{tA})=e^{tr(tA)}$ in this case. 
    \end{itemize}
\end{enumerate}

\chapter{Classification of Lie Algebras}

As discussed in the previous chapter, mutually commuting generators of a Lie algebra of rank $l$ is called Cartan subalgebra. The generators in this subalgebra are Hermitian $(H_i = H_i^{\dagger}$, mutually commuting $([H_i, H_j]=0))$ and form a basis for linear vector space of dimension $l$. Choose the normalization to be $Tr(H_i H_j) = \lambda \delta_{ij}$ where $i, j \in {0, 1, \cdots , l}$. \par
In this vector space, denote a state corresponding to a generator $X_a$ as $\vert X_a \rangle$. In this notation, action of a generator $X_b$ on this state is defined to be $X_b \vert X_a \rangle= \vert [X_b, X_a] \rangle$. It is clearly visible that for the elements of Cartan subalgebra, $\vert[H_i, H_j]\rangle=0$. Define scalar product in this space to be $\langle X_a \vert X_b \rangle= \frac{1}{\lambda} Tr(X_a^{\dagger}X_b)$. Hence, elements of Cartan subalgebra are orthonormal. \par
In the adjoint representation, all the states $\vert E_a \rangle$ that are not in Cartan subalgebra satisfy $H_i \vert E_a \rangle= a_i \vert E_a \rangle$, which, from the notation introduced, says that; $[H_i, E_a]=a_i E_a$.  Taking hermitian conjugate: $[H_i, E_a^{\dagger}]=-a_i E_{\alpha}^{\dagger}$. Hence, $E_a^{\dagger}= E_{-a}$. \par
As Cartan subalgebra generators (practically all generators) form an orthonormal basis, a state can be labelled as $\vert m, D \rangle$ where $H_i \vert m, D \rangle = m_i \vert m, D \rangle$. $D$ labels the representation. The vector $m=(m_1, m_2, \cdots, m_l)$ is the Weight Vector. Now, consider the following: 
\begin{align}
    H_i (E_{\pm a}\vert m, D \rangle) &= [H_i, E_{\pm a}]\vert m, D \rangle + E_{\pm a} H_i \vert m, D \rangle\\ 
    &= (m_i \pm a_i) E_{\pm a} \vert m, D \rangle
\end{align}
From this, it is easy to see that $E_a \vert E_{-a} \rangle$ has weight 0. In the adjoint representation, it must be some linear combination of Cartan subalgebra generators. From this, it can be immediately proved that: 
\begin{equation}
    [E_a, E_{-a}] = a_i H_i
\end{equation}
The equations discussed above define the \textit{Cartan-Weyl Basis}. Let us see this for the example of SU(3):\par
We start with introducing the $SU(3)$ algebra in the 3-dimensional representation as below, famously known as \textit{Gell-Mann Matrices}: 
\begin{lstlisting}[style=in]
ElementName={$T_x,T_y,T_z,V_x,V_y,U_x,U_y,Y$};
Elements={1/2 {{0,1,0},{1,0,0},{0,0,0}},1/2 {{0,-I,0},{I,0,0},{0,0,0}},1/2 {{1,0,0},{0,-1,0},{0,0,0}},1/2 {{0,0,1},{0,0,0},{1,0,0}},1/2 {{0,0,-I},{0,0,0},{I,0,0}},1/2 {{0,0,0},{0,0,1},{0,1,0}},1/2 {{0,0,0},{0,0,-I},{0,I,0}},1/3 {{1,0,0},{0,1,0},{0,0,-2}}};
NameToElem=Thread[ElementName->Elements];
MatRep[x_]:=x/.NameToElem;
ElemToName=Thread[Elements->ElementName];
NameElem[x_]:=x/.ElemToName;
Commutator[x_,y_]:=(MatRep[x].MatRep[y]-MatRep[y].MatRep[x]);
\end{lstlisting}
The first 3 elements are just Pauli Matrices with extra rows and columns. Hence, it is conventional to define $T_+$ and $T_-$ as $T_\pm = T_x \pm i T_y$. Besides, we also define matrices: 
$V_{\pm}=V_x \pm i V_y$ and $U_{\pm} = U_x \pm i U_y$. As this representation is traceless, diagonal elements will have two independent components; hence, there is no $V_z$ or $U_z$, and only two independent diagonal generators can be there, which are historically chosen to be $T_z$ and Y. 
\begin{lstlisting}[style=in]
    FinalNames = {$T_+, T_-, T_z, U_+, 
   U_-, V_+, V_-, Y$};
FinalElements = {Elements[[1]] + I*Elements[[2]], 
   Elements[[1]] - I*Elements[[2]], Elements[[3]], 
   Elements[[6]] + I*Elements[[7]], Elements[[6]] - I*Elements[[7]], 
   Elements[[4]] + I*Elements[[5]], Elements[[4]] - I*Elements[[5]], 
   Elements[[8]]};
NameToElem = Thread[FinalNames -> FinalElements];
MatRep[x_] := x /. NameToElem;
ElemToName = Thread[FinalElements -> FinalNames];
NameElem[x_] := x /. ElemToName;
\end{lstlisting}
Below, we present the commutation table for $SU(3)$. The row-label and the column-label are the first and second elements of the commutator, respectively. 
\begin{lstlisting}[style=in]
    Coeff[i_, j_] := 
 Solve[Commutator[FinalNames[[i]], FinalNames[[j]]] == 
   Sum[a[i, j, k] MatRep[FinalNames[[k]]], {k, 1, 8}], {a[i, j, 1], 
   a[i, j, 2], a[i, j, 3], a[i, j, 4], a[i, j, 5], a[i, j, 6], 
   a[i, j, 7], a[i, j, 8]}]
DecompElem[i_, j_] := 
 Sum[a[i, j, k] FinalNames[[k]], {k, 1, 8}] /. (Coeff[i, j] // 
    Flatten)
    CommTable = Table[DecompElem[i, j], {i, 1, 8}, {j, 1, 8}];
CommTable = 
  Prepend[CommTable, {$"T_+", "T_-", "T_z", "U_+", 
   "U_-", "V_+", "V_-", "Y"$}];
CommTable = 
  MapThread[
   Prepend, {CommTable, {"", $"T_+", "T_-", "T_z", "U_+", 
   "U_-", "V_+", "V_-", "Y"$}}];
Grid[CommTable, Frame -> All]
\end{lstlisting}

\begin{table}[]
    \centering
    $\begin{array}{c|c|c|c|c|c|c|c|c}
 \text{} & T_+ & T_- & T_z & U_+ & U_- & V_+ & V_- & \text{Y} \\
 \hline
 T_+ & 0 & 2 T_z & -T_+ & V_+ & 0 & 0 & -U_- & 0 \\
 T_- & -2 T_z & 0 & T_- & 0 & -V_- & U_+ & 0 & 0 \\
 T_z & T_+ & -T_- & 0 & -\frac{U_+}{2} & \frac{U_-}{2} & \frac{V_+}{2} & -\frac{V_-}{2} & 0 \\
 U_+ & -V_+ & 0 & \frac{U_+}{2} & 0 & \frac{3 Y}{2}-T_z & 0 & T_- & -U_+ \\
 U_- & 0 & V_- & -\frac{U_-}{2} & T_z-\frac{3 Y}{2} & 0 & -T_+ & 0 & U_- \\
 V_+ & 0 & -U_+ & -\frac{V_+}{2} & 0 & T_+ & 0 & T_z+\frac{3 Y}{2} & -V_+ \\
 V_- & U_- & 0 & \frac{V_-}{2} & -T_- & 0 & -T_z-\frac{3 Y}{2} & 0 & V_- \\
 \text{Y} & 0 & 0 & 0 & U_+ & -U_- & V_+ & -V_- & 0 \\
    \end{array}$
    \caption{Commutators of generators of $SU(3)$}
    \label{tab:commutators-SU3}
\end{table}

\pagebreak

It can be clearly seen that $T_+$, $T_-$, and $T_z$ are closed under commutation. This is a \textbf{Subalgebra}, which is indeed isomorphic to SU(2). An \textbf{Ideal} is a special kind of subalgebra such that, if $x \in J$, an ideal, and $y \in L$, the whole lie algebra, then $[x,y] \in J$. A lie algebra with no non-trivial ideals is called \textbf{Simple}. An algebra with no \textbf{Abelian} ideal is called \textbf{Semi-simple Lie algebra}. $SU(3)$ is a simple lie algebra.\par 

The formulation for Adjoint Representations in terms of Structure constants that was introduced in the last chapter can be implemented for $SU(3)$ to get the following: 
\begin{lstlisting}[style = in]
ElemToIdx[x_] := x /. Thread[FinalNames -> Range[1, 8]]
StrCoeff[i_, j_, k_] := a[i, j, k] /. (Coeff[i, j] // Flatten)
AdjRep[x_] := 
 Table[StrCoeff[ElemToIdx[x], j, k], {k, 1, 8}, {j, 1, 8}]
\end{lstlisting}
which gives us the Adjoint representation matrices for all the generators: 
\begin{lstlisting}[style = in]
AdjRep[$T_+$] // MatrixForm
\end{lstlisting}
\begin{lstlisting}[style = out]
    $\pmqty{0&0&-1&0&0&0&0&0\\ 0&0&0&0&0&0&0&0 \\ 0&2&0&0&0&0&0&0 \\ 0&0&0&0&0&0&0&0 \\ 0&0&0&0&0&0&-1&0 \\ 0&0&0&1&0&0&0&0 \\ 0&0&0&0&0&0&0&0 \\ 0&0&0&0&0&0&0&0}$
\end{lstlisting}

and so on. As expected, it will be observed that only $T_z$ and $Y$ will be diagonal operators. Rest are eigen 'vectors' of these two in adjoint representation, which can be, in general, written as $X = a T_z+ b Y$. Here, the binary operation is the commutator of the two generators. In Mathematica, finding out the eigenvalues for these other ('ladder') operators is a matter of a minute. These values can be found to be:\\ \\
\begin{center}
    \begin{tabular}{|c|c|c|c|c|c|c|c|c|}
    \hline
     Operators & $T_+$ & $T_-$ & $T_z$ & $U_+$ & $U_-$ & $V_+$ & $V_-$ & $Y$\\ 
     \hline
     Eigenvalues & a & -a & 0 & $-\frac{a}{2}+b$ & $\frac{a}{2}-b$ & $\frac{a}{2}+b$ & $-\frac{a}{2}-b$ & 0\\
     \hline
\end{tabular}
\end{center}

\begin{lstlisting}[style = in]
X = a AdjRep[$T_z$] + b AdjRep[Y];
$\alpha$[x_] := Solve[Commutator[X, AdjRep[x]] == p AdjRep[x], p];
$\alpha$[$T_+$]
\end{lstlisting}
\begin{lstlisting}[style = out]
    {{p -> a}}
\end{lstlisting}
This can be written in the following manner: 
\begin{equation}
    \alpha_{U_+}(aT_z+bY) = -\frac{a}{2}+b
\end{equation}
etc. $\alpha_{i}$ are linear functions that act over elements of Cartan subalgebra and give complex numbers as outputs (i.e. these are \textbf{Functionals}). They are called \textbf{Roots}, and corresponding vectors in the root space are called \textbf{Root Vectors}.\par
Now, say a and b are elements of a Lie Algebra $L$, then \textbf{Killing Form} is defined to be: 
\begin{equation}
    (a, b) = Tr(a, b)
\end{equation}

in Adjoint Representation. This can be implemented as follows: 

\begin{lstlisting}[style = in]
KillForm[x_, y_] := Tr[x . y];
KillForm[AdjRep[$T_z$], AdjRep[$T_z$]]
KillForm[AdjRep[Y], AdjRep[Y]]
KillForm[AdjRep[$T_z$], AdjRep[Y]]
\end{lstlisting}
\begin{lstlisting}[style = out]
    3 
    4
    0
\end{lstlisting}
Defining killing form enables us to make a connection between the Cartan subalgebra, $H$, and its dual $H^*$ (though this holds only for Semi-simple Lie Algebra, which we are mostly concerned with). If $\rho \in H^*$, there exists a unique element $h_{\rho} \in H$ such that for every $k \in H$, $\rho(k) = (h_{\rho}, k)$. This one-to-one relationship between $H$ and $H^*$ can be illustrated for $SU(3)$ as follows: \par
If we are given some expression that looks like $\rho(k)$, we will recover $h_{\rho}$. Let us write the three non-zero roots of $SU(3)$ as: 

\begin{align*}
    \alpha_1(aT_z+bY) &= a\\
    \alpha_2(aT_z+bY) &= -\frac{a}{2}+b\\
    \alpha_3(aT_z+bY) &= \frac{a}{2}+b
\end{align*}

and the corresponding elements in $H$ as 

\begin{equation}
    h_{i} = c_i T_z + d_i Y \qquad i \in {1,2,3}
\end{equation}
\begin{lstlisting}[style = in]
$\alpha_1$[a_, b_] := a;
$\alpha_2$[a_, b_] := -$\half$ a + b;
$\alpha_3$[a_, b_] := $\half$ a + b;
h[$\alpha_1$] = c[1] AdjRep[$T_z$] + d[1] AdjRep[Y];
h[$\alpha_2$] = c[2] AdjRep[$T_z$] + d[2] AdjRep[Y];
h[$\alpha_3$] = c[3] AdjRep[$T_z$] + d[3] AdjRep[Y];
\end{lstlisting}
We set a and b values to 0 and 1 alternatively to get $c_i$ and $d_i$. 
\begin{lstlisting}[style = in]
Coeffs = Solve[{$\alpha_1$[1, 0] == KillForm[h[$\alpha_1$],AdjRep[$T_z$]], $\alpha_1$[0, 1] == KillForm[h[$\alpha_1$], AdjRep[Y]], $\alpha_2$[1, 0] == KillForm[h[$\alpha_2$], AdjRep[$T_z$]], $\alpha_2$[0, 1] == KillForm[h[$\alpha_2$], AdjRep[Y]], $\alpha_3$[1, 0] == KillForm[h[$\alpha_3$], AdjRep[$T_z$]], $\alpha_3$[0, 1] == KillForm[h[$\alpha_3$], AdjRep[Y]]}, {c[1], d[1], c[2], d[2], c[3], d[3]}]
\end{lstlisting}
\begin{lstlisting}[style = out]
{{c[1] -> 1/3, d[1] -> 0, c[2] -> -1/6, d[2] -> 1/4, c[3] -> 1/6,       d[3] -> 1/4}}
\end{lstlisting}
Now, we define the inner product in this root space: 
\begin{equation}
    \langle \alpha, \beta \rangle = (h_{\alpha}, h_{\beta})
\end{equation}

\begin{lstlisting}[style = in]
InnerProd[x_, y_] := KillForm[h[x], h[y]] /. (Coeffs // Flatten)   
InnerProd[$\alpha_1, \alpha_1$]
InnerProd[$\alpha_2, \alpha_2$]
InnerProd[$\alpha_3, \alpha_3$]
InnerProd[$\alpha_1, \alpha_2$]
InnerProd[$\alpha_2, \alpha_3$]
InnerProd[$\alpha_1, \alpha_3$]
\end{lstlisting}
\begin{lstlisting}[style = out]
    $\frac{1}{3}$
    $\frac{1}{3}$
    $\frac{1}{3}$
   -$\frac{1}{6}$
    $\frac{1}{6}$
    $\frac{1}{6}$
\end{lstlisting}

This enables us to represent the roots geometrically by vectors of length $\frac{1}{\sqrt{3}}$ and separated by angle $\theta$ such that $\cos{\theta}= \pm \frac{1}{2}$. \par

Semi-simple lie algebras are sums of simple ideals, so the properties of semi-simple algebras can be obtained by generalizing the simple algebra rules. We will generalize the properties of $SU(3)$ that are discussed above: 
\begin{itemize}
    \item Every semi-simple lie algebra is a sum of simple ideals, each of which can be treated as simple lie algebra. 
    \item An element $h \in L$ is said to be \textbf{Regular} if $ad(h)$ (the adjoint representation of $h$) has the minimal number of zero eigenvalues. For SU(3), $ad(T_z)$ has two zero eigenvalues, and $ad(Y)$ has four zero eigenvalues. So $T_z$ is regular, but Y is not. A Cartan subalgebra is obtained by finding maximal commutative subalgebra containing regular elements. 
    \item The remaining generators are eigenvectors of $ad(h)$ $\forall h \in H$. They are called \textbf{Root Vectors}. We associate with each of these latter generators a linear functional as follows: 
    \begin{equation}
        (ad (h)) e_{\alpha} =[ad(h), ad(e_{\alpha})]= \alpha(h)e_{\alpha}
    \end{equation}
    which gives us the scalar eigenvalue $\alpha(h)$ corresponding to that particular generator with respect to the element in $H$. 
    \item We can associate a particular ordering with the root space. Let $\alpha_1, \alpha_2, \cdots, \alpha_n$ be a fixed basis of roots so that every element of root space can be written as $\rho = \sum_{i}c_i \alpha_i$. $\rho$ is called \textbf{positive} if $c_1>0$. If $c_1=0$, then we consider the sign of $c_2$, and so on. We will say $\rho > \sigma$ if $\rho - \sigma$ is positive. A \textbf{Simple Root} is a positive root that cannot be written as the sum of two positive roots.
    \item Properties of roots of $SU(3)$ demonstrate some properties for semi-simple lie algebras in general, e.g., if $\alpha$ is a root, so is $-\alpha$. For each root, there is only one linearly independent generator with that root. If $\alpha$ is a root, $2\alpha$ is not a root. 
\end{itemize}
If we take our basis of the lie algebra as the elements of Cartan subalgebra H i.e., $h_1, h_2, \cdots $ and $e_{\alpha_1}, e_{\alpha_2}, \cdots$, then we can write the adjoint representation of elements in H as $Diag(0, 0, \cdots, \alpha_1(h), \alpha_2(h), \cdots)$ where the number of zeros in the diagonal is equal to the number of elements in the Cartan subalgebra. From this, it is readily visible that the killing form of two elements in $H$ can be expanded as: 
\begin{equation}
    (h_1, h_2)= \sum_{\alpha \in \Sigma} \alpha(h_1)\alpha(h_2)
\end{equation}
where $\Sigma$ is the set of all \textbf{roots}. This explains our previously assumed definition of Killing Form (i.e., $Tr(x.y)$ in Adjoint Representation).  \par
Now, we know the commutation relations between elements of H and the roots, but we do not know the commutation relation between roots themselves. Let us use Jacobi identity (mentioned in the previous chapter) to find $[e_{\alpha}, e_{\beta}]$ in adjoint representation: 
\begin{align*}
   [h, [e_{\alpha}, e_{\beta}]] & = -[e_{\alpha}, [e_{\beta}, h]]-[e_{\beta}, [h, e_{\alpha}]]\\
   & = \beta(h) [e_{\alpha}, e_{\beta}]+ \alpha(h) [e_{\alpha}, e_{\beta}]\\
   & = (\alpha(h)+\beta(h))[e_{\alpha}, e_{\beta}]
\end{align*}
This has three possibilities: 
\begin{itemize}
    \item $[e_{\alpha}, e_{\beta}]$ is zero. 
    \item $[e_{\alpha}, e_{\beta}]$ is a root vector with root $(\alpha + \beta)$. 
    \item $\alpha + \beta =0$, in which case, $[e_{\alpha}, e_{\beta}]$ commutes with every h and is thus an element of Cartan Subalgebra. 
\end{itemize}
It is easy to show that $(e_{\alpha}, e_{\beta})=0$ unless $\alpha + \beta =0$. 
\begin{lstlisting}[style = in]
KillForm[AdjRep[$T_+$], AdjRep[$V_+$]]
KillForm[AdjRep[$T_+$], AdjRep[$U_-$]]
\end{lstlisting}
\begin{lstlisting}[style = out]
0
0
\end{lstlisting}
and so on. \par
It also implies that \textbf{$\left[e_{\alpha}, e_{-\alpha}\right]$ is an element of Cartan subalgebra}. We can make it more explicit as follows: \par
It is straightforward to see that (the notation we will be using is $ad(a) = A$):
\begin{align*}
    (a, [b, c]) &= Tr A [B,C]\\
    & = Tr (ABC - ACB)\\
    & = Tr (ABC) - Tr (ACB)\\
    & = Tr (ABC) - Tr(BAC) \qquad \textit{Cyclicity of Trace}\\
    & = Tr [A, B] C \\
    & = ([a, b], c)
\end{align*}
Hence, $([e_{\alpha}, e_{-\alpha}], h)=\alpha(h) [e_{\alpha}, e_{-\alpha}]$. Comparing with the definition $\rho(k)= (h_{\rho}, k)$, we get 
\begin{equation}
    [e_{\alpha}, e_{-\alpha}]= (e_{\alpha}, e_{-\alpha}) h_{\alpha}
\end{equation}
For example, taking $e_{\alpha}= U_{+}$ (we have seen before that $h_{U_{+}}=-\frac{1}{6}T_z+\frac{1}{4}Y$):
\begin{lstlisting}[style=in]
    Commutator[AdjRep[$U_+$], AdjRep[$U_-$]] // MatrixForm
\end{lstlisting}
\begin{lstlisting}[style = out]
    $\pmqty{-1&0&0&0&0&0&0&0\\0&1&0&0&0&0&0&0\\0&0&0&0&0&0&0&0\\0&0&0&2&0&0&0&0\\0&0&0&0&-2&0&0&0\\0&0&0&0&0&1&0&0\\0&0&0&0&0&0&-1&0\\0&0&0&0&0&0&0&0}
    $
\end{lstlisting}
\begin{lstlisting}[style = in]
KillForm[AdjRep[$U_+$], AdjRep[$U_-$]] (-1/6 AdjRep[$T_z$] + 1/4 AdjRep[Y]) // MatrixForm
\end{lstlisting}
\begin{lstlisting}[style = out]
    $\pmqty{-1&0&0&0&0&0&0&0\\0&1&0&0&0&0&0&0\\0&0&0&0&0&0&0&0\\0&0&0&2&0&0&0&0\\0&0&0&0&-2&0&0&0\\0&0&0&0&0&1&0&0\\0&0&0&0&0&0&-1&0\\0&0&0&0&0&0&0&0}
    $
\end{lstlisting}
Finally, we get the following relations: 
\begin{align}
    [h_1, h_2] & = 0 \qquad {h_1, h_2 \in H}\\
    [h, e_{\alpha}] &= \alpha(h)e_{\alpha} \\ 
    [e_{\alpha}, e_{\beta}] &= N_{\alpha \beta}e_{\alpha + \beta} \qquad \alpha + \beta \in \Sigma\\
    & = (e_{\alpha}, e_{-\alpha})h_{\alpha} \qquad \alpha + \beta = 0\\
    & = 0 \qquad \alpha + \beta \neq 0, \alpha+\beta \notin \Sigma
\end{align}
$N_{\alpha \beta}$ is yet to be determined.

\section{Weyl Reflections}
In this section, we will generalise a few concepts from the relatively simple $SU(2)$ to general Lie algebras and use this knowledge to develop the idea of Weyl reflections.

We start with a general $N$-dimensional representation of the algebra in terms of $N$ basis vectors ${\phi^a}$, let's say they form the standard basis as:
\begin{equation}
    \phi^1=\pmqty{1\\0\\0\\\vdots\\0},\qquad \phi^2=\pmqty{0\\1\\0\\\vdots\\0}, \ldots
\end{equation}

Now, whenever we choose any matrix (say $H$) representing a generator $h$, from the Cartan subalgebra, we must choose a basis in which this matrix is diagonal. So we have:
\begin{equation}
    H\phi^a=M^a\phi^a
\end{equation}
with $M^a$ (the eigenvalue of $H$ corresponding to the eigenvector $\phi^a$) depending on the element $h$ from the Cartan subalgebra, which is, in general, a linear combination of the matrices in the subalgebra. The set $\{M^a(H)\}$ with $H$ being a basis of the Cartan subalgebra, is a vector of numbers and is called the weight of the vector $\phi^a$. In fact, if $H_i$ are the elements of the Cartan subalgebra such that $H_i\phi^a=\lambda_i^a\phi^a$, and $H=\sum_i c_i H_i$, then 
\begin{equation}
M^a(H)=\sum_i c_i \lambda_i^a.
\end{equation}

We have the commutation relation:
\begin{equation}
    [H, E_\alpha]=\alpha(h) E_\alpha
\end{equation}
which leads to the result:
\begin{equation}\label{eq:HE}
    H~E_\alpha\phi^a=\qty(M^a(h)+\alpha(h))~E_\alpha\phi^a
\end{equation}
Thus, the generator $E_\alpha$, when applied to a basis element with weight vector $M^a(h)$, leads to another basis element, corresponding to a weight vector $M^a(h)+\alpha(h)$. This, when applied repeatedly, leads to an array of such weight vectors in the form of a uniformly spaced 'lattice`.

Now, consider a pair of conjugate generators $(E_\alpha, E_{-\alpha})$ which generate an array of weight vectors in the representation \[\phi_0, \phi_1, \ldots , \phi_j, \ldots, \phi_q\]
Note that the subscripts $1, 2$ etc. are not the weights, rather, they are just labels to the weights that we will be using for this section.
So that \begin{equation}\label{eq:E_minus_alpha}
    \phi_j = E_{-\alpha} \phi_{j-1}\forall j=1, 2, \ldots , q
\end{equation}
with $E_{-\alpha}\phi_q=0$. This way, calling the highest weight $M^*$, we can assign weights to them: $M^*, M^*-\alpha, \ldots, M^*-j\alpha, \ldots, M^*-q\alpha$ respectively. Then, the matrices $E_\alpha, E_{-\alpha}$ act as raising and lowering operators on a representation of $SU(2)$ in the form of the array of weight vectors. One can prove that
\begin{equation}\label{eq:q-su2}
    q=\frac{2\expval{M^*, \alpha}}{\expval{\alpha, \alpha}}
\end{equation}
The proof can be found in the book by \cite{Cahn}, chapter V, on page 35.

Lie algebras in more than one dimension often have an $SU(2)$ subalgebra. The above equation can be used to find the dimensionality of the $SU(2)$ subalgebras.

Now, we use the $SU(2)$ array of weights $M-m\alpha, \ldots, M-\alpha, M, M+\alpha, \ldots, M+p\alpha$. Here the highest weight is $M^*=M+p\alpha$ and the lowest is $M-m\alpha$. Thus, for this representation, $q=p+m$. Using \autoref{eq:q-su2}, we have:
\[p+m=\frac{2\expval{M+p\alpha, \alpha}}{\expval{\alpha, \alpha}}=\frac{2\expval{M, \alpha}}{\expval{\alpha, \alpha}}+2p\]
Hence, we have:
\begin{equation}\label{eq:m-p}
m-p=\frac{2\expval{M, \alpha}}{\expval{\alpha, \alpha}}
\end{equation}
This implies that for any weight $M$ whatsoever, the quantity $\frac{2\expval{M, \alpha}}{\expval{\alpha, \alpha}}$ must be an integer. This reveals an important symmetry of the system of weights of the representation. Consider the transformation:
\[S_\alpha: M\to M'=M-\frac{2\expval{M, \alpha}}{\expval{\alpha, \alpha}}\alpha\]
If we describe the weights in terms of vectors in a weight-space (which has the same number of dimensions as the Cartan subalgebra), then the transformation represents a reflection of the weight vector about the line orthogonal to the corresponding root $\alpha$ and passing through the origin. Note that the image $M'$ is still a weight of the representation because it belongs to the same $SU(2)$ subalgebra with its corresponding integers $m'$ and $p'$, with the highest weight being $M'+p'\alpha$, lowest being $M'-m'\alpha$ such that $m'=p$ and $p'=m$. Thus, we can reflect a given weight vector $M$ using any root vector $\alpha$ to get another root vector in the same representation. These reflections are called \textbf{Weyl Reflections} and the group formed by them is termed the \textbf{Weyl Group}.

For example, consider the group $SU(3)$. We saw earlier in this chapter, there are two simple roots, $\alpha_1$ and $\alpha_2$ which are of equal length and inclined at an angle of 120\textdegree. Now, this means
\[\expval{\alpha_1, \alpha_2}=-\half\expval{\alpha_1, \alpha_1}=-\half\expval{\alpha_2, \alpha_2}\]
So if we start with a weight vector $M=c_1\alpha_1+c_2\alpha_2$, and apply the reflections $S_{\alpha_1}$ and $S_{\alpha_2}$:
\begin{align}
S_{\alpha_1}: M\to M-\frac{2\expval{M, \alpha_1}}{\expval{\alpha_1, \alpha_1}}\alpha_1=(c_2-c_1)\alpha_1+c_2\alpha_2\\
S_{\alpha_2}: M\to M-\frac{2\expval{M, \alpha_2}}{\expval{\alpha_2, \alpha_2}}\alpha_1=c_1\alpha_1+(c_1-c_2)\alpha_2
\end{align}

We denote the reflections as $s_1, s_2$ respectively. Then all the outcomes of applying different combinations of the reflections on the vector form a group of order 6:
\begin{align*}
    M&=c_1\alpha_1+c_2\alpha_2\\
    s_1M&=(c_2-c_1)\alpha_1+c_2\alpha_2\\
    s_2M&=c_1\alpha_1+(c_1-c_2)\alpha_2\\
    s_1s_2M&=-c_2\alpha_1+(c_1-c_2)\alpha_2\\
    s_2s_1M&=(c_2-c_1)\alpha_1-c_1\alpha_2\\
    s_1s_2s_1M&=-c_2\alpha_1-c_1\alpha_2
\end{align*}
The generators $s_1, s_2$ satisfy $s_1^2=s_2^2=\e$, $s_2s_1s_2=s_1s_2s_1$.

\begin{exercise}
    Show that the Weyl group of $SU(3)$ described above is equivalent to the group $\Scal_3$ discussed earlier in the book, by obtaining the generators $s_1$, $s_2$ in terms of the generators $a$ and $b$ (See \autoref{ex:S3} of \autoref{ch:group_basics}).
\end{exercise}
\begin{exercise}
    Find the Weyl group of the Lie algebra $SU(4)$.
\end{exercise}

Let us see how a simple application of \autoref{eq:m-p} gives rise to a constraint on the kinds of roots a semi-simple Lie algebra can have. We look at a pair of roots $\alpha$ and $\beta$. Since the number $m-p$ as in \autoref{eq:m-p} has to be an integer, we call it $n$. So, we have:
\[n=\frac{2\expval{\beta, \alpha}}{\expval{\alpha, \alpha}}\]
where we took $M=\beta$. Similarly, since $\beta$ is also a root, we can replace $\alpha$ by $\beta$ in \autoref{eq:m-p} and use $M=\alpha$ to get:
\[n'=\frac{2\expval{\alpha, \beta}}{\expval{\beta, \beta}}\]
Combining these two equations, we get:
\begin{equation}
   \frac{\expval{\alpha, \beta}\expval{\beta, \alpha}}{\expval{\alpha, \alpha}\expval{\beta, \beta}}=\frac{1}{4}nn'
\end{equation}
The expression on the left-hand side is the \textit{cosine-squared} of the angle between $\alpha$ and $\beta$, and hence is always between 0 and 1, while $n$ and $n'$ on the right-hand side must be integers. This largely limits the kinds of semi-simple Lie algebras. In fact, the expression on the right-hand side can only be one of $\qty(0, \frac{1}{4}, \frac{1}{2}, \frac{3}{4})$. It cannot be unity since $\alpha$ and $\beta$ are independent and hence not parallel.

As a result, the angle between two distinct root vectors can only be one of $\qty(\frac{\pi}{6},\frac{\pi}{4}, \frac{\pi}{3}, \frac{\pi}{2}, \frac{2\pi}{3}, \frac{3\pi}{4}, \frac{5\pi}{6})$. This gives rise to the different classes or families of semi-simple Lie algebra.

\chapter{More On the Representation}

As we have already seen, \textbf{Simple Root} is a positive root that cannot be written as the sum of two positive roots. As we had seen before for SU(3): 
\begin{align}
    \alpha_1(aT_z+bY) &= a\\
    \alpha_2(aT_z+bY) &= -\frac{1}{2}a+b\\
    \alpha_3(aT_z+bY) &= \frac{1}{2}a+b
\end{align}
Suppose we choose the basis of root space to be $\alpha_1$ and $\alpha_3$ in that order. Since $\alpha_2 = - \alpha_1 + \alpha_3$, $\alpha_2$ is negative, the set of positive roots are $\alpha_1, -\alpha_2, \alpha_3$. As $\alpha_1 = \alpha_3 + (-\alpha_2)$, $\alpha_1$ is the sum of two positive roots and thus is not a simple root. Alternatively, we can take $\alpha_1, \alpha_2$ to be the basis of the root space when $\alpha_3$ would not be a simple root. \par
Denote the set of all roots as $\Sigma$ and the set of all simple roots as $\Pi$. If $\alpha, \beta \in \Pi$, suppose $\alpha-\beta$ is root. Then either $\alpha-\beta$ or $\beta - \alpha$ should be a positive root. Then $\alpha = (\alpha-\beta)+\beta$ or $\beta = (\beta-\alpha)+\alpha$, which is a contradiction to the initial fact that $\alpha, \beta \in \Pi$. So, \textbf{difference of two simple roots is not a root at all.}\par
Previously, we had seen that $m-p = 2\frac{\langle M, \alpha \rangle}{\langle \alpha, \alpha \rangle}$. Now, if $\alpha,\beta$ are simple roots, $\beta-\alpha$ is not a root. Thus, p=0. Hence, $\langle \alpha, \beta \rangle \leq 0$. \par
We can define the \textbf{Cartan Matrix} to be 
\begin{equation}
    A_{ij}= 2\frac{\langle\alpha_i, \alpha_j \rangle}{\langle \alpha_i, \alpha_j \rangle}
\end{equation}
where i goes over all the simple roots. It can be immediately seen that the diagonal elements are 2, and off-diagonal elements can take values 0, -1, -2, and -3. Taking the positive basis for $SU(3)$ as $\alpha_1, \alpha_2$, we can write down the Cartan matrix accordingly. 
\begin{lstlisting}[style = in]
CartanMat = Table[2 InnerProd[$\alpha_i,\alpha_j$]/InnerProd[$\alpha_j, \alpha_j$], {i,1, 2}, {j, 1, 2}] // MatrixForm
\end{lstlisting}
\begin{lstlisting}[style = out]
    $\pmqty{2&-1\\-1&2}$
\end{lstlisting}
\begin{exercise}
    Find the Cartan matrix for G2, continuing the analysis from the exercise of the last chapter. Find the generators of G2 from any source that you may prefer.
\end{exercise}
All the positive roots can be recovered from this matrix, each of which can be written in the form $\beta = \sum_{i}k_i\alpha_i$. $\sum_{i}k_i$ is called \textbf{level} of the root $\beta$. Hence, simple roots are at the first level. If we have all the roots up to n-th level, we can check if (n+1)-th level root exists or not by checking for each $\beta + \alpha_i$. Below, we present it as an algorithm:\par
Say a root at n-th level is $\beta$. Hence, we know how far we can extend back up to $\beta - m \alpha_i$. It gives how far we can extend forward up to $\beta + p \alpha_i$: 
\begin{align}
    m-p &= 2 \frac{\langle \beta, \alpha_i \rangle}{\langle \alpha_i, \alpha_i \rangle}\\
    & = \sum_{j} 2 k_j \frac{\langle \alpha_j, \alpha_i \rangle}{\langle \alpha_i, \alpha_i \rangle}\\
    & = \sum_{j} k_j A_{ji}
\end{align}
$p$ has to be more than 0 to have a root. 
\begin{example}{\textbf{SU(3):}}
    The Cartan matrix being $A=\begin{pmatrix}
        2 &-1\\
        -1 & 2
    \end{pmatrix}$: 
    \begin{itemize}
        \item Write down the two simple roots as the two rows of this matrix: 
    $$\alpha_1 = (2, -1), \alpha_2=(-1,2)$$
    \item To check if $\alpha_1+\alpha_2$ is a root, starting from $\alpha_1$: $k_1=1, k_2 = 0, m=0$. Hence $p=0-(1 \times A_{12}+0 \times A_{22})=1>0$, $\alpha_1+\alpha_2$ is a root, denoted by (1, 1). This is the only possible root at level 2. 
    \item To check if $\alpha_1+2\alpha_2$ is a root, starting from $\alpha_1+\alpha_2$: $k_1=1, k_2 = 1, m=1$. Hence, $p = 1 - (1 \times A_{12}+ 1 \times A_{22})=0$. So, it is not a root. Similarly, for $2\alpha_1+\alpha_2$, $k_1=1, k_2=1, m=1$ and $p = 1 - (1 \times A_{11}+ 1 \times A_{21})=0$. So, it is also not a root. Hence, there is no root at level 3 or above. 
    \end{itemize}
    Hence, we found three positive roots for SU(3), as expected. All other roots are just the negative of the positive roots (because we saw earlier that if $\alpha$ is a root, so is $-\alpha$). 
\end{example}
\begin{example}{\textbf{G(2):}}

The Cartan matrix is $A=\begin{pmatrix}
    2 & -3 \\
    -1 & 2
\end{pmatrix}$
\begin{itemize}
    \item Two simple roots are 
$$\alpha_1 = (2, -3), \alpha_2=(-1, 2)$$. 
\item To check if $\alpha_1+\alpha_2$ is a root, starting from $\alpha_1$: $k_1=1, k_2 = 0, m=0$. Hence $p=0-(1 \times A_{12}+0 \times A_{22})=3>0$, $\alpha_1+\alpha_2$ is a root, denoted by (1, -1). This is the only possible root at level 1. 
\item To check if $\alpha_1+2\alpha_2$ is a root, starting from $\alpha_1+\alpha_2$: $k_1=1, k_2 = 1, m=1$. Hence, $p = 1 - (1 \times A_{12}+ 1 \times A_{22})=2>0$. So, it is a root. Similarly, for $2\alpha_1+\alpha_2$, $k_1=1, k_2=1, m=1$ and $p = 1 - (1 \times A_{11}+ 1 \times A_{21})=0$. So, it is not a root. Hence, there is only one root at level 3, i.e., $\alpha_1+2\alpha_2=$(0, 1). 
\item To check if $\alpha_1+3\alpha_2$ is a root, starting from $\alpha_1+2\alpha_2$: $k_1=1, k_2 = 2, m=2$. Hence, $p = 2 - (1 \times A_{12}+ 2 \times A_{22})=1>0$. So, it is a root. Similarly, for $2\alpha_1+2\alpha_2$, $k_1=1, k_2=2, m=0$ and $p = 0 - (1 \times A_{11}+ 2 \times A_{21})=0$. So, it is not a root. Hence, there is only one root at level 4, i.e., $\alpha_1+3\alpha_2=$(-1, 3). 
\item To check if $\alpha_1+4\alpha_2$ is a root, starting from $\alpha_1+3\alpha_2$: $k_1=1, k_2 = 3, m=3$. Hence, $p = 3 - (1 \times A_{12}+ 3 \times A_{22})=0$. So, it is not a root. Similarly, for $2\alpha_1+3\alpha_2$, $k_1=2, k_2=3, m=1$ and $p = 0 - (1 \times A_{11}+ 2 \times A_{21})=1>0$. So, it is a root. Hence, there is only one root at level 5, i.e., $2\alpha_1+3\alpha_2=$(1, 0). 
\item To check if $2\alpha_1+4\alpha_2$ is a root, starting from $2\alpha_1+3\alpha_2$: $k_1=2, k_2 = 3, m=0$. Hence, $p = 0 - (2 \times A_{12}+ 3 \times A_{22})=-2$. So, it is not a root. Similarly, for $3\alpha_1+3\alpha_2$, $k_1=2, k_2=3, m=0$ and $p = 0 - (2 \times A_{11}+ 3 \times A_{21})=-3$. So, it is also not a root. Hence, there is no root at level 6 and above.

\end{itemize}
All the commutation relations among generators of a group can also be recovered from the Cartan matrix. Define the following: 
\begin{align}
    e_i &= e_{\alpha_{i}}\\
    f_i &= e_{-\alpha_i}2 \left[(e_{\alpha_i}, e_{-\alpha_i})\langle \alpha_i, \alpha_i \rangle\right]^{-1}\\
    h_i &= h_{\alpha_i}\frac{2}{\langle \alpha_i, \alpha_i \rangle}
\end{align}
From here, it directly follows that: 
\begin{align}
    \left[e_i,  f_j\right]  &= \delta_{ij}h_j\\
    \left[h_i, e_j\right] &= A_{ji} e_j\\
    \left[h_i, f_j\right] &= -A_{ji}f_j
\end{align}
All raising operators can be written like $e_{i_1}, \left[e_{i_1}, e_{i_2}\right], \left[e_{i_1}, [e_{i_2}, e_{i_3}]\right]$ and so on. Similarly all lowering operators  can be written as $f_{i_1}, \left[f_{i_1}, f_{i_2}\right], \left[f_{i_1}, [f_{i_2}, f_{i_3}]\right]$ and so on.
\end{example}
\begin{example}{\textbf{SU(3):}}
\begin{lstlisting}[style = in]
$e_1$ = AdjRep[$T_+$];
$e_2$ = AdjRep[$U_+$];
$e_3$ = AdjRep[$V_+$]; 
$e_{-1}$ = AdjRep[$T_-$];
$e_{-2}$ = AdjRep[$U_-$];
$e_{-3}$ = AdjRep[$V_-$]; 
\end{lstlisting}

We saw previously that $h_{\alpha_1}= \frac{1}{3}T_z$, $h_{\alpha_2}=-\frac{1}{6}T_z+\frac{1}{4}Y$ and $h_{\alpha_3}=\frac{1}{6}T_z+\frac{1}{4}Y$. Hence, 
\begin{lstlisting}[style = in]
CartanMat = {{2, -1}, {-1, 2}};
h[$\alpha_1$] = 1/3 AdjRep[$T_z$];
h[$\alpha_2$] = -1/6 AdjRep[$T_z$] + 1/4 AdjRep[Y];
h[$\alpha_3$] = 1/6 AdjRep[$T_z$] + 1/4 AdjRep[Y];
H[i_] := h[$\alpha_i$] 2/KillForm[h[$\alpha_i$], h[$\alpha_i$]];
Commutator[H[1], $e_1$] == CartanMat[[1, 1]] $e_1$
\end{lstlisting}
\begin{lstlisting}[style = out]
    True
\end{lstlisting}
Similarly, it is trivial to check the other relations on Mathematica. Below, the method of getting raising and lowering operators is demonstrated as well. 
\begin{lstlisting}[style = in]
    $e_1$==AdjRep[$T_+$]
    $e_{-1}$==AdjRep[$T_-$]
\end{lstlisting}
\begin{lstlisting}[style = out]
    True
    True
\end{lstlisting}
\begin{lstlisting}[style = in]
    $e_2$==AdjRep[$U_+$]
    $e_{-2}$==AdjRep[$U_-$]
\end{lstlisting}
\begin{lstlisting}[style = out]
    True
    True
\end{lstlisting}
\begin{lstlisting}[style = in]
   Commutator[$e_1, e_2$] == AdjRep[$V_+$]
   Commutator[$e_{-2}, e_{-1}$] == AdjRep[$V_-$]
\end{lstlisting}
\begin{lstlisting}[style = out]
    True
    True
\end{lstlisting}

\end{example}
The Cartan matrix can be represented diagrammatically, which is called \textbf{Dynkin diagram}. We place a dot for each simple root, and the dots are joined by a number of lines equal to $A_{ij}A_{ji}$. For examples of Dynkin diagrams, refer to the notebook \texttt{SimpleGroup demo.nb} \par
We will formally talk about \textbf{Dynkin Coefficients} now. We will apply it extensively using the Mathematica package in the next chapter: 
From the relation,
\begin{equation}
    m-p = 2 \frac{\langle M, \alpha \rangle}{\langle \alpha, \alpha \rangle}
\end{equation}
where the complete string of roots is $M+p\alpha, \cdots, M-m \alpha$. Let $\{\alpha_i\}$ be a basis of simple roots, and let $\Lambda$ be the highest weight of an irreducible representation. Then $\Lambda+\alpha_i$ is not a weight, so $m>0, p=0$ implying: 
\begin{equation}
    \Lambda_i = 2 \frac{\langle \Lambda, \alpha_i \rangle}{\langle \alpha_i, \alpha_i \rangle} \geq 0
\end{equation}
These are called Dynkin coefficients. Given the highest weight, we can find all the weights in terms of Dynkin coefficients. The algorithm is to find for each step if $M-\alpha_i$ is a weight. As we will be \textit{subtracting} $\alpha_i$ in each step, we will know the value of $p$. Keep track of $M_i = 2 \frac{\langle M, \alpha_i \rangle}{\langle \alpha_i, \alpha_i \rangle}$. If 
\begin{equation}
    m_j = p_j + M_j >0
\end{equation}
the subtraction result is another root. 
\begin{example}{\textbf{SU(3):}}
    The Cartan matrix is $\begin{pmatrix}
        2&-1\\
        -1&2
    \end{pmatrix}$. Let us consider \textit{a particular representation} where $\Lambda = (1,0)$ is the highest weight. To expand it in terms of simple roots, write $\Lambda = a \alpha_1 + b \alpha_2$. Hence: 
    \begin{align}
        a + b \frac{\langle \alpha_2, \alpha_1 \rangle}{\langle \alpha_1, \alpha_1 \rangle} &= \frac{1}{2}\\
        a + b \frac{\langle \alpha_2, \alpha_2 \rangle}{\langle \alpha_1, \alpha_2 \rangle} &= 0
    \end{align}
    We had the relations: $\langle \alpha_2, \alpha_2 \rangle=\frac{1}{3}=\langle \alpha_1, \alpha_1 \rangle, \langle \alpha_1, \alpha_2 \rangle = -\frac{1}{6}$. Hence, we get: $\Lambda = \frac{2}{3}\alpha_1 + \frac{1}{3}\alpha_2$. 
    \begin{itemize}
        \item The highest weight is $\Lambda = (1, 0)$. 
        \item For checking if $\Lambda - \alpha_1$ is a weight: $p_1=0, M_1 = 1, m_1=1>0$. So it is a weight. To check if $\Lambda - \alpha_2$ is a weight: $p_2=0, M_2 = 0, m_2=0$. So it is not a weight. In this level, the only weight is $(1, 0)-(2, -1)= (-1, 1)$. 
        \item For checking if $\Lambda - 2 \alpha_1$ is a weight: $p_1=1, M_1 = -1, m_1=0$. So it is not a weight. To check if $\Lambda - \alpha_1-\alpha_2$ is a weight: $p_2=0, M_2 = 1, m_2=1>0$. So it is a weight. In this level, the only weight is $(-1, 1)-(-1, 2)= (0, -1)$. 
        \item Similarly, it can be checked that there are no other weights in this representation of SU(3). 
    \end{itemize}
    
\end{example}

We will further look into it in the next chapter. Now let us look into \textbf{product of representations:} \par 
Previously, we saw the example of the product of representations for $SU(2)$. In general:  
\begin{itemize}
    \item If the two highest weights are $\Lambda_1$ and $\Lambda_2$ for two representations, the highest weight for the product representation is $\Lambda_1 + \Lambda_2$. For example, the product of the representations (1, 0) and (0, 1) for $SU(3)$ is (1, 1). 
    \item If we take the product of the same representation, product space can be broken into a symmetric part with the highest weight to be $2\Lambda$ and another anti-symmetric part found by the sum of the highest and the second-to-highest weight. For example, in $SU(3)$ (1, 0) representation (written as \textbf{3}), the second highest weight is (-1, 1). Hence, the anti-symmetric part is (0, 1) (written as $\Bar{\textbf{3}}$), and the symmetric part is (2, 0) (written as \textbf{6}). 
\end{itemize}
More about it in the next chapter. 
\chapter{The Final Touch}

\section{Weyl Dimension Formula}

    The Weyl dimension formula allows one to obtain the dimensionality of an irreducible representation of a simple Lie algebra in terms of its highest weight. Following Cahn (pg.102 onwards), the over-parametrization (over-parametrization because we choose to not write it only in terms of the simple roots) of elements of $H_{0}^{*}$ (for functions defined on $H_{0}^{*}$) is conveniently given as (Note here that $H_{0}^{*}$ refers to the space of real, linear combinations of the roots of the algebra (essentially the space of weights), similar to the dual space $H^{*}$ to the Cartan subalgebra $H$):
    
    \begin{equation}
    \begin{aligned}
\rho = \sum_{\alpha \in \Sigma} \rho_{\alpha} \alpha
\end{aligned}
    \end{equation}

    An action of an element $S$ of the Weyl group, on a function of $\rho$ is then defined as:

    \begin{equation}
    \begin{aligned}
        SF(\rho) = F(S^{-1} \rho)
\end{aligned}
    \end{equation}

    As is shown in Cahn (pg.103), calculating this action on the character of a representation $\chi(\rho)$ yields a form of invariance under the Weyl group i.e., $S_{\chi} = \chi$. A few other examples are elaborately worked out in Cahn, which then leads us to the key intellectual content of Weyl's character formula, namely:

    \begin{equation}
    \begin{aligned}
\chi(\rho) = \frac{\sum_{S \in W}^{} (det S) e^{\langle \Lambda + \delta, S\rho \rangle}}{\sum_{S \in W}^{} (det S) e ^{ \langle \delta, S\rho \rangle}}
\end{aligned}
    \end{equation}

    The dimension of the irrep is then obtained by calculating $\chi(\rho = 0)$, which needs to be computed as a limit and not by setting $\rho = 0$ directly. We let $\rho = t \delta$ and work in the limit $t \rightarrow 0$. The expression then becomes:

\begin{align}
    \chi(t \delta) &= \frac{\sum_{S \in W}^{} (det S) e^{\langle S(\Lambda + \delta), t \delta \rangle}}{\sum_{S \in W}^{} (det S) e ^{ \langle S \delta, t \delta \rangle}} \\
    &= \frac{Q(t(\Lambda + \delta))}{Q(t \delta)} \\
    &= e^{\langle -\delta, t\Lambda \rangle} \prod_{\alpha > 0}^{} \frac{e^{\langle \alpha, t(\Lambda + \delta) \rangle} - 1}{e^{\langle \alpha, t\delta \rangle} - 1}
\end{align}

    Taking the small $t$ limit, we compute $dim(R) = \chi(0)$ as:
    \begin{equation}
        dim R = \prod_{\alpha > 0}^{} \frac{\langle \alpha, \Lambda + \delta \rangle}{\langle \alpha, \delta \rangle}
    \end{equation}
    Rewriting the positive root(s) $\alpha$ in terms of simple roots, we admit the following decomposition:
    \begin{equation}
        \alpha = \sum_{i}^{} k_{\alpha}^{i} \alpha_{i}
    \end{equation}
    In terms of the Dynkin coefficients $\Lambda_{i}$ of $\Lambda$, we obtain:

    \begin{equation}
        dim R = \prod_{\alpha > 0} \frac{\sum_{i}^{} k_{\alpha}^{i} (\Lambda_{i} + 1) \langle \alpha_{i}, \alpha_{i} \rangle}{\sum_{i}^{} k_{\alpha}^{i} \langle \alpha_{i}, \alpha_{i} \rangle}
    \end{equation}
%

%
%
We compute $dim R$ explicitly for a few groups, namely $SU(3) \equiv A_{2}$, $G_{2}$ and $SO(10) \equiv D_{5}$\,.
We realize that in $SU(3)$, for an irrep which has a highest weight with Dynkin coefficients $(m_{1},m_{2})$, the dimensionality is computed as:
\begin{align}
dim R &= \left(\frac{m_{1} + 1}{1} \right) \cdot \left(\frac{m_{2} + 1}{1} \right) \cdot \left(\frac{m_{1} + m_{2} + 2}{2} \right) \notag\\
%
&= \left(\frac{m_{1} + 1}{1} \right) \cdot \left(\frac{m_{2} + 1}{1} \right) \cdot \left(\frac{m_{1} + 1 +  m_{2} + 1}{1 + 1} \right)
    \end{align}

Computing the same for $G_{2}$ yields the following table (the dimensions of the reps with highest weights (0,1) and (1,0) are computed where the first entry corresponds to $\alpha_{1}$ and the second corresponds to $\alpha_{2}$):
    \begin{center}
\begin{tabular}{||c c c||} 
 \hline
 Positive roots & (0,1) & (1,0) \\ [0.5ex] 
 \hline\hline
 (1),(2) & $\frac{3}{3} \cdot \frac{2}{1}$ & $\frac{3(2)}{3} \cdot \frac{1}{1}$ \\ 
 \hline
 (12) & $\frac{3+2}{3+1}$ & $\frac{3(2) + 1}{3+1}$  \\
 \hline
 (12$^{2}$) & $\frac{3+2+2}{3 + 1 + 1}$ & $\frac{3(2) + 1 + 1}{3 + 1 + 1}$  \\
 \hline
 (12$^{3}$) & $\frac{3 + 2 + 2 + 2}{3 + 1 + 1 + 1}$ & $\frac{3(2) + 1 + 1 + 1}{3 + 1 + 1 + 1}$  \\
 \hline
 (1$^{2}$2$^{3}$) & $\frac{3 + 3 + 2 + 2 + 2}{3 + 3 + 1 + 1 + 1}$ & $\frac{3(2) + 3(2) + 1 + 1 + 1}{3+3+1+1+1}$  \\
 \hline
  & $dim R = 7$ & $dim R = 14$ \\[1ex] 
 \hline
\end{tabular}
\end{center}

Note: The dimensions of the rep with highest weight (0,1) are obtained by multiplying all the factors in the column pertaining to (0,1). An exactly identical procedure yields the dimension of the rep with highest weight (1,0).

\vspace{1mm}

We illustrate this procedure for the group $SO(10) \equiv D_{5}$ (we compute the dimensions of the reps with highest weights (1,0,0,0,0), (0,0,0,0,1) and (0,0,0,0,2) respectively)

\begin{center}
\begin{tabular}{||c c c c||} 
 \hline
 Positive roots & (1,0,0,0,0) & (0,0,0,0,1) & (0,0,0,0,2) \\ [0.5ex] 
 \hline\hline
 (1),(2),(3),(4),(5) & 2 & 2 & 3 \\ 
 \hline
 (12),(23),(34),(35) & $\frac{3}{2}$ & $\frac{3}{2}$ & $\frac{4}{2}$ \\
 \hline
 (123),(234),(235),(345) & $\frac{4}{3}$ & $\frac{4}{3} \frac{4}{3}$ & $\frac{5}{3} \frac{5}{3}$ \\
 \hline
 (1234),(2345),(1235)  & $\frac{5}{4} \frac{5}{4}$ & $\frac{5}{4} \frac{5}{4}$ & $\frac{6}{4} \frac{6}{4}$ \\
 \hline
 (12345),(23$^{2}$45) & $\frac{6}{5}$ & $\frac{6}{5} \frac{6}{5}$ & $\frac{7}{5} \frac{7}{5}$ \\ 
 \hline
 (123$^{2}$45) &$\frac{7}{6}$ & $\frac{7}{6}$ & $\frac{8}{6}$\\
 \hline
 (12$^{2}$3$^{2}$45) &$\frac{8}{7}$ &$\frac{8}{7}$ &$\frac{9}{7}$\\
 \hline
  & $dim R = 10$ & $dim R = 16$ & $dim R = 126$ \\ [1ex]
  \hline
\end{tabular}
\end{center}

Using this same procedure and generalizing, we can obtain the dimensions of $SU(N)$ irreps. The dimension(s) for the rep with highest weight $(m_{1},m_{2},\dots m_{n})$ is(are) computed via:

\begin{center}
\begin{tabular}{ c c}
 $(1),(2), \dots (n)$ & $\frac{m_{1} + 1}{1}$ $\frac{m_{2} + 1}{1}$ \dots $\frac{m_{n} + 1}{1}$ \\ 
 $(12),(23),\dots (n-1 \vspace{0.5mm} n)$ & $\frac{m_{1} + m_{2} + 2}{2} \frac{m_{2} + m_{3} + 2}{2} \dots \frac{m_{n-1} + m_{n} + 2}{2}$ \\  
 $(12\dots n)$ & $\frac{m_{1} + m_{2} \dots m_{n} + n}{n}$ \\
\end{tabular}
\end{center}

As earlier, multiplying all the factors together yields the dimension of the desired rep.

    \section{Subalgebra and Branching}

    We employ the Mathematica packages LieArt 2.0 and GroupMath to highlight and comprehensively illustrate to an extent, the process of finding maximal subalgebras and branching rules.


    We review the notion of subalgebras in brief. Suppose that we start with an algebra $G$ and then impose additional constraints/restrictions on it to obtain a "subalgebra" $G'$ (one could imagine $G$ to be a set of traceless matrices of some arbitrary fixed dimension and $G'$ to be a restricted set of matrices (a subset of G) with the diagonals being non-zero and the off-diagonals being zero; note that this is not true for all Lie algebras, only some). Denoting the space proportional to $e_{\alpha}$ to be $G_{\alpha}$, we have:
    \begin{equation}
G = H + \sum_{\alpha \in \Sigma}^{} G_{\alpha}
    \end{equation}
    and for some set $\Sigma' \subset \Sigma$, we have:
    \begin{equation}
G' = H' + \sum_{\alpha \in \Sigma'}^{} G_{\alpha}
    \end{equation}
    where $H',H$ are the Cartan subalgebras of $G',G$ respectively.

Subalgebras that satisfy the above property are called \textbf{regular}. However, not all subalgebras are regular and it requires additional work to obtain and extract the subalgebras that are non-regular. We also realize that algebras of intermediate size (size here refers to the dimension of the algebra/subalgebra) may possess a huge number of subalgebras. To obtain them in an orderly fashion, we review the notion of a \textbf{maximal subalgebra}. $G'$ is said to be a \textbf{maximal subalgebra} of $G$ if there does not exist any larger subalgebra that contains it, except $G$ itself. For the sake of not complicating the notion of a maximal subalgebra, we restrict ourselves to finding maximal semi-simple subalgebras. 


\vspace{1mm}

We now review the concept of branching rules to better understand how an irrep of an algebra becomes a rep of the subalgebra. We begin with a subalgebra $G'$ that is embedded in the algebra $G$ via a mapping (homomorphism) $f:G' \rightarrow G$ such that the commutation relations are preserved:
\begin{equation}
f([x',y']) = [f(x'),f(y')], x',y' \in G'
\end{equation}
We can arrange the homomorphism such that the Cartan subalgebra $H' \subset G'$ is mapped by $f$ into the Cartan subalgebra $H \subset G$. We note that if $\phi$ is a representation of $G$ then $\phi \circ f$ is a representation of $G'$. \\

We now consider the root spaces $H_{0}^{*'}$ and $H_{0}^{*}$ and given the mapping $f$, we define the mapping $f^{*}: H_{0}^{*} \rightarrow H_{0}^{*'} $ via:

\begin{equation}
f^{*} \circ \rho = \rho \circ f
\end{equation}

where $\rho \in H_{0}^{*}$. This essentially means that if $h' \in H'$, then:

\begin{equation}
(f^{*} \circ \rho) (h') = \rho(f(h'))
\end{equation}

Thinking of $G'$ to be already within $G$ allows us to view that the mapping $f:H' \rightarrow H$ just maps $H'$ onto itself as the identity. We remember that there exists a one-to-one correspondence between the elements of the root space $\rho \in H_{0}^{*}$ and the elements $h_{\rho}$ of the Cartan subalgebra. This injective mapping connects to $H'$ a space that we consider as $H_{0}^{*'}$. A subsequent construction which involves the decomposition of $H_{0}^{*}$ as a sum of $H_{0}^{*'}$ and a space $H^{*}$ orthogonal to it, allows us to write $f^{*} \circ \rho = \rho_{1}$, where $\rho \in H_{0}^{*}$ and $\rho_{1} \in H_{0}^{*'}$. \\

We then let $M$ be a weight of a representation $\phi$ of $G$:

\begin{equation}
\phi(h) \zeta_{M} = M(h) \zeta_{M}
\end{equation}

If $h' \in H' \subset H$, then:

\begin{equation}
\phi(f(h)) \zeta_{M} = M(f(h)) \zeta_{M} = f^{*} \circ M(h) \zeta_{M}
\end{equation}

It is then sufficiently clear that if $M$ is a weight of $\phi$ then $f^{*} \circ M$ is a weight of the representation $\phi \circ f$ of $G'$. To put a slightly more pictorial note to it, the weights of the representation of the subalgebra are obtained by projecting the weights of the algebra from $H_{0}^{*}$ onto $H_{0}^{*'}$. Finding these mappings $f^{*}$ for regular subalgebras is what we call the \textbf{branching rules} of the algebra.

\section{Casimir Operator and Freudenthal's Formula}

We emphasize on the fact that $SU(2)$ analysis is facilitated by the existence of a "Casimir" operator $T^{2} = T_{x}^{2} + T_{y}^{2} + T_{z}^{2}$ which commutes with all the generators $T_{x},T_{y},T_{z}$ individually. We note that $T^{2}$ makes sense only for representations and not as an actual element of the Lie algebra. We now attempt to generalize the $T^{2}$ operator for an arbitrary simple Lie algebra. For this, we use the well-known decomposition:
\begin{equation}
T^{2} = \frac{1}{2} (T_{+}T_{-} + T_{-}T_{+}) + T_{z}T_{z}
\end{equation}
We suspect the generalization to be of the form:
\begin{equation}
C = \sum_{j,k}^{} H_{j}M_{jk}H_{k} + \sum_{\alpha \neq 0} E_{\alpha}E_{-\alpha}
\end{equation}
where $H_{j} = H_{\alpha_{j}}$ and the $\alpha_{j}$ form a basis of simple roots. The matrix $M$ is then obtained by enforcing that $C$ commute with all the generators of the algebra. The $e_{\alpha}$ normalizations are chosen such that:
\begin{equation}
(e_{\alpha}, e_{\beta}) = \delta_{\alpha,-\beta}
\end{equation}
which allow us to obtain the following commutation relations:
\begin{equation}
[e_{\alpha},e_{-\alpha}] = h_{\alpha}
\end{equation}
\begin{equation}
[E_{\alpha},E_{-\alpha}] = H_{\alpha}
\end{equation}
$C$ is then computed explicitly, as is shown in Cahn (pgs 85 - 87). We demonstrate the computation of $C$ for $SU(2)$ as an example.


\vspace{1mm}

The standard commutators are
\begin{equation}
[t_{z},t_{\pm}] = \pm t_{\pm} ; [t_{+},t_{-}] = 2t_{z}
\end{equation}
We find
\begin{equation}
(t_{+}, t_{-}) = Tr (ad t_{+} \cdot ad t_{-}) = 4
\end{equation}
The correct normalization is then obtained by setting
\begin{equation}
t_{+}^{'} = \frac{1}{2}t_{+} ;\qquad t_{-}^{'} = \frac{1}{2}t_{-}
\end{equation}
which ensures $(t_{+}^{'}, t_{-}^{'}) = 1\,.$ 

We then let $e_{\alpha} = t_{+}^{'}$. The corresponding $h_{\alpha}$ is then
\begin{equation}
h_{\alpha} = [t_{+}^{'}, t_{-}^{'}] = \frac{1}{2} t_{z}
\end{equation}
We then compute
\begin{equation}
\langle \alpha, \alpha \rangle = (h_{\alpha}, h_{\alpha}) = \frac{1}{4} (t_{z},t_{z}) = \frac{1}{2}
\end{equation}
It is then clear that the matrix $M = A^{-1}$ (which is $1 \times 1$ for this case) is just $2$ (note that the $A$ defined here is different from the $A$ defined earlier). We then obtain:
\begin{equation}
C = 2H_{\alpha} H_{\alpha} + E_{\alpha}E_{-\alpha} + E_{-\alpha}E_{\alpha} = \frac{1}{2} T_{z}T_{z} + \frac{1}{4} (T_{+}T_{-} + T_{-}T_{+})
\end{equation}

We stress on the importance of the Casimir operator as it commutes with all the generators, including the raising and lowering operators. Hence, it has the same value on every vector of an irrep because every vector can be obtained by making lowering operators act on the highest weight vector. The value of the Casimir operator on an irrep can be obtained by considering its action on the highest weight vector. 

\vspace{1mm}

We suppose that the highest weight is $\Lambda$ and $\phi_{\Lambda}$ is a vector which has this weight. Now, for every positive root $\alpha$ we know that $E_{\alpha} \phi_{\Lambda} = 0$ (it would otherwise have weight $\Lambda + \alpha$, contradicting the supposition that $\Lambda$ is the highest weight). We then compute
\begin{equation}
E_{\alpha}E_{-\alpha}\phi_{\Lambda} = (E_{-\alpha}E_{\alpha} + H_{\alpha}) \phi_{\Lambda} = \Lambda(h_{\alpha}) \phi_{\Lambda} = \langle \Lambda, \alpha \rangle \phi_{\Lambda}
\end{equation}
if $\alpha$ is positive. Then, following Cahn (pages 88 and 89), we find:
\begin{equation}
C = \sum_{j,k}^{} \langle \Lambda ,\alpha_{j} \rangle A_{jk}^{-1} \langle \Lambda , \alpha_{k} \rangle + \sum_{\alpha > 0}^{} \langle \Lambda, \alpha \rangle
\end{equation}
\begin{equation}
= \langle \Lambda, \Lambda \rangle + \langle \Lambda, 2\delta \rangle 
\end{equation}
on this irrep, where we have defined an element $\delta$ of $H_{0}^{*}$ as
%
\begin{equation}
    \delta = \frac{1}{2} \sum_{\alpha > 0}^{} \alpha 
\end{equation}

We now employ the Casimir operator to obtain the Freudenthal recursion formula, for the dimensionality of a weight space. (The procedure and results along with an example for $SU(3)$  are highlighted in Cahn, pgs 91 - 96; we attempt to illustrate an example for $SU(2)$ that highlights a correspondence which yields the correct normalization.) 

The idea is to essentially consider an irrep with highest weight $\Lambda$ and attempt to obtain the dimensionality of the space with weight $M$. We know what the value of $C$ is on the entire carrier space of the representation (a constant). Subsequently, we calculate its trace, when restricted to the space with weight $M$:

\begin{equation}
Tr_{M} C = n_{M} \langle \Lambda, \Lambda + 2 \delta \rangle
\end{equation}

For $SU(2)$, the normalization we employ is
\begin{equation}
(e_{\alpha}, e_{-\alpha}) = 1
\end{equation}
so that
\begin{equation}
[E_{\alpha}, E_{-\alpha}] = H_{\alpha}, \qquad [H_{\alpha}, E_{\alpha}] = \langle \alpha, \alpha \rangle E_{\alpha}
\end{equation}
The commutation relations between the generators of $SU(2)$ are:
\begin{equation}
[T_{+}, T_{-}] = 2T_{z}, [T_{z},T_{\pm}] = \pm T_{\pm}
\end{equation}
and the Casimir operator is written down as:
\begin{equation}
T^{2} = T_{z}^{2} + \frac{1}{2} [T_{+}T_{-} + T_{-}T_{+}]
\end{equation}

Now, to consider the $SU(2)$ generated by $E_{\alpha}, E_{-\alpha}, H_{\alpha}$, we use the following correspondence for the correct normalization:
\begin{equation}
T_{z} = \frac{H_{\alpha}}{\langle \alpha, \alpha \rangle},\quad T_{+} = \sqrt{\frac{2}{\langle \alpha, \alpha \rangle}} E_{\alpha},\quad T_{-} = \sqrt{\frac{2}{\langle \alpha, \alpha \rangle}} E_{-\alpha}
\end{equation}

From the $SU(2)$ example that we illustrated above, we make use of the correspondence (for the correct normalization) and consider the weight space associated with the weight $M$. We realize that the full irrep contains more than one irrep of $SU(2)$ associated with the root $\alpha$. We exercise choice and pick a basis for the weight space for $M$ such that each basis vector belongs to a distinct $SU(2)$ irrep. Each such irrep is then characterized by an integer or a half-integer, $t$ which happens to be $T_{z}$'s maximal eigenvalue. The Casimir operator then has corresponding eigenvalues $t(t+1)$. Now, if $\phi_{t}$ is an appropriate weight vector, we write
\begin{equation}
\left[\frac{H_{\alpha} H_{\alpha}}{{\langle \alpha, \alpha \rangle}^{2}} + \frac{E_{\alpha}E_{-\alpha}}{\langle \alpha, \alpha \rangle} + \frac{E_{-\alpha}E_{\alpha}}{\langle \alpha, \alpha \rangle} \right] \phi_{t} = t(t+1) \phi_{t}
\end{equation}
so that 
\begin{equation}
[E_{\alpha}E_{-\alpha} + E_{-\alpha}E_{\alpha}] \phi_{t} = \langle \alpha, \alpha \rangle t(t+1) \phi_{t} - \frac{{\langle M, \alpha \rangle}^{2}}{\langle \alpha, \alpha \rangle} \phi_{t} 
\end{equation}

We make use of the fact that $\phi_{t}$ has weight $M$. The weight vector $\phi_{t}$ belongs to a series of weight vectors which form a basis for an irrep of $SU(2)$ (as demonstrated in the example above). We now assume the highest weight in this series to be $M + k\alpha$ and the associated weight vector to be $\phi_{M + k\alpha}$. Then we get
\begin{equation}
T_{z} \phi_{M + k\alpha} = t \phi_{M + k\alpha} = \frac{H_{\alpha}}{\langle \alpha, \alpha \rangle} \phi_{M + k\alpha} = \frac{\langle \alpha, M + k\alpha \rangle}{\langle \alpha, \alpha \rangle} \phi_{M + k\alpha}
\end{equation}
We then find
\begin{equation}
t = \frac{\langle \alpha, M + k\alpha \rangle}{\langle \alpha, \alpha \rangle}
\end{equation}
Inserting this in Eqn(7.29), we obtain
\begin{equation}
[E_{\alpha}E_{-\alpha} + E_{-\alpha}E_{\alpha}] \phi_{t} = [k(k+1) \langle \alpha, \alpha \rangle + (2k + 1) \langle M, \alpha \rangle] \phi_{t}
\end{equation}

Following Cahn (pgs 93 - 94), we obtain
\begin{equation}
Tr_{M} C = n_{M} \langle M, M \rangle + \sum_{\alpha > 0}^{} \left [n_{M} \langle M, \alpha \rangle + \sum_{k = 1}^{\infty} 2 n_{M + k\alpha} \langle M + k\alpha, \alpha \rangle \right]
\end{equation}
This is Freudenthal's dimensional formula.

From this, the dimension $n_{M}$ is obtained (in terms of higher weights) as:

\begin{equation}
n_{M} = \frac{\sum_{\alpha > 0} \sum_{k = 1}^{\infty} 2 n_{M + k\alpha} \langle M + k\alpha, \alpha \rangle}{\langle \Lambda + M + 2\delta, \Lambda - M \rangle}
\end{equation}

We note that the highest weight always has a one-dimensional space. Freudenthal's formula can be then used to determine the dimensionality of the spaces of lower weights. The denominator is most simply evaluated via expressing the first factor by its Dynkin coefficients and the second factor in terms of simple roots. If the Dynkin coefficients of $\Lambda + M + 2\delta$ are:

\begin{equation}
(\Lambda + M + 2\delta) = (a_{1}, a_{2}, \dots)
\end{equation}

and 

\begin{equation}
\Lambda - M = \sum_{i} k_{i} \alpha_{i} 
\end{equation}

Then, we get:

\begin{equation}
\langle \Lambda + M + 2\delta, \Lambda - M \rangle = \sum_{i} a_{i} k_{i} \frac{1}{2} \langle \alpha_{i}, \alpha_{i} \rangle 
\end{equation}

The numerator can also be evaluated with a decent amount of ease. For a particular given positive root $\alpha$, we employ a check to ascertain if $M' = M + k\alpha$ is a weight as well. If $M'$ is also a weight, then $M'$ and $M$ lie in a linear string of weights separated by the $\alpha$'s. Letting the highest and lowest weights in the string be $M' + p\alpha$ and $M' - m\alpha$, respectively, we find (using a previous result):

\begin{equation}
2 \langle M', \alpha \rangle = (m - p) \langle \alpha, \alpha \rangle 
\end{equation}

\section{Using Mathematica packages for Group Theory Calculations}

We have developed the theory of compact Lie groups in the last few chapters. Now, we will demonstrate some packages that a user can safely use as black boxes for illustration purposes and research.
\texttt{GroupMath} is such a package for the demonstration of properties of Lie groups. As we saw, the Cartan matrix contains all the information about the group. This package identifies a group to its own Cartan matrix.  
\begin{lstlisting}[style = in]
    << GroupMath`
    SU3 
\end{lstlisting}
\begin{lstlisting}[style = out]
    {{2, -1},{-1, 2}}
\end{lstlisting}
\begin{lstlisting}[style = in]
    SO5 
\end{lstlisting}
\begin{lstlisting}[style = out]
    {{2, -2},{-1, 2}}
\end{lstlisting}
There also exists an inverse function that takes input as a Cartan matrix and gives out the name of the group:
\begin{lstlisting}[style = in]
    CMtoName[{{2}}]
\end{lstlisting}
\begin{lstlisting}[style = out]
    SU2
\end{lstlisting}
\begin{lstlisting}[style = in]
    CMtoName[{{2, -1, 0, 0}, {-1, 2, -1, 0}, {0, -1, 2, -1}, {0, 0, -1, 2}}]
\end{lstlisting}
\begin{lstlisting}[style = out]
    SU5
\end{lstlisting}
The function \texttt{RepName} takes output as the group name and the representation weight and gives output as the representation name: 
\begin{lstlisting}[style = in]
    RepName[SU3, {1, 0}]
    RepName[SU3, {0, 1}]
    RepName[SU3, {1, 1}]
    RepName[SU3, {2, 0}]
\end{lstlisting}   
\begin{lstlisting}[style = out]
    3
    $\Bar{3}$
    8
    $\Bar{6}$
\end{lstlisting}
For discussion on the convention of bar representation, refer to the book by Cahn that is mentioned in the references.\par
We can get the Dynkin coefficients for the adjoint representation as follows: 
\begin{lstlisting}[style = in]
    Adjoint[SO10]
\end{lstlisting}
\begin{lstlisting}[style = out]
    {0, 1, 0, 0, 0}
\end{lstlisting}
We can extract all the positive roots with the following useful function: 
\begin{lstlisting}[style = in]
    PositiveRoots[G2]
    PositiveRoots[SU3]
\end{lstlisting}
\begin{lstlisting}[style = out]
    {{2, -3}, {-1, 2}, {1, -1}, {0, 1}, {-1, 3}, {1, 0}}
    {{2, -1}, {-1, 2}, {1, 1}}
\end{lstlisting}
as we already saw. The function \texttt{DimR} gives the dimension of a representation of a lie group using the Weyl character formula: 
\begin{lstlisting}[style = in]
    DimR[SU3, {2, 0}]
    DimR[SU3, {1, 1}]
    DimR[SO10, {0, 0, 0, 0, 2}]
    DimR[SO10, {0, 0, 0, 0, 1}]
\end{lstlisting}
\begin{lstlisting}[style = out]
    6
    8
    126
    16
\end{lstlisting}
For more such commands, we recommend to the reader the original document by Fonseca that is available on arXiv \citep{GroupMath}. There is another excellent package named \texttt{LieArt}. We have uploaded some notebooks demonstrating its use in Group theory on the GitHub page. Please refer to those notebooks and the original documents by \cite{LieArt1, LieArt2}. There is another package, Affine, \citep{Affine} which we have not used to demonstrate anything because it fulfills the same purpose as the other two packages. 

\part{Non-Compact Groups}
\chapter{Towards Non-compactness: Euclidean Group in 2 Dimensions}

The physical world, being homogeneous and isotropic, translation, and rotation are symmetries of the Euclidean world for trivial reasons. It consists of two transformations: translation along the direction $\hat{b}$ for a distance $b$, written as $\mathbf{T(b)}$, and rotation around an axis $\hat{n}$ with an angle $\theta$, written as $\mathbf{R_n(\theta)}$. These two combined are the Euclidean group of $n$-dimensions $E_n$. It acts upon $n$-dimensional Euclidean space $R^n$, which has all the points with coordinates $\{x^i: i = 1,2,\cdots, n\}$. Length of a vector with endpoints $\mathbf{x}$ and $\mathbf{y}$ is $l = [\sum_{i}(x^i-y^i)^2]^{1/2}$. \par
Now, a general linear transformation takes the form $\mathbf{x}\rightarrow\mathbf{x'}$ with $x'^i=\tensor{R}{^i_j}x^j+b^i$. It is trivial to show that for this metric to be \textit{length-preserving}, the tensor $\tensor{R}{^i_j}$ has to be \textbf{Orthogonal}.\par
In $2$-dimensional space, these transformations can be parameterized as: 
\begin{align}
    x'^1 &= x^1\cos{\theta}-x^2\sin{\theta}+b^1\\
    x'^2 &= x^1\sin{\theta}+x^2\cos{\theta}+b^2
\end{align}
Denote this transformation as $g(\mathbf{b},\theta)$, where $\mathbf{b}=(b^1,b^2)$. Then:
\begin{equation}
    g(\mathbf{b},\theta)=\begin{pmatrix}
        \cos{\theta} & -\sin{\theta} & b^1\\
        \sin{\theta} & \cos{\theta} & b^2\\
        0 & 0 & 1
    \end{pmatrix}
\end{equation}
If $\mathbf{J}$ is the generator of rotation and $P_i$ is the generator of translation along $i$-th direction, $R(\theta)=e^{-i \theta J}$ and $T(\mathbf{b})=e^{-i\mathbf{b}\cdot \mathbf{P}}$. It can be observed from the expression for $g(\mathbf{b},\theta)$ that
\begin{align}
    J=\begin{pmatrix}
        0 & -i & 0\\
        i & 0 & 0\\
        0 & 0 & 0
    \end{pmatrix}
    \qquad
    P_1 = \begin{pmatrix}
        0 & 0 & i\\
        0 & 0 & 0\\
        0 & 0 & 0
    \end{pmatrix}
    \qquad
    P_2= \begin{pmatrix}
        0 & 0 & 0\\
        0 & 0 & i\\
        0 & 0 & 0
    \end{pmatrix}
\end{align}
In this representation, each point $\mathbf{x}$ is a $3$-component vector $(x^1, x^2, 1)$. Also, the composition rule for group elements is
\begin{equation}
    g(b_1, \theta_1)g(b_1, \theta_1)=g(b_3, \theta_3)
\end{equation}
where $\theta_3=\theta_1+\theta_2$ and $b_3=R(\theta_2)b_1+b_2$. Also, $R(\theta)=g(0, \theta)$ and $T(\mathbf{b})=g(\mathbf{b},0)$. Now, 
\begin{align}
    g(\mathbf{b},\theta)R(\theta)^{-1}&=g(\mathbf{b},\theta)g(0,-\theta)\\
    & = g(\mathbf{b},0)\\
    & = T(\mathbf{b})
\end{align}
Hence, $g(\mathbf{b}, \theta)=T(\mathbf{b})R(\theta)$. Also, 
\begin{align}
    [P_1, P_2] &= 0\\
    [J, P_k] &= i \varepsilon^{km}P_m
\end{align}
where $\tensor{\varepsilon}{^k_m}$ is $2 \times 2$ anti-symmetric tensor $\begin{pmatrix}
    0 & -1 \\
    1 & 0
\end{pmatrix}$. \par
Using this, we can see that $\{P_k\}$ transforms as a vector operator, i.e., 
\begin{equation}
    e^{-i \theta J}P_k e^{i \theta J}=P_m(\tensor{\delta}{^m_k}+ \theta \tensor{\varepsilon}{^m_k})=P_m \tensor{R(\theta)}{^m_k}
\end{equation}
Also, \begin{equation}
    e^{-i \theta J}\mathbf{P} \cdot \mathbf{b} e^{i \theta J}= P_m \tensor{R(\theta)}{^m_k}b^k=\mathbf{P} \cdot \mathbf{b'}
\end{equation}
Hence, \begin{equation}
    e^{-i \theta J}T(\mathbf{b}) e^{i \theta J}= T [R(\theta)\mathbf{b}]
\end{equation}
It can now be easily seen that $\{T(\mathbf{b})\}$ form an invariant subgroup $T_2$ of $E_2$, and the factor group $E_2/T_2$ is isomorphic to $SO(2)$. 
\begin{align}
    g(\mathbf{b}, \theta)T(\mathbf{a})g(\mathbf{b}, \theta)^{-1} &= T(b)R(\theta)T(a)R(-\theta)T(-\mathbf{b})\\
    &= T(R(\theta)\mathbf{a})
\end{align}
$SO(2)$ are labeled by an integer $m = 0, \pm 1, \pm 2$ as they are all one-dimensional representations, and the homomorphism goes like $R(\theta) \rightarrow e^{-I m \theta}$. So, the 'degenerate' representation of $E_2$ induced by the factor group will be $U_m(\mathbf{b}, \theta)=e^{-i m \theta}$. \par
Now, in order to form a general unitary representation of $E_2$, we define the raising and lowering operators as was done for $SU(3)$ before. 
\begin{equation}
    P_{\pm}=P_1 \pm i P_2
\end{equation}
They commute with each other and $[J, P_{\pm}]= \pm P_{\pm}$. Define the Casimir operator as $P^2 = P_{+}P_{-}$. It can be checked that it commutes with all the generators of the group (i.e., it is indeed a Casimir operator). Hence, the value of $P^2$ is unique for an irreducible representation. We will denote the representation space of $E_2$ by the direct sum of one-dimensional subspaces labeled by eigenvalue $m$ of $J$ and the eigenvalue $p=p^2$ of $P^2$: 
\begin{align}
    P^2 \vert p, m \rangle &= p^2 \vert p, m \rangle\\
    J \vert p, m \rangle &= m \vert p, m \rangle
\end{align}
We know that $P_{\pm}\vert p, m \rangle$ is eigenstate of J, with eigenvalue $(m \pm 1)$. 
\begin{align}
    \langle p, m \vert P_{\pm}^{\dagger}P_{\pm}\vert p, m \rangle &= \langle p. m \vert P^2 \vert p, m \rangle\\
    &= p^2 \langle p. m \vert p, m \rangle\\
    &=p^2
\end{align}
When $p^2 = 0$, it boils down to the above representation induced by the factor group as the representation becomes one-dimensional. But for $p^2 > 0$, define: 
\begin{equation}
    \vert p, m \pm 1 \rangle \equiv (\pm \frac{i}{p}) P_{\pm}\vert p, m \rangle
\end{equation}
Hence, in a particular representation, the matrix elements for the generators will become:
\begin{align}
    \langle p, m' \vert J \vert p, m \rangle &= m \tensor{\delta}{^{m'}_m}\\
    \langle p, m' \vert P_{\pm} \vert p, m \rangle &= \mp ip \tensor{\delta}{^{m'}_{m\pm 1}}
\end{align}
So, 
\begin{equation}
    \langle p, m' \vert e^{-i \theta J}\vert p. m \rangle = e^{-i m \theta} \tensor{\delta}{^{m'}_m}
\end{equation}
Also, $T(b, \theta)= R(\theta)T(b, 0) R(\theta)^{-1}$. Hence, it is sufficient to consider translations along only $x$-direction.
\begin{equation}
    \langle p, m' \vert T(\mathbf{b}) \vert p, m \rangle = e^{i(m-m')\theta}\langle p, m' \vert T(b, 0) \vert p, m \rangle
\end{equation}
If $b$ is aligned to x-direction (which we can assume without loss of any generality), $T(b, 0)=e^{-i b P_1}= e^{-i b (P_+ + P_-)/2}=\sum_{k, l}(iP_+)^k (-iP_-)^l(-1)^k\frac{(b/2)^{k+l}}{k! l!}$. Between the states, $\vert p, m \rangle$ and $\langle p, m' \vert$, only those terms with $k-l = m'-m$ will contribute. Hence, we replace $n=k+l$ and the expression becomes: 
\begin{align}
    \langle p, m' \vert T(b, 0) \vert p, m \rangle &= \sum_{n} (-1)^{(n+m'-1)/2)} \frac{(pb/2)^n}{\frac{n+m'-m}{2}! \frac{n-m'+m}{2}!}\\
    &= J_{m-m'}(p b)
\end{align}
where $J_{m-m'}(p b)$ is the Bessel function of the first kind. 
\begin{equation}
    \langle p, m' \vert T(b, \theta) \vert p, m \rangle = e^{i(m-m')\theta}J_{m-m'}(p b)
\end{equation}
This basis ${\vert p, m \rangle}, m=0, \pm 1, \cdots$ is called angular momentum basis. By applying $P_{\pm}$ successively, we can get the basis with any integer value of $m$, starting from any initial integer. So, the representation is infinite-dimensional.

\chapter{Rotation Group in 3-Dimensions}
The $SO(3)$ group consists of all continuous linear transformations in $3$- dimensions that preserve the length of coordinate vectors. Consider coordinate vectors $\hat{e_i}$ for $i = 1,2,3$. Under rotation, 
\begin{equation}
    \hat{e_i}\rightarrow \hat{e'_i}=\tensor{R}{^j_i}\hat{e_j}
\end{equation}
The preservation of length $(x_ix^i = x'_i x'^i)$ yields $R^TR=RR^T=\mathbb{I}$. Hence, $det(R)=\pm 1$. Now, as all physical rotations can be reached from identity by applying rotation matrices, the determinant should be $+1$. Hence, another condition on $R$ is $det(R)=1$. These two conditions can be written as follows: 
\begin{align}
    \tensor{R}{^i_k}\tensor{R}{^j_l}\tensor{\delta}{^k^l} &= \tensor{\delta}{^i^j}\\
    \tensor{R}{^i_l}\tensor{R}{^j_m}\tensor{R}{^k_n}\tensor{\varepsilon}{^l^m^n} &= \tensor{\varepsilon}{^i^j^k}
\end{align}
Hence, rotation matrices keeps the two tensors $\tensor{\delta}{^k^l}$ and $\tensor{\varepsilon}{^i^j^k}$ invariant. \par
A general SO(3) group element depends on three continuous group parameters. One of them is using Euler angles. It is well known that 
\begin{equation}
    R(\alpha, \beta, \gamma) = R_3(\alpha)R_2(\beta)R_3(\gamma)
\end{equation}
i.e., every rotation can be decomposed into a product of simple rotations around two fixed axes. Hence, to construct a matrix for any general rotation, we need expressions for $R_2$ and $R_3$, which, by intuitions from the last chapter, can be directly written as: 
\begin{align}
    R_3(\theta) &= \begin{pmatrix}
        \cos{\theta} & -\sin{\theta} & 0\\
        \sin{\theta} & \cos{\theta} & 0\\
        0 & 0 & 1
    \end{pmatrix}\\
    R_2(\theta) &= \begin{pmatrix}
        \cos{\theta} & 0 & \sin{\theta}\\
        0 & 1 & 0\\
        -\sin{\theta} & 0 & \cos{\theta}\\
        \end{pmatrix}
\end{align}
There is another parametrization for rotation, called Angle-and-Axis parametrization, where the angle around the axis of rotation is denoted by $\psi$, and the orientation of the axis is given by two angles (polar and azimuthal) $(\theta, \phi)$. Here $0 \leq \psi, \theta \leq \pi$, $0 \leq \phi \leq 2 \pi$. 
\begin{exercise}
    Think about a group formed by all rotation around a particular axis $\hat{n}$, which forms a subgroup of SO(3) and is isomorphic to $SO(2)$. These generators are the same $J_n$ from the last chapter $((J_k)^l_m=-i \tensor{\varepsilon}{_k_l_m})$. This can be related to Angle-and-Axis parametrisation as $R_n(\psi)=e^{-i \psi J_k n^k}$.  
\end{exercise}
\begin{exercise}
    Find the exact relations between $(\alpha, \beta, \gamma)$ and $(\psi, \theta, \phi)$.
\end{exercise}
\begin{exercise}
    If a rotation is designated  by $R_{\hat{n}}(\psi)$, realise that $R_{-\hat{n}}(\psi)=R_{\hat{n}}(-\psi)$. Hence, there are two classes of closed curves on the SO(3) group manifold. 
\end{exercise}

If a physical system is represented by a Hamiltonian H and it is invariant under rotations, $[H, R_n(\psi)]=0$ for all $\hat{n}$ and $\psi$. This is equivalent to the condition $[H, J_k]= 0$ for $k=1,2,3$. This means physical quantities corresponding to the generators are conserved quantities, which, in this case, is the angular momentum.  

To form an irreducible representation of SO(3), like before, we define the Casimir operator as $J^2 = \sum_{i=1}^3 J_i^2$. It commutes with all the generators of the group, and from Schur's lemma, it is a multiple of the identity operator of any irreducible representation. In other words, all vectors in a particular representation are eigenvectors of $J^2$ with the same eigenvalue. Define the raising and lowering operators as $J_{\pm}=J_1 +i J_2$. It satisfies $[J_3, J_{\pm}]=\pm J_{\pm}$, $[J_+, J_-]=2J_3$. It can be seen from the exercise above that the group manifold is compact. So, there must be a particular l for which
\begin{equation}
    0 = \langle l \vert J_+ J_- \vert l \rangle = j(j+1)-l(l-1)
\end{equation}
Here, $j$ denotes the particular representation where the eigenvalue for $J^2$ for all the vectors is $j(j+1)$. Hence, we get $l = -j$. Similarly, we can see that the highest value of $l$ will be $j$. Hence, there are $2j+1$ states. In a particular representation denoted by $j$, we have the following matrix elements: 
\begin{align}
    J^2 \vert j, m \rangle &= j(j+1)\vert j, m \rangle\\
    J_3 \vert j, m \rangle &= m \vert j, m \rangle\\
    J_{\pm} \vert j, m \rangle &= (j(j+1)-m(m \pm 1))^{1/2}\vert j, m \pm 1 \rangle
\end{align}
This is realized in Mathematica in Section 4.1. In a particular representation, write $R(\alpha, \beta, \gamma)= U(\alpha, \beta, \gamma)$, then
\begin{equation}
    U(\alpha, \beta, \gamma) \vert j, m \rangle = D^j (\alpha, \beta, \gamma)^{m'}_m \vert j, m' \rangle
\end{equation}
As from the Euler angle parametrisation, we can see that $R(\alpha, \beta, \gamma)= e^{-i \alpha J_3}e^{-i \beta J_2} e^{-i \gamma J_3}$, we can write
\begin{equation}
     \tensor{D^j (\alpha, \beta, \gamma)}{^{m'}_m} = e^{-i \alpha m'}\tensor{d^j(\beta)}{^{m'}_m} e^{-i \gamma m}
\end{equation}
where $\tensor{d^j(\beta)}{^{m'}_m} = \langle j, m' \vert e^{-i \beta J_2} \vert j, m \rangle$. In the universally adopted convention, $J_2$ is taken to be imaginary antisymmetric, resulting $\tensor{d^j(\beta)}{^{m'}_m}$ to be real and orthogonal. 
\begin{example}
    For $j = \frac{1}{2}$, $d^{1/2}(\beta)=e^{-i \beta \sigma_2 /2}=\begin{pmatrix}
        \cos{(\beta/2)} & -\sin{(\beta/2)}\\
        \sin{(\beta/2)} & \cos{(\beta/2)}
    \end{pmatrix}$
    Hence, \begin{equation}
        D^{1/2}(\alpha, \beta, \gamma)= \begin{pmatrix}
        e^{-i (\alpha/2)}\cos{(\beta/2)}e^{-i (\gamma/2)} & -e^{-i (\alpha/2)}\sin{(\beta/2)}e^{i (\gamma/2)}\\
        e^{i (\alpha/2)}\sin{(\beta/2)}e^{-i (\gamma/2)} & e^{i (\alpha/2)}\cos{(\beta/2)}e^{i (\gamma/2)}
    \end{pmatrix}
    \end{equation}
\end{example}
\begin{exercise}
    Using $R_{n'}(\psi)=R R_{n} R^{-1}$ where R is an arbitrary rotation and $\hat{n'}$ is the unit vector obtained from $\hat{n}$ by the rotation R, show that $D[R_n(2 \pi)]= -\mathbb{I}$, i.e., $SU(2)$ is a double cover of $SO(3)$. 
\end{exercise}
As of now, we have realized that $SO(3)$ can be written as $2 \times 2$ unitary matrix with unit determinant $D^{1/2}(\alpha, \beta, \gamma)$; it is $SU(2)$. A particular convenient form to write $SU(2)$ is $A = \begin{pmatrix}
    r_0 - i r_3 & -r_2 - i r_1 \\
    r_2 - i r_1 & r_0 + i r_3
\end{pmatrix}$
with the constraint $\sum_{i=0}^3 r_i^2 = 1$. Hence, in a 3-dimensional Euclidean space, it is simply the surface of a unit sphere, which is both compact and simply connected. Now, associate every coordinate vector $\mathbf{x}=(x^1, x^2, x^3)$ to a $2 \times 2$ traceless hermitian matrix $X = \sum_{i=1}^3 \sigma_i x^i$, it is seen that $\det(X)= \vert \mathbf{x} \vert ^2$. Now, if A is an arbitrary SU(2) matrix, linear transformation on X will be $X \rightarrow X' = A X A^{-1}$. As a result, $X'$ will also be hermitian and traceless, and $\det(X)= \det(X')$. Physically, it is exactly the rotation in $3$-dimensions or $SO(3)$. This mapping from $SU(2)$ to $SO(3)$ is two-to-one as $\pm A$ corresponds to the same transformation.

\chapter{Non-compact Group: Lorentz and Poincare}
Following up from the rotation group and Euclidean group that we studied in the previous two chapters, we will now discuss their `analogues' in 4-dimensions, the Lorentz and Poincare groups respectively, which describe rotations, velocity boosts and translations in $1+3$-dimensional spacetime, and are widely used in the subject of special theory of relativity. We will be working with the metric $\tensor{g}{_{\mu}_{\nu}}=diag(-1,1,1,1)$. 

Homogeneous Lorentz transformation $\Lambda$ is the continuous linear transformation: 
\begin{align}
    e_{\mu} \rightarrow e'_{\mu} &= e_{\nu} \tensor{\Lambda}{^{\nu}_{\mu}}\\
    x^{\mu} \rightarrow x'^{\mu} &= \tensor{\Lambda}{^{\mu}_{\nu}}x^{\nu}
\end{align}
which preserves the length of the 4-vector. Similar to the arguments in the last chapter, we can write the constraints on $\Lambda$ in terms of invariant tensors as follows:
\begin{align}
    \tensor{\Lambda}{^{\mu}_{\lambda}}\tensor{\Lambda}{^{\nu}_{\sigma}}\tensor{g}{^{\lambda}^{\sigma}} &= \tensor{g}{^{\mu}^{\nu}}\\
    \tensor{\Lambda}{^{\mu}_{\alpha}}\tensor{\Lambda}{^{\nu}_{\beta}}\tensor{\Lambda}{^{\lambda}_{\gamma}}\tensor{\Lambda}{^{\sigma}_{\delta}}\tensor{\varepsilon}{^{\alpha}^{\beta}^{\gamma}^{\delta}} &= \tensor{\varepsilon}{^{\mu}^{\nu}^{\lambda}^{\sigma}}
\end{align}
Setting $\mu = \nu =0$ in the first equation, we get $(\tensor{\Lambda}{^0_0})^2 \geq 1$, implying $\tensor{\Lambda}{^0_0} \geq 1$ or $\tensor{\Lambda}{^0_0} \leq -1$, denoting two disjoint regions. Since we take $\tensor{\Lambda}{^0_0}=1$ for identity transformation, and any transformation can be reached from identity transformation, we take $\tensor{\Lambda}{^0_0} \geq 1$. This is called \textit{Proper} Lorentz transformation. The matrices for rotation will be just the rotation matrix as it is, with an additional $g_{00}=1$ and $g_{0i}=0$. The Lorentz boosts along different space directions will look like 
\begin{equation}
\tensor{(L_1)}{^{\mu}_{\nu}}= \begin{pmatrix}
    \cosh{\zeta} &  \sinh{\zeta} & 0 & 0\\
    \sinh{\zeta} &  \cosh{\zeta} & 0 & 0\\
    0 & 0 & 1 & 0\\
    0 & 0 & 0 & 1
\end{pmatrix}
\end{equation}
Here $\tanh{\zeta} = \frac{v}{c}$, where v is the velocity of the transformed frame along the $X$-axis. As $\zeta \in (-\infty, \infty)$, the Lorentz group is non-compact, as opposed to the rotation group.

Similar to the relation between SO(3) and SU(2), there is a relation between the Lorentz group and SL(2). We associate each spacetime point $x^{\mu}$ with a $2 \times 2$ hermitian matrix $X$ as 
\begin{equation}
    x^{\mu}\rightarrow X = \sigma_{\mu}x^{\mu} = \begin{pmatrix}
        x^0+x^3 & -x^1 -i x^2\\
        -x^1 + i x^2 & x^0 - x^3
    \end{pmatrix}
\end{equation}
A vector $x^{\mu}$, if transformed by a Lorentz transformation $\Lambda$, the corresponding matrix will transform like $X' = A X A^{\dagger}$, where the determinant remains the same. It means $det(A)det(A^{\dagger})=\vert det(A) \vert^2=1$. It is customary to choose $det(A)=1$, i.e., the $SL(2, \mathbb{C})$ group.

When we add translation to the story (just like getting the Euclidean group from the rotation group), we get the Poincar\'e group. 
\begin{equation}
    x'^{\mu}= \tensor{\Lambda}{^{\mu}_{\nu}}x^{\nu}+b^{\mu}
\end{equation}
The composition rule can also be written from the analogy of Euclidean group: $x''=\Lambda'x'+b'=\Lambda'(\Lambda x +b)+b' = (\Lambda' \Lambda)x+(\Lambda'b+ b')$. A general element of Poincar\'e transformation can be written as $g(b, \Lambda)= T(b) \Lambda$. Translations again form an invariant subgroup of the Poincar\'e group. \par
The generators of translation are $P_{\mu}$. Contravariant translation generators are defined as $P^{\mu}= \tensor{g}{^{\mu}^{\nu}}P_\nu\,$. The generator of time translation $P^0$ is the Hamiltonian. Generators of Lorentz transformation are $\tensor{J}{_{\mu}_{\nu}}$. Corresponding contravariant generator is defined as $\tensor{J}{^{\mu}^{\nu}}= \tensor{g}{^{\mu}^{\lambda}}\tensor{J}{_{\lambda}_{\sigma}}\tensor{g}{^{\sigma}^{\nu}}$. If the generator of rotation in the mn plane is $J_{mn}$ and the generator of rotation around the k axis is $J_k$, where $(k, m, n)$ is some cyclic permutation of (1, 2, 3), we get: 
\begin{equation}
    J_k = \frac{1}{2}\tensor{\varepsilon}{^k^m^n}\tensor{J}{_m_n} \qquad J_{mn} = \tensor{\varepsilon}{^m^n^k}J_k
\end{equation}
From previous chapters, the matrix form for the generators of rotation can be trivially extended to 4 dimensions. For boost, from the matrix for $L_1$, it can be clearly seen that the generator for boost along $x$ direction will be: 
\begin{equation}
    K_1 = \begin{pmatrix}
        0 & i & 0 & 0 \\
        i & 0 & 0 & 0\\
        0 & 0 & 0 & 0\\
        0 & 0 & 0 & 0
    \end{pmatrix}
\end{equation}
The generators of boosts along the three axes are given by $K_m\equiv J_{m0}$.

\begin{exercise}
    With the same technology, write the generators of Poincare transformation as 5-dimensional matrices and write the translation generators. 
\end{exercise}
\begin{exercise}
    Check the following commutation relations among the matrices found in the previous question using Mathematica:
    \begin{enumerate}
        \item $[P_{\mu}, P_{\nu}]=0$
        \item $[P^0, J_n] = 0$
        \item $[P_m, J_n]=i \tensor{\varepsilon}{^m^n^l}P_l$
        \item $[P_m, K_n]=i \delta_{mn}P^0$
        \item $[P^0, K_n]= i P_n$
        \item $[J_m, J_n]=i \tensor{\varepsilon}{^m^n^l}J_l$
        \item $[K_m, J_n]= i \tensor{\varepsilon}{^m^n^l}K_l$
        \item $[K_m, K_n]=-i \tensor{\varepsilon}{^m^n^l}J_l$
    \end{enumerate}
\end{exercise}
Lie algebra of the Lorentz group can be reduced to the direct product of two subalgebras as follows: 
\begin{align}
    M_i &= \frac{J_i + i K_i}{2}\\
    N_i &= \frac{J_i - i K_i}{2}
\end{align}
Then, it is straightforward to check that: 
\begin{align}
    [M_i, M_j] &= i \tensor{\varepsilon}{^i^j^k}M_k\\
    [N_i, N_j] &= i \tensor{\varepsilon}{^i^j^k}N_k\\
    [M_i, N_j] &= 0
\end{align}
for $i, j = 1,2,3$. This algebra is referred to as $SU(2)_M \times SU(2)_N$. Though $SU(2) \times  SU(2)$ is compact, Lorentz is not. The former involves exponentiation of ${iM_i, iN_i}$ whereas the later involves ${iJ_i, iK_i}$. Although an irreducible representation of one would be an irreducible representation of the other, some property (e.g., Unitarity) may be violated. For unitarity, we need generators to be hermitian. Because there is an $i$ in the definition of M and N, the two sets of generators (M, N) and (J, K) cannot simultaneously be hermitian. This non-unitary nature of the finite-dimensional representation of the Lorentz group, thus, cannot correspond to any physical state. \textit{But they definitely have very crucial physical applications in physics.} \par
If we denote the two SU(2)'s with u and v, with Casimir operator eigenvalues as u(u+1) and v(v+1), respectively, Then the two-dimensional representations would be $(\frac{1}{2}, 0)$ and $(0, \frac{1}{2})$. These are the SL(2) matrices that we have already talked about. If one of them is written as ${A}$, another one would be ${A^*}$. These two are called \textbf{Fundamental Representations}. \par
The 4-dimensional representation would be $(\frac{1}{2}, \frac{1}{2})$. The coordinate vector $X^{\mu}$, 4-momentum $P^{\mu}$, Electromagnetic 4-potential $A^{\mu}$ etc transform as this representation. \par
A second-rank tensor $\tensor{t}{^{\mu}^{\nu}}$ in Minkowski space can be decomposed into 3 parts. 6 independent anti-symmetric components $t^{[\mu, \nu]}$ transforms as $(1, 0) \oplus (0, 1)$. The electromagnetic field tensor $F^{\mu \nu}$ is this one. Then there are the 10 symmetric components, which can be broken into 9 independent traceless components and a trace. The former transforms as a $(1, 1)$ representation whereas the later transforms as the $(0, 0)$ representation. The energy-momentum tensor is an example of a symmetric second-rank tensor. \par
We will not talk about the unitary representations of the Lorentz group in this letter. There are many available resources that an interested reader may refer to. Some of them are mentioned in the bibliography. \par
Two historical papers can be mentioned in this regard- one by \cite{Dirac1945} and another by \cite{HarishChandra1947}.  

The readers are advised to visit the notebooks \texttt{Lorentz\_4D.nb} for a review of the 4-dimensional representation of the Lorentz group discussed earlier in this chapter, and \texttt{Lorentz\_Irreps.nb} for a demonstration on building any finite irrep of the group. The readers are also encouraged to attempt these on the Poincare group on their own. (See \texttt{Poincare.nb} for some hints on this.)

\section*{Exercises}
\textit{Hints for * marked questions are given in the notebook \texttt{Exercise 2.nb}}
\begin{enumerate}
    \item (*) Consider $2 \times 2$ traceless matrices $\begin{pmatrix}
        a & b\\
        c & -a
    \end{pmatrix}$, where all entries are real. Show that they are closed under the commutator bracket. 

    \item (*) Real Heisenberg group is defined as $H = \begin{pmatrix}
        1 & a & b \\
        0 & 1 & c \\
        0 & 0 & 1
    \end{pmatrix}$ where all entries are real.    
    \begin{itemize}
        \item Show that matrices of the above form are a group under matrix multiplication. 
        \item Find the tangent space at the identity of the Heisenberg group. 
        \item Find a basis for tangent space of $H$. 
        \item Is the tangent space closed under the matrix bracket? 
    \end{itemize}

    \item (*) We will look into the group $SL(2, \mathbb{R})$ using the basis $E=\begin{pmatrix}
        0 & 1\\
        0 & 0
    \end{pmatrix}$, $H=\begin{pmatrix}
        1 & 0 \\
        0 & -1
    \end{pmatrix}$, $F=\begin{pmatrix}
        0 & 0\\
        1 & 0
    \end{pmatrix}$. 
    \begin{itemize}
        \item Find the commutator brackets among the basis elements. \par

        For a matrix $GL(n, \mathbb{R})$ for integer n, define a function:
        \begin{equation*}
            ad(M)(X) = [A,X] = AX-XA
        \end{equation*}
        \item Assume A is in tangent space of $SL(2, \mathbb{R})$, say $L(G)$. Explain why $ad(A) : L(G) \rightarrow L(G)$. 
        \item Find $\mathcal{E}=ad(E), \mathcal{H}=ad(H), \mathcal{F}=ad(F)$. Those will be a $3 \times 3$ matrices. 
    \end{itemize}

    \item Repeat the same formalism of last question to find $e^{\mathcal{E}}, e^{\mathcal{H}}, e^{\mathcal{F}}$. 

    \item (*) Find $e^{tA}$ for the following: 
    \begin{equation*}
        A = \begin{pmatrix}
            i & 0\\
            0 & 0
        \end{pmatrix} \qquad
        A = \begin{pmatrix}
            0 & i\\
            0 & 0
        \end{pmatrix}
    \end{equation*}

    \item Consider $A_j = i \sigma_j$ for $j=1,2,3$. $\sigma$'s are Pauli matrices. Find $e^{t A_j}$ and show that they are in SU(2).

    \item (*)Do the same analysis that we did for SU(3) in previous chapters for $SL(3, \mathbb{C})$. 

    \item Repeat the same analysis for $SL(2, \mathbb{C})$. 
\end{enumerate}



\bibliography{grp_th}

\begin{thebibliography}{12}
\providecommand{\natexlab}[1]{#1}
\providecommand{\url}[1]{\texttt{#1}}
\expandafter\ifx\csname urlstyle\endcsname\relax
  \providecommand{\doi}[1]{doi: #1}\else
  \providecommand{\doi}{doi: \begingroup \urlstyle{rm}\Url}\fi

\bibitem[Cacciatori and Cerchiai(2009)]{Cacciatori2009}
Sergio~L. Cacciatori and Bianca~L. Cerchiai.
\newblock Exceptional groups, symmetric spaces and applications, 2009.

\bibitem[Cahn(2014)]{Cahn}
R.N. Cahn.
\newblock \emph{Semi-Simple Lie Algebras and Their Representations}.
\newblock Dover Publications, 2014.
\newblock ISBN 9780486150314.

\bibitem[Dirac(1945)]{Dirac1945}
P.~A.~M. Dirac.
\newblock Unitary representations of the lorentz group.
\newblock \emph{Proceedings of the Royal Society of London. Series A, Mathematical and Physical Sciences}, 183\penalty0 (994):\penalty0 284--295, 1945.
\newblock ISSN 00804630.
\newblock URL \url{http://www.jstor.org/stable/97721}.

\bibitem[{Feger} and {Kephart}(2012)]{LieArt1}
Robert {Feger} and Thomas~W. {Kephart}.
\newblock {LieART -- A Mathematica Application for Lie Algebras and Representation Theory}.
\newblock \emph{arXiv e-prints}, art. arXiv:1206.6379, June 2012.
\newblock \doi{10.48550/arXiv.1206.6379}.

\bibitem[{Feger} et~al.(2020){Feger}, {Kephart}, and {Saskowski}]{LieArt2}
Robert {Feger}, Thomas~W. {Kephart}, and Robert~J. {Saskowski}.
\newblock {LieART 2.0 - A Mathematica application for Lie Algebras and Representation Theory}.
\newblock \emph{Computer Physics Communications}, 257:\penalty0 107490, December 2020.
\newblock \doi{10.1016/j.cpc.2020.107490}.

\bibitem[{Fonseca}(2021)]{GroupMath}
Renato~M. {Fonseca}.
\newblock {GroupMath: A Mathematica package for group theory calculations}.
\newblock \emph{Computer Physics Communications}, 267:\penalty0 108085, October 2021.
\newblock \doi{10.1016/j.cpc.2021.108085}.

\bibitem[Harish-Chandra(1947)]{HarishChandra1947}
Harish-Chandra.
\newblock Infinite irreducible representations of the lorentz group.
\newblock \emph{Proceedings of the Royal Society of London. Series A, Mathematical and Physical Sciences}, 189\penalty0 (1018):\penalty0 372--401, 1947.
\newblock ISSN 00804630.
\newblock URL \url{http://www.jstor.org/stable/97833}.

\bibitem[Nazarov(2012)]{Affine}
Anton Nazarov.
\newblock Affine.m—mathematica package for computations in representation theory of finite-dimensional and affine lie algebras.
\newblock \emph{Computer Physics Communications}, 183\penalty0 (11):\penalty0 2480–2493, November 2012.
\newblock ISSN 0010-4655.
\newblock \doi{10.1016/j.cpc.2012.06.014}.
\newblock URL \url{http://dx.doi.org/10.1016/j.cpc.2012.06.014}.

\bibitem[Pollatsek and of~America(2009)]{pollatsek2009lie}
H.S.K. Pollatsek and Mathematical~Association of~America.
\newblock \emph{Lie Groups: A Problem Oriented Introduction Via Matrix Groups}.
\newblock MAA textbooks. Mathematical Association of America, 2009.
\newblock ISBN 9780883857595.

\bibitem[Ramadevi and Dubey(2019)]{Ramadevi}
P.~Ramadevi and V.~Dubey.
\newblock \emph{Group Theory for Physicists: With Applications}.
\newblock Cambridge University Press, 2019.
\newblock ISBN 9781108429474.

\bibitem[Ramond(2010)]{Ramond}
P.~Ramond.
\newblock \emph{Group Theory: A Physicist's Survey}.
\newblock Cambridge University Press, 2010.
\newblock ISBN 9781139489645.

\bibitem[Tung(1985)]{Tung}
W.K. Tung.
\newblock \emph{Group Theory in Physics}.
\newblock G - Reference, Information and Interdisciplinary Subjects Series. World Scientific, 1985.
\newblock ISBN 9789971966560.

\end{thebibliography}

\end{document}